\documentclass[preprint,3p,12pt]{elsarticle} 
\biboptions{sort&compress}

\usepackage{amssymb}
\usepackage{amsmath}
\usepackage{bm}
\usepackage{subcaption}
\usepackage{graphicx}
\usepackage{latexsym}
\usepackage{hyperref}
\usepackage{cleveref}
\usepackage{booktabs}
\usepackage{color}
\usepackage{longtable}

\begin{document}

\begin{frontmatter}

\title{An immersed boundary method for the discrete velocity model of the Boltzmann equation}

\author[pku]{Longqing Ge}
\author[pku]{Qingdong Cai}
\author[imech,ucas]{Yonghao Zhang}
\author[imech,ucas]{Tianbai Xiao\corref{cor}}

\cortext[cor]{Corresponding author}

\address[pku]{Department of Mechanics, School of Mechanics and Engineering Science, Peking University, Beijing, China}
\address[imech]{Centre for Interdisciplinary Research in Fluids, Institute of Mechanics, Chinese Academy of Sciences, Beijing, China}
\address[ucas]{School of Engineering Science, University of Chinese Academy of Sciences, Beijing, China}

\begin{abstract}
Computational modeling and simulation of fluid-structure interactions constitute a fundamental cornerstone for advancing aerospace engineering endeavors.
This paper addresses the notion and implementation of the immersed boundary method for the discrete velocity model of the Boltzmann equation.
The method incorporates the Maxwell gas–surface interaction model into the construction of ghost-cell particle distribution functions, facilitating meticulous characterization of velocity slip and temperature jump effects within a Cartesian grid framework, which ultimately achieves accurate prediction of aerodynamic parameters.
This study presents two principal advancements.
First, an upwind-weighted compact interpolation strategy is developed in physical space, which ensures numerical stability and robustness for arbitrary geometries without relying on large stencils or normal-direction projections.
Second, a cut-cell correction methodology is proposed in velocity space to address the degradation of quadrature accuracy caused by surface discontinuities.
The resulting framework is equally applicable to both two- and three-dimensional problems without requiring any dimension-specific modifications.
Rigorous analysis is provided to prove that the approach maintains second-order accuracy across both physical and velocity space, while ensuring robust numerical stability.
Comprehensive numerical experiments demonstrate that the solution algorithm achieves the designed accuracy and delivers precise predictions comparable to body-conformal solvers, while retaining the simplicity, flexibility, and scalability of the Cartesian grid method.
The proposed approach provides a unified and physically consistent immersed boundary framework for simulating dynamic interactions between non-equilibrium flows and structural components across a wide range of flow regimes.
The open-source codes to reproduce the numerical results are available under the MIT license \footnote{\url{https://github.com/CFDML/IBDVM-Paper}}.
\end{abstract}

\begin{keyword}
computational fluid dynamics, kinetic theory, gas kinetic scheme, immersed boundary method, gas-surface interaction
\end{keyword}

\end{frontmatter}

\begin{longtable}{p{3.2cm} p{11.5cm}}
\caption{Nomenclature.}\label{table:nomenclature}\\
\hline
\textbf{Symbol} & \textbf{Description} \\
\hline
\endfirsthead

\hline
\textbf{Symbol} & \textbf{Description} \\
\hline
\endhead

\hline
\multicolumn{2}{r}{\small Continued on next page}\\
\endfoot

\hline
\endlastfoot

IBM & immersed boundary method \\
DVM & discrete velocity method \\
$t$ & time variable \\
$\bm x$ & space variable \\
$\bm \xi$ & velocity variable (with its magnitude denoted as $\xi$) \\
$\bm c$ & peculiar velocity (with its magnitude denoted as $c$) \\
$f$ & particle distribution function \\
$\mathcal Q$ & collision operator of kinetic equation \\
$g$ & Maxwellian distribution function \\
$g^S$ & Shakhov equilibrium distribution function \\
$\tau$ & relaxation time \\
$\mu$ & dynamic viscosity \\
$p$ & pressure \\
$k$ & Boltzmann constant \\
$\rho$ & density \\
$n$ & number density \\
$\mathbf U$ & macroscopic velocity \\
$T$ & temperature \\
$E$ & energy \\
$\psi$ & collision invariants \\
$\bm W$ & macroscopic conservative variables $(\rho,\rho \mathbf U,\rho E)$ \\
$\bm q$ & heat flux \\
$N_V$ & number of collocation points \\
$\omega$ & quadrature weight \\
$\phi$ & arbitrary integrable function of $\bm\xi$ \\
$\mathbf \Omega$ & computational domain \\
$V$ & volume of domain \\
$\mathbf F^f$ & numerical flux for distribution function \\
$S$ & area of cell face \\
$\bm n$ & outward normal vector of surface \\
$\Theta$ & Heaviside step function \\
$\boldsymbol\sigma$ & spatial slope of distribution function \\
$f_{\mathrm w}^*$ & surface distribution function considering wall effects \\
$f_{\mathrm w}^i$ & incident particle distribution \\
$f_{\mathrm w}^r$ & reflected particle distribution \\
$\alpha$ & accommodation coefficient \\
$g_{\mathrm w}$ & surface Maxwellian distribution \\
$\rho_{\mathrm w}$ & density in $g_{\mathrm w}$ \\
$\bm U_{\mathrm w}$ & preconfigured wall velocity \\
$T_{\mathrm w}$ & preconfigured wall temperature \\
$\xi_n$ & projected particle velocity normal to the boundary \\
gc, dc, ac & ghost cell, donor cell, auxiliary cell \\
$N_S$ & number of surface cells \\
c.c. & cut cell \\
$\mathcal C$ & set of the cut cells’ indices \\
$\mathcal G$ & subset of $\mathcal C$ with $\xi_n\le0$ \\
$\mathcal S$ & subset of $\mathcal C$ with $\xi_n>0$ \\
$\lambda$ & particle mean free path \\
$R_{\mathrm w}$ & boundary curvature radius \\
$\theta$ & proportion of particle velocity exhibiting discontinuity \\
$\left[f\right]$ & jump of distribution function \\
$E^*$ & quadrature error \\
$h,b$ & reduced distribution functions \\
$H,B$ & reduced Maxwellian distributions \\
$H^S,B^S$ & reduced equilibrium distributions \\
$L$ & characteristic length \\
$\gamma$ & specific heat ratio \\
$C$ & dimensionless sound speed \\
$C_p$ & pressure coefficient \\

\end{longtable}

\newpage

\section{Introduction}

Gaseous flows demonstrate distinct behavioral patterns across different regimes.
The Knudsen number, defined as the ratio of molecular mean free path to a characteristic length scale, serves as a fundamental parameter for distinguishing flow regimes based on the degree of dilution.
Continuum mechanics, e.g., the Euler and Navier-Stokes equations, is rigorously applicable solely under the condition that the Knudsen number is asymptotically small.
With the advancement of modern industries (e.g., aerospace technology and advanced chip packaging) towards extreme conditions, rarefied gas dynamics has become increasingly critical, with applications ranging from very-low-earth-orbit flight (when the mean free path is considerably large) \cite{llop2014very} to micro-electromechanical system (when the scale of interest becomes small) \cite{gad2005mems}.
Various numerical methods, e.g., stochastic particle methods \cite{bird2013dsmc} and deterministic solvers for the Boltzmann equation \cite{broadwell_study_1964}, have been developed to simulate the evolution of rarefied gases.

While numerical methods employing body-conformal meshes remain predominant in flow simulations, the Cartesian grid methodology has been attracting growing attention.
In this approach, the computational domain is discretized uniformly using structured grids aligned with Cartesian coordinates.
By avoiding cumbersome (re)meshing for specific geometries, the simulation pipeline can be significantly simplified and automated.
The regular, structured grids allow straightforward indexing, thus directly supporting higher-order stencil computations \cite{peng2009high}, optimized memory access, and load balancing.
The method is naturally compatible with convergence acceleration techniques, e.g., the multi-grid method \cite{trottenberg2001multigrid} and adaptive mesh refinement \cite{plewa_adaptive_2005}.
Among others, accurate computational modeling of the interaction between gases and immersed objects is a prerequisite for the faithful prediction of flow dynamics and aerodynamic parameters by Cartesian grid methods.

Several approaches have been proposed to model the fluid-structure interaction using the Cartesian grid method.
The cut-cell method embeds a curved boundary into the Cartesian mesh by trimming the cell that intersects the geometry into an irregular fluid cell, and computing its reduced volume, partial face areas, and centroids \cite{ingram2003developments}.
The governing equations are then applied in these new cells, and the boundary conditions can be directly enforced on the newly created cut edges.
The technique is non-intrusive to the solution algorithm and thus easier to implement.
It has been applied to particle methods (represented by the direct simulation Monte Carlo) \cite{zhang2012robust,jin2017new,plimpton2019direct} and discrete velocity methods of the Boltzmann equation \cite{dechriste_cartesian_2016} in rarefied gas dynamics.
The main challenges of the cut‐cell method arise from the tiny, irregular cells created at the boundary.
They enforce a severely restrictive time step (often requiring intricate cell‐merging \cite{seo_sharp-interface_2011} or flux‐redistribution \cite{schneiders2013accurate} stabilization), demand elaborate pre-processing (calculating geometric parameters for newly created cells), and break the uniform Cartesian stencils near the wall (resulting in lower-order or one-sided interpolations).

The immersed boundary method (IBM) provides an alternative model for solving fluid-structure coupling.
It employs a fixed Cartesian mesh throughout the solution process, thereby avoiding the additional computational complexity caused by the grid cutting.
Technically, this approach can be broadly divided into two categories.
The first category incorporates the notion of introducing a continuous force term within the governing equations to emulate the influence of boundary effects.
Since Peskin's original method \cite{peskin_flow_1972}, a series of such methods have been proposed to solve continuum flows, typically represented by (but not limited to) the direct-forcing method \cite{uhlmann2005immersed,su2007immersed,yang2009smoothing} and penalization method \cite{arquis1984conditions,angot1999penalization,kevlahan2001computation}.
This type of method is also known as the diffused interface method, as the immersed interface can be smeared across multiple meshes due to the use of the discrete delta function or mask function.
The second category of methods turns to implicitly model boundary effects by modifying discretization stencils in the solution algorithm, thus eliminating this generally undesirable feature.
Representative efforts include the immersed interface method \cite{leveque1994immersed,li2001immersed} and the ghost-cell-based method \cite{mohd1997simulations,ye1999accurate,fadlun2000combined,gibou2002second,ghias2007sharp,han2025navier}.

A significant challenge inherent to Cartesian grid-based IBM lies in the fact that the number of grids increases at a substantially faster rate with rising Reynolds numbers, compared to a structured body-conformal mesh.
This results in prohibitively high computational overheads when applying it to solve extremely high‐Reynolds, boundary‐layer-dominated flows.
In rarefied gas dynamics, the Reynolds number typically remains moderate, and the flow structure is relatively easier to resolve, making it particularly well-suited to leveraging the IBM for addressing multi‐body or moving‐geometry problems \cite{cercignani2000rarefied}.
However, research on IBM for rarefied flows remains notably scarce.
Despite considerable efforts to advance the integration of the IBM with the lattice Boltzmann method (LBM, as a simplified discrete velocity model of the Boltzmann equation) and its extended models \cite{feng2004immersed,dupuis2008immersed,wu2009implicit,zhu2011immersed,kang2011direct,favier2014lattice,tao_combined_2018}, these studies focus primarily on the implementation of no-slip boundary conditions in the continuum limit, thereby failing to predict the phenomena of velocity slip and temperature jump in rarefied flows.

In studies of the IBM for rarefied gaseous flows,
Xu et al. improved the suspending grid method and developed an immersed boundary-LBM method that accommodates slip boundaries \cite{xu2020immersed}.
Wang et al. integrated the slip model through the penalty feedback immersed boundary technique, thereby expanding the applicability of the Navier-Stokes equations in the slip flow regime \cite{wang2024immersed}.
These two methods can effectively predict moderately non-equilibrium flows with Knudsen numbers that are not excessively large.
For highly rarefied flows, Pekardan et al. presented three IBM implementations for the discrete velocity method (DVM) of the Boltzmann equation \cite{pekardan2012immersed}.
This work was a pioneering exploration, and its performance has been tested only on right-angle boundaries (and not on curvilinear geometries).
Ragta et al. developed a ghost-cell-based IBM for the unified gas-kinetic scheme (as an asymptotic-preserving DVM) \cite{ragta_unified_2017}.
This method employs a relatively large flow-field interpolation stencil, potentially incurring substantial computational overhead and presenting challenges when handling complex three-dimensional geometries.
Meanwhile, it utilizes the wall Maxwellian distribution as a supplemental condition for least-squares-based interpolation, which may impair its accuracy in solving strongly non-equilibrium flows.
A comprehensive numerical analysis evaluating both the solution accuracy and computational efficiency remains absent.

To address the limitations of current methodologies in accurately capturing non-equilibrium fluid-structure interactions, this study presents an innovative ghost-cell immersed boundary method (GCIBM) for the discrete velocity model of the Boltzmann equation.
By adjusting flow variables in ghost cells, it implicitly incorporates the Maxwell gas-surface interaction model, which enables automatic adaptation of slip velocities and temperature jumps in response to local flow conditions.
A particle-velocity-based upwind compact interpolation stencil is proposed to determine wall quantities in the gas-surface interaction model, from which the ghost-cell values are determined.
For the space-velocity coupling problem in the Boltzmann equation, a velocity-space cut-cell method is developed to model the discontinuous particle distribution function at the boundary.
The proposed interpolation scheme stays well-conditioned for arbitrary geometries.
The developed methodology demonstrates universal applicability across all flow regimes for fluid-structure interactions.
Comprehensive numerical experiments are performed to validate the proposed method.

The rest of this paper is organized as follows. 
\Cref{sec:2} concisely overviews the kinetic theory of gases and the computational framework associated with the discrete velocity method for the Boltzmann model equation.
\Cref{sec:3} provides a detailed description of the ghost-cell immersed boundary method, along with a corresponding numerical analysis of the proposed method.
\Cref{sec:4} includes numerical examples to demonstrate the performance of the current method.
The last section presents concluding remarks. 
The nomenclature is presented in Table \ref{table:nomenclature}.

















\section{Discrete Velocity Method}
\label{sec:2}
\subsection{Kinetic theory}

The discrete velocity method (DVM) is dedicated to numerically solving the Boltzmann equation.
In the absence of an external force, the Boltzmann equation for monatomic gases writes
\begin{equation}
    \frac{\partial f}{\partial t}+\bm{\xi}\cdot\nabla_{\bm x} f=\mathcal Q(f),
    \label{eq:boltzmann}
\end{equation}
where $f(t,\bm x,\bm \xi)$ denotes the particle distribution function with respect to time $t\in \Gamma \subseteq \mathbb R^+$, space $\bm x\in \mathcal D \subseteq \mathbb R^D$ and molecular velocity $\bm\xi\in\mathbb R^D$, with $D$ the dimension of the problem.
The collision term $\mathcal Q$ models binary molecular interactions with a five-fold integral operator.
Simplified model equations have been developed to reduce the computational complexity for solving the Boltzmann equation.
In the Bhatnagar-Gross-Krook (BGK) model \cite{bhatnagar1954model}, the collision integral is replaced by a relaxation operator, i.e.,
\begin{equation}
    \mathcal Q=\frac{g-f}{\tau},
\end{equation}
where $g$ refers to the Maxwellian distribution, i.e.,
\begin{equation}
    g=\frac{\rho}{(2\pi RT)^{3/2}}\exp{\left(-\frac{c^2}{2RT}\right)}.
\end{equation}
Here $\bm{c}=\bm{\xi}-\bm{U}$ is the peculiar velocity, $c$ denotes its magnitude, $\bm{U}$ is the macroscopic bulk velocity, $R$ is the specific gas constant, and $\{\rho,T\}$ denotes the macroscopic density and temperature, respectively.
The relaxation time is defined as $\tau=\mu/p$, where $\mu$ and $p$ are the dynamic viscosity and the pressure.

The BGK model establishes the fundamental framework for developing relaxation-type simplifications of the Boltzmann equation.
The inherent limitations of the model can be mitigated through the construction of extended models based on it.
As an example, to overcome the constraint of a fixed unit Prandtl number in the BGK model, the Shakhov model introduces higher-order correction terms into the equilibrium distribution function, i.e.,
\begin{equation}
    g^S=g+g^+=g\left[ 1 + (1 - \mathrm{Pr}) \frac{\bm{c} \cdot \bm{q}}{5pRT} \left( \frac{c^2}{RT} - 5 \right) \right],
\end{equation}
where $\mathrm{Pr}$ is the Prandtl number and $\bm{q}$ is the heat flux.

%

A particle distribution function is related to a unique macroscopic state.
The macroscopic conservative variables can be determined by taking moments of the distribution function over velocity space, i.e.,
\begin{equation}
    \bm{W}=\left(
    \begin{array}{c}
    \rho\\
    \rho\bm U\\
    \rho E
    \end{array}\right)=\int_{\mathbb R^D} f\psi d\bm\xi,
    \label{eq:moments1}
\end{equation}
where $\psi=(1,\bm\xi,\xi^2/2)^{\rm T}$ is a vector of collision invariants. 
The temperature is defined as
\begin{equation}
    \frac{D}{2}kT=\frac{1}{2n} \int_{\mathbb R^D} c^2 fd\bm\xi,
    \label{eq:moments2}
\end{equation}
where $k$ is the Boltzmann constant, $n$ is the number density.
The pressure and heat flux can be obtained via
\begin{equation}
        p=\frac{1}{D}\int_{\mathbb R^D} c^2fd\bm\xi, \quad 
        \bm q=\frac 12\int_{\mathbb R^D} \bm cc^2fd\bm\xi.
    \label{eq:moments3}
\end{equation}

\subsection{Solution algorithm}
\label{sec:solution algorithm}

The Shakhov model is employed to concisely elucidate the core principles of the DVM.
With a series of collocation points in a truncated velocity space, the discrete governing equation can be written as
\begin{equation}
    \frac{\partial f_k}{\partial t}+\bm{\xi}_k\cdot\nabla_{\bm x} f_k=\frac{g^S_k-f_k}{\tau},\quad k=1,2,\dots,N_V, 
    \label{eq:shakhov}
\end{equation}
where the subscript $k$ represents the index of the collocation point in the discrete velocity space, $N_V$ is the total number of collocation points, and $g_k^S$ denotes the modified equilibrium distribution function at the $k^{th}$ velocity point.
The integral over velocity space can be approximated using numerical quadrature, i.e.,
\begin{equation}
    \int_{\mathbb R^D} \phi(\bm \xi) d\bm\xi \approx \sum_{k=1}^{N_V}\omega_k \phi_k,
    \label{eq:quadrature}
\end{equation}
where $\omega_k$ is the quadrature weight of the $k^{th}$ velocity point, and $\phi$ represents an arbitrary integrable function with respect to $\bm\xi$, such as that presented from \Cref{eq:moments1} to \Cref{eq:moments3}.
Different quadrature rules can be adopted, e.g., the Newton-Cotes and Gauss-Hermite formulas \cite{davis2007methods}.

\Cref{eq:shakhov} requires further discretizations in space and time.
We first employ the finite volume method for spatial discretization.
The cell-averaged distribution function in the control volume $\Omega_j$ is defined as
\begin{equation}
    f_{j,k} = \frac{1}{V_j} \int_{\Omega_j} f_k dV,
\end{equation}
where $V_j$ denotes the volume of $\Omega_j$.
Integrating \Cref{eq:shakhov} over each control volume yields
\begin{equation}
    \frac{\partial f_{j,k}}{\partial t}=-\frac{1}{V_j}\int_{\partial \Omega_j}\bm{n}\cdot \bm{\xi}_{k}f_{\partial \Omega_j,k}dS+\frac{g_{j,k}^S-f_{j,k}}{\tau_j},
    \label{eq:half-discrete}
\end{equation}
where $\bm n$ is the outward normal vector of the surface of the control volume. 
The integrand in the above equation can be written as a numerical flux function, i.e.,
\begin{equation}
    \bm{F}^{f}_{\partial \Omega_{j},k}=\bm{\xi}_k f_{\partial \Omega_{j},k}.
\end{equation}
\Cref{eq:half-discrete} can be solved using a time discretization method, e.g., the explicit Euler and multi-stage Runge-Kutta methods \cite{xiao2021flux}.

The core task of a finite volume method is the computation of numerical fluxes. 
Taking the face located at $\bm x_{j+1/2}$, i.e., the one shared by $j$-th and $(j+1)$-th cells, as an example, the distribution function can be constructed based on particle velocity information,
\begin{equation}
\begin{aligned}
    f_{j+\frac{1}{2},k}=&\Theta\left(\xi_{n,k}\right)\left[f_{j,k}+\left(\bm x_{j+\frac{1}{2}}-\bm x_j\right)\cdot \bm\sigma_{j,k}\right] \\
    &+\left[1-\Theta\left(\xi_{n,k}\right)\right]\left[f_{j+1,k}-\left(\bm x_{j+1}-\bm x_{j+\frac{1}{2}}\right)\cdot\bm\sigma_{j+1,k}\right],
\end{aligned}
\label{eq:reconstruction}
\end{equation}
where $\Theta$ is the Heaviside step function and $\xi_{n,k}=\bm\xi_k \cdot \bm n_{j+1/2}$.
The spatial slopes $\{\boldsymbol\sigma_{j},\boldsymbol\sigma_{j+1}\}$ can be obtained using different interpolation stencils with limiters. 
As an example, a second-order interpolation with the van Leer limiter results in
\begin{equation}
\begin{aligned}
    &\sigma_{j,k}^n=\left( \mathrm{sign}(s_1) + \mathrm{sign}(s_2) \right) \frac{|s_1||s_2|}{|s_1| + |s_2|},\\
    &s_1 = \frac{(f_{j,k}^n - f_{j-1,k}^n)}{(x_j - x_{j-1})}, \quad s_2 = \frac{(f_{j+1,k}^n - f_{j,k}^n)}{(x_{j+1} - x_{j})}.
\end{aligned}
\label{eq:limiter}
\end{equation}
Taking the $x$-axis direction as an example here, the slopes along other directions can be calculated sequentially.
After the distribution function at the face is constructed, the numerical flux can thus be determined,
\begin{equation}
    \bm{F}_{j+1/2,k}^f=\bm{\xi}_k f_{j+1/2,k}.
\end{equation}


\subsection{Gas-surface interaction}
\label{sec:BC}

When solving the Boltzmann equation in the presence of boundaries, the particle distribution function at the wall can generally be formulated based on the particle velocity direction relative to the surface, i.e.,
\begin{equation}
    f_{\mathrm{w},k}^* = 
    \left\{
    \begin{array}{ll}
        f_{\mathrm{w},k}^i, \quad \bm\xi_k\cdot\bm{n}_{\rm w}\leq 0,\\
        f_{\mathrm{w},k}^r, \quad \bm\xi_k\cdot\bm{n}_{\rm w}>0,
    \end{array}
    \right. 
    \label{eq:general bc}
\end{equation}
where $f^*_{\mathrm w}$ denotes the distribution function at the surface considering wall effects,
$f_{\mathrm{w}}^i$ and $f_{\mathrm{w}}^r$ represent incident and reflected particle distributions, respectively, and $\bm n_{\rm w}$ is the outward normal vector of the boundary.

The Maxwell model is the most widely adopted gas-surface interaction model to characterize the distribution of reflected particles.
In this model, gas molecules are postulated to undergo both diffuse and specular reflections when interacting with solid boundaries.
By introducing the accommodation coefficient $\alpha$, which takes the values 1 and 0 for complete thermal accommodation and perfect momentum conservation limits, respectively, the reflected distribution function can be modeled as
\begin{equation}
    f_{{\rm w},k}^r = \alpha g_{{\rm w},k}+(1-\alpha)f_{{\rm w},k}^s, \quad \bm\xi_k\cdot\bm{n}_{\rm w}>0.
    \label{eq:Maxwell BC} 
\end{equation}

The contribution of diffuse reflections is modeled as a Maxwellian distribution at the wall, i.e.,
\begin{equation}
    g_{{\rm w},k}=\frac{\rho_{\rm w}}{\left(2 \pi R T_{\rm w} \right)^{3/2}} e^{-\frac{1}{2 R T_{\rm w}} (\bm{\xi}_k - \bm{U}_{\rm w})^2},
    \label{eq:boundary maxwell}
\end{equation}
where $\bm U_{\rm w}$ and $T_{\rm w}$ are the preconfigured macroscopic velocity and temperature of the solid wall.
The contribution of specular reflection can be obtained through
\begin{equation}
    f_{{\rm w},k}^s = f^s(\bm x_{\rm w},\bm\xi_k,t)=f(\bm x_{\rm w},\bm\xi_k',t),
\end{equation}
where $f(\bm x_{\rm w},\bm\xi_k',t)$ denotes the distribution function of the incoming flow and $\bm\xi_k'$ represents the particle velocity symmetrically mirrored about the tangential plane of the surface with respect to $\bm\xi_k$, which takes the form
\begin{equation}
    \bm\xi_k'=\bm\xi_k-2(\bm\xi\cdot \bm n_{\rm w})\bm n_{\rm w}.
\end{equation}

In this study, we restrict ourselves to cases with fully diffuse boundary conditions, i.e., $\alpha=1$. 
To apply \Cref{eq:Maxwell BC}, the density can be determined using the no penetration condition, i.e.,
%
\begin{equation}
    \int_{\xi_{n}<0}\xi_{n}g_{\rm w}d\bm{\xi}+\int_{\xi_n>0}\xi_nf_{\rm w}^i d\bm\xi=0,
    \label{eq:npc}
\end{equation}
for which the discrete form can be written as
\begin{equation}
    \sum_{\xi_{n,k}<0}\omega_k\xi_{n,k}g_{{\rm w},k}+\sum_{\xi_{n,k}>0}\omega_k\xi_{n,k}f_{{\rm w},k}^i=0,
    \label{eq:dnpc}
\end{equation}
where $f_{{\rm w},k}(t)$ denotes the distribution function of the gas at the boundary, and $\xi_{n,k}=\bm{\xi}_k \cdot \bm{n}_\mathrm{w}$ is the projected particle velocity normal to the boundary.
\section{Immersed Boundary Method}
\label{sec:3}
\subsection{Problem statement}
\label{sec:problem formulation}

The gas-surface interaction model in \Cref{eq:general bc} implies a nonlinear Dirichlet problem, where the boundary data depend nonlinearly on the solution of the distribution function.
Thus, the numerical implementation of $f_{\mathrm{w},k}$ and potential error origins comprise two principal components:
\begin{enumerate}
    \item The construction of the distribution of incident particles at the surface ($\bm\xi_k\cdot \bm n_{\rm w}<0$) requires extrapolation of the interior solution;
    \item The determination of the distribution of reflected particles at the surface ($\bm\xi_k\cdot\bm n_{\rm w}>0$) involves multiple numerical quadratures.
\end{enumerate}
The numerical errors inherent in formulating the implicit Dirichlet boundary condition are dominated by the lowest-order terms processed in the two components.

A noteworthy characteristic of the kinetic gas-surface interaction model in \Cref{eq:general bc} is the discontinuity near the wall surface.
Taking the Maxwell model in \Cref{eq:Maxwell BC} as an illustrative case, this discontinuity clearly arises from the disparity between incident and reflected particle distributions, and persists in both physical domain and particle velocity space.
Theoretical analysis and numerical investigations have been performed for surfaces of different shapes \cite{sone_discontinuity_1992,kim_formation_2011}.
The existence of the discontinuity necessitates judicious selection of interpolation schemes and weighting parameters within the boundary formulations to achieve a balance between accuracy, robustness, and flexibility.

The immersed boundary method (IBM) devised in this study furnishes meticulous, point-specific resolutions to the aforementioned challenges.
First, the main framework of the IBM method will be introduced in \Cref{sec:gc}.
The construction of the distribution function values in the ghost cells meticulously incorporates the effects of surface discontinuities.
The methodologies of upwind-weighted interpolation within physical space for the incident particle distribution, alongside the cut-cell correction in velocity space for the reflected particle distribution, will be elucidated in \Cref{sec:upwind-weighted} and \Cref{sec:CVC}, correspondingly.
The numerical analysis of the developed IBM, encompassing interpolation inaccuracies, integration discrepancies, and numerical stability, will be addressed in \Cref{sec:analysis}.

\subsection{Solution algorithm}
\subsubsection{Main framework}\label{sec:gc}

The core task of the ghost-cell immersed boundary method (GCIBM) is to integrate the gas-surface interaction model specified by \Cref{eq:general bc} into the solution algorithm through appropriate assignment of ghost-cell values.
Note that GCIBMs designed for continuum flows typically arrange image points along the normal direction of the boundary curve to interpolate ghost-cell values \cite{tseng_ghost-cell_2003,yousefzadeh_high_2019}.
Since the Maxwell gas-surface interaction model in \Cref{eq:Maxwell BC} does not explicitly incorporate normal derivatives, 
we propose a new approach for evaluating ghost-cell values.
This technique performs interpolation and construction directly along the Cartesian coordinate axes,
circumventing the need for normal projections.
It streamlines the numerical implementation and facilitates extension to three-dimensional and complex geometric topologies.

\begin{figure}[htbp!]
    \centering
    \begin{subfigure}[t]{0.55\textwidth}
        \includegraphics[width=\textwidth]{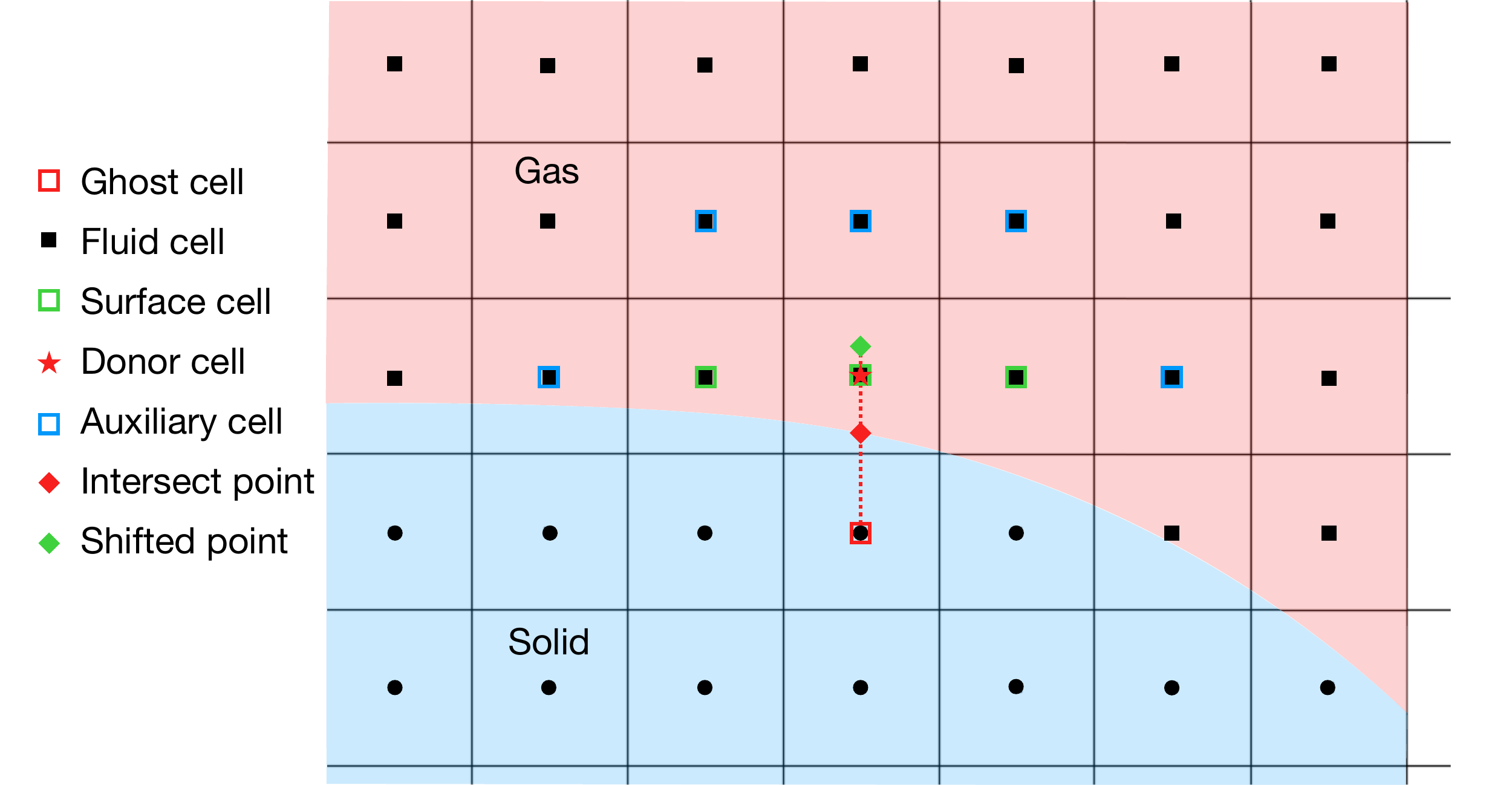}
        \caption{Cells and points involved in a solution stencil}
        \label{fig:ghost cell}
    \end{subfigure}
    \hfil
    \begin{subfigure}[t]{0.37\textwidth}
        \includegraphics[width=\textwidth]{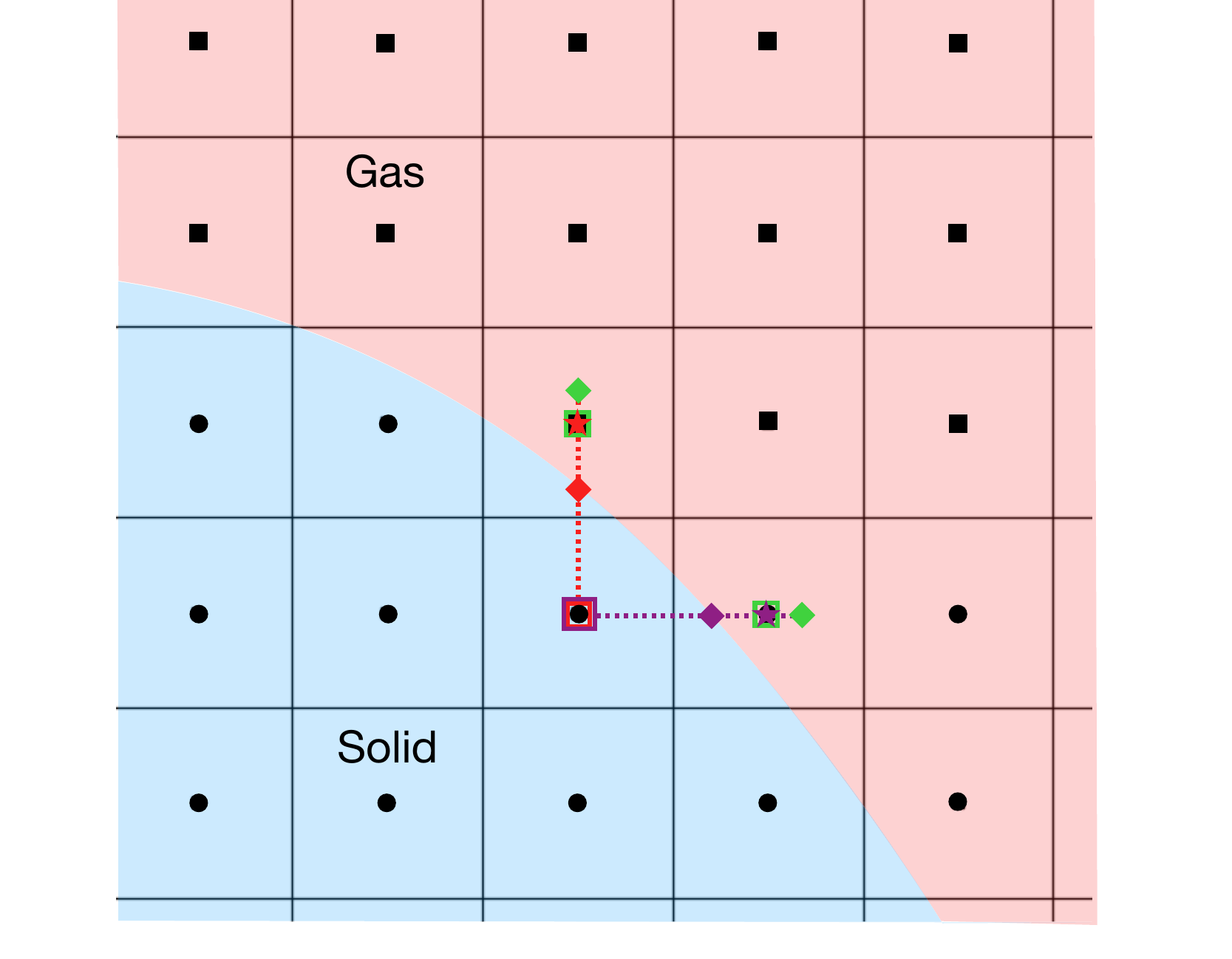}
        \caption{A special case that one ghost cell corresponds to different intersecting points}
        \label{fig:ghost cell2}
    \end{subfigure}
    \caption{Schematic of the computational module near the gas-solid surface.}
    \label{fig:ghost cell total}
\end{figure}

The computational module for achieving this objective is illustrated in \Cref{fig:ghost cell total}.
The computational mesh near the gas-solid surface is divided into five types of cells, which can be classified into
\begin{enumerate}
    \item \textbf{Ghost cell}: A cell whose center lies inside the solid region but has at least one neighboring cell, through a shared face, whose center lies in the gas region;
    \item \textbf{Fluid cell}: A cell whose center lies in the gas region;
    \item \textbf{Surface cell}: A fluid cell that directly neighbors a ghost cell, either through a face or a corner;
    \item \textbf{Donor cell}: A surface cell used directly to construct the values in the ghost cell;
    \item \textbf{Auxiliary cell}: A fluid cell for assisting in the evaluation of spatial gradients in the donor cell;
\end{enumerate}
and two types of points, i.e.,
\begin{enumerate}
    \item \textbf{Intersect point}: A point where the line connecting the center of the ghost cell to the center of its donor cell intersects the solid boundary;
    \item \textbf{Shifted point}: A point located along the extension of the line from the ghost cell center through the intersect point, positioned at a specified distance beyond the intersection.
\end{enumerate}

Using \Cref{fig:ghost cell} as an example, we describe the solution algorithm along the $y$-axis.
The same construction approach applies along the rest axes.
The method is equally applicable to both two-dimensional and three-dimensional configurations without requiring any dimension-specific adjustments.

1. \textbf{Shifted point interpolation}

It can be demonstrated that if the intersection point potentially overlaps with the center of the donor cell, i.e., the distance between these two points is extremely small, the numerical error resulting from spatial interpolation based on the donor cell center will be amplified
(see \Cref{sec:accuracy} for detailed numerical analysis).
Therefore, in each donor cell, a shifted point is defined along the line connecting the centers of the ghost and the donor cells.
In this study, the position of the shifted point is defined as
\begin{equation}
    y_{\rm sp}=y_{\rm w}+(y_{\rm dc}-y_{\rm gc})/2,
    \label{eq:sp}
\end{equation}
where $y_{\rm dc}$ and $y_{\rm gc}$ denote the center positions of the donor and ghost cells, and $y_{\rm w}$ is the location of the intersect point.

The particle distribution function at the shifted point can be approximated using a linear interpolation within the donor cell, i.e.,
\begin{equation}
    f_{{\rm sp},k}=f_{{\rm dc},k}+\sigma^y_{{\rm dc},k}(y_{\rm sp}-y_{\rm dc}).
    \label{eq:fsp}
\end{equation}
The spatial slope can be determined as
\begin{equation}
    \sigma_{\mathrm{dc},k}^y=\frac{f_{\mathrm{ac},k}-f_{\mathrm{dc},k}}{y_{\mathrm{ac}}-y_{\mathrm{dc}}},
\end{equation}
where $f_{\mathrm{ac}}$ is the distribution function in the auxiliary cell that is directly adjacent to the donor cell in the $y$-direction, and $y_{\rm ac}$ denotes its center location.

2. \textbf{Ghost cell interpolation}

Subsequently, we construct the intermediate distribution function in the ghost cell, utilizing the information from the surface cells in \Cref{fig:ghost cell}.
The interpolation formula writes
\begin{equation}
    f_{{\rm gc},k}=\sum_{j=1}^{N_S} w_{j,k}\left(f_{j,k}-\bm{\sigma}_{j,k}\cdot\bm{r}_{j}\right).
    \label{eq:extrapolation 1}
\end{equation}
where $f_{j}$ and $\bm\sigma_{j}$ denote the distribution function and its spatial gradient in the $j$-th surface cell, and $N_S$ is the total number of surface cells.
Note that no superscript $*$ is added in $f_{\rm gc}$, indicating that it is an intermediate distribution that does not account for the presence of boundaries.
The interpolation weights $w_{j,k}$ are constructed in an upwind manner based on particle velocities.
The detailed implementation of this construction will be elaborated in the following \Cref{sec:upwind-weighted}, and the corresponding numerical analysis will be conducted in \Cref{sec:stability}.

3. \textbf{Intersect point construction}

The two intermediate distribution functions constructed above are then used to interpolate the incident distribution function at the intersect point via
\begin{equation}
    f_{{\rm w},k}^i=f_{{\rm sp},k}+\frac{f_{{\rm gc},k}-f_{{\rm sp},k}}{y_{\rm gc}-y_{\rm sp}}(y_{\rm w}-y_{\rm sp}).
    \label{eq:fw}
\end{equation}
This interpolation method incorporates the combined influence of a hierarchy of surface cells.
In contrast to a naive linear extrapolation (as shown by the dashed line in \Cref{fig:ghost cell}), this approach is anticipated to mitigate the zigzag effect at Cartesian grid-represented boundaries and reduce non-physical oscillations in the solution.
A numerical verification will be presented later in \Cref{fig:cylinder slip}.

The incident distribution function in \Cref{eq:fw} will be used to determine the gas density at the wall surface $\rho_{\rm w}$ using \Cref{eq:dnpc}, and the Maxwellian distribution $g_{\rm w}$ can be computed with \Cref{eq:boundary maxwell}.
The distribution function at the surface considering wall effects $f^*_{\rm w}$ can then be constructed following \Cref{eq:general bc}.
The correction for discontinuities in the distribution function in the velocity space induced by the wall surface will be presented in \Cref{sec:CVC}.

4. \textbf{Ghost cell construction}

With all necessary information acquired, the particle distribution function in the ghost cell can be constructed as
\begin{equation}
    f_{{\rm gc},k}^*=f_{{\rm sp},k}+\frac{f_{{\rm w},k}^*-f_{{\rm sp},k}}{y_{\rm w}-y_{\rm sp}}(y_{\rm gc}-y_{\rm sp}).
    \label{eq:extrapolation 2}
\end{equation}
The derived distribution function subsequently enables the computation of the donor cell's numerical fluxes, as demonstrated in \Cref{sec:solution algorithm}.

\subsubsection{Upwind weighting in physical space}
\label{sec:upwind-weighted}

\begin{figure}[htbp!]
    \centering
    \begin{subfigure}[t]{0.49\textwidth}
        \includegraphics[width=\textwidth]{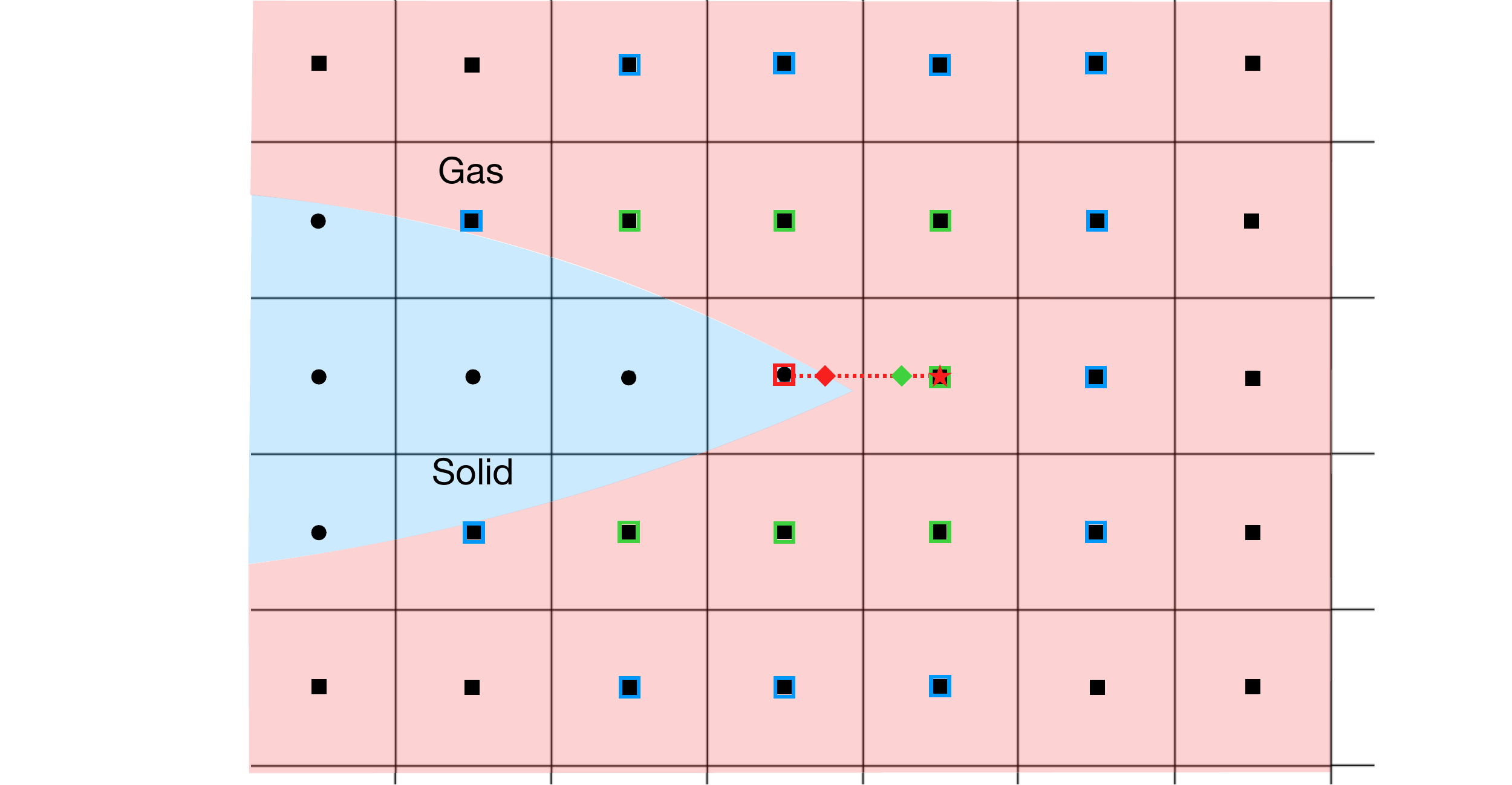}
        \caption{Extremely convex solid surface ($N_S=7$)}
        \label{fig:convex tip}
    \end{subfigure}
    \hfil
    \begin{subfigure}[t]{0.49\textwidth}
        \includegraphics[width=\textwidth]{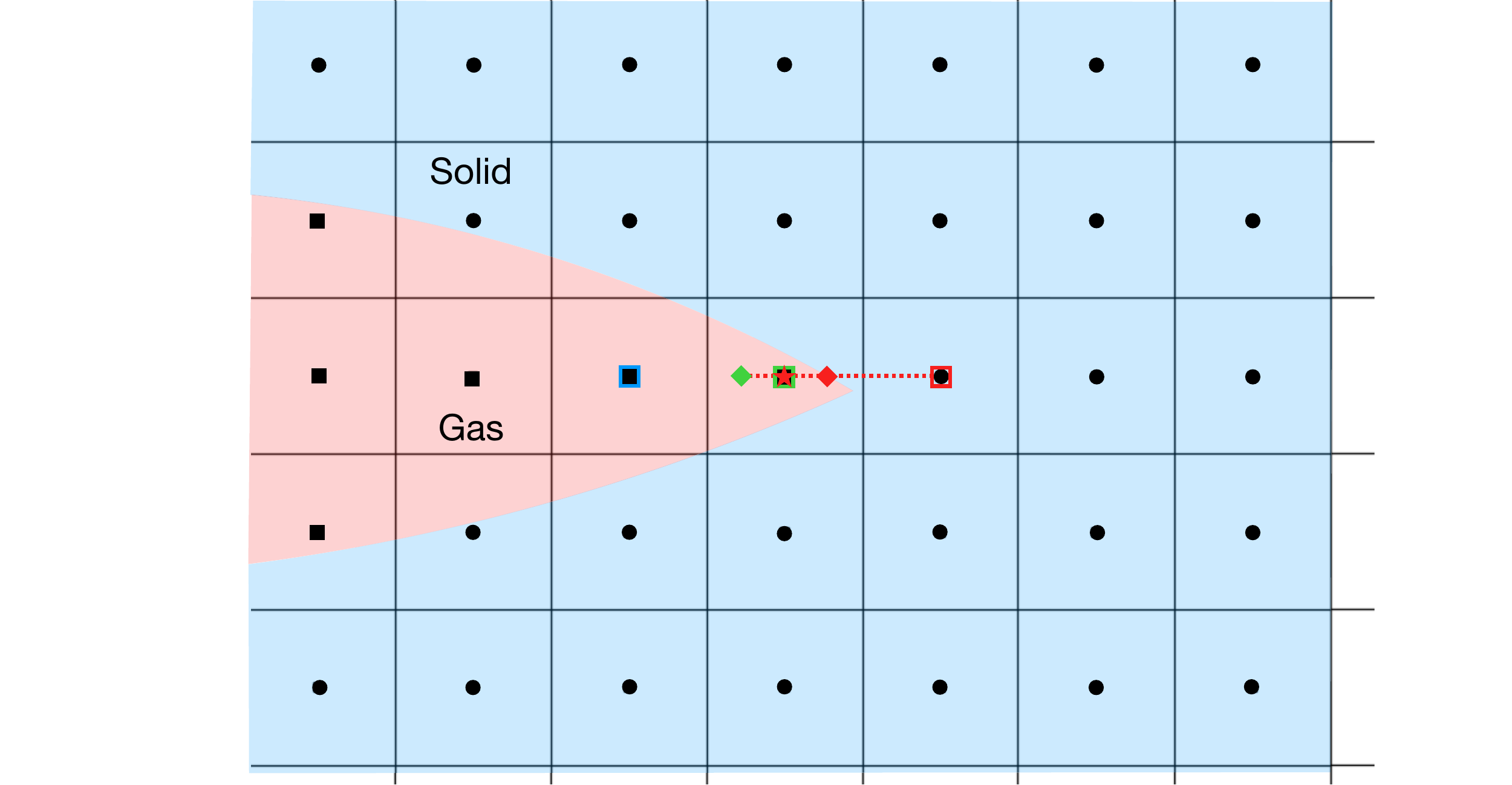}
        \caption{Extremely concave solid surface ($N_S=1$)}
        \label{fig:concave tip}
    \end{subfigure}
    \caption{Schematic of the solution stencil when $N_S$ reaches its maximum and minimum.}
    \label{fig:extreme cases}
\end{figure}

For each discrete velocity $\bm\xi_k$, the advection operator of \Cref{eq:shakhov} is hyperbolic over $(t,\bm x)$ \cite{alldredge2019regularized,xiao2023relaxnet}.
Thus, the upwind approach is selected to establish the interpolation weights in \Cref{eq:extrapolation 1} based on particle velocities, aiming to achieve enhanced numerical stability.
Specifically, the interpolation weights are determined via
\begin{equation}
    w_{j,k} = 
    \left\{
    \begin{array}{ll}
        \max (0, \bm{\xi}_k \cdot \bm{\hat r}_j)^2/\sum_{s=1}^{N_S} \max (0, \bm{\xi}_k \cdot \bm {\hat r}_s)^2, \quad \sum_{s=1}^{N_S} \max (0, \bm{\xi}_k \cdot \bm {\hat r}_s)^2 \neq 0,\\
        1/N_S, \quad \sum_{s=1}^{N_S} \max (0, \bm{\xi}_k \cdot \bm {\hat r}_s)^2 = 0.
    \end{array}
    \right. 
    \label{eq:upwind}
\end{equation}
Here, the subscript $j$ denotes the index of a surface cell, and $\bm {\hat r}_j$ is the unit vector pointing from the center of the ghost cell to $j$-th fluid cell's center. 
All the $N_S$ surface cells (highlighted with green quadrilateral markers in \Cref{fig:ghost cell}) participate in determining the weights.
\Cref{fig:extreme cases} illustrates the extreme scenarios where $N_S$ reaches its maximum ($N_S=7$ for the two-dimensional case; $N_S=25$ for the three-dimensional case) and minimum ($N_S=1$) resulting from varying curvature geometries.
Note that \Cref{eq:extrapolation 1} indicates that different surface cell numbers and interpolation weight selections should solely influence the numerical stability of the solution algorithm, without altering the second-order accuracy.

The fundamental concept embodied in the first line of \Cref{eq:upwind}  is: as the direction of $\bm\xi_k$ becomes increasingly aligned with that of $\bm r_j$, the molecules at the wall surface with the velocity $\bm\xi_k$ are more likely to be transported from this direction.
I.e., this surface cell becomes the predominant contributor to $f_{{\rm gc},k}$.
In the second line, the relation $\sum_{j=1}^N \max (0, \bm{\hat \xi}_k \cdot \bm {\hat r}_j)^2 = 0$ implies that these particles are emitted from the solid material rather than originating from the internal flow field.
Therefore, even though a direct averaging is employed here, as demonstrated in the intersect point construction in \Cref{sec:gc}, the values of the distribution function $f_{{\rm w},k}^*$ at these collocation points will subsequently be superseded by the Maxwellian distribution $g_{\mathrm{w},k}$, rendering its direct influence on the final solution negligible.

To demonstrate the effectiveness of the proposed upwind interpolation approach, a comparative study will be conducted in \Cref{sec:cylinder}, where \Cref{fig:cylinder slip} serves as empirical validation.




\subsubsection{Cut-cell correction in velocity space}
\label{sec:CVC}

In this work, the composite midpoint rule is employed for numerical quadrature.
The quadrature error is generally of second order for a sufficiently continuous integrand in \Cref{eq:quadrature}.
However, discontinuities induced by the wall boundary may potentially result in a reduction in order of accuracy.
A detailed mathematical proof will be presented in \Cref{sec:quadrature accuracy}.
To address this issue, we develop a cut-cell correction strategy in particle velocity space.


\begin{figure}[htbp]
    \centering
    \begin{subfigure}[c]{0.4\textwidth}
        \centering
        \includegraphics[width=\textwidth]{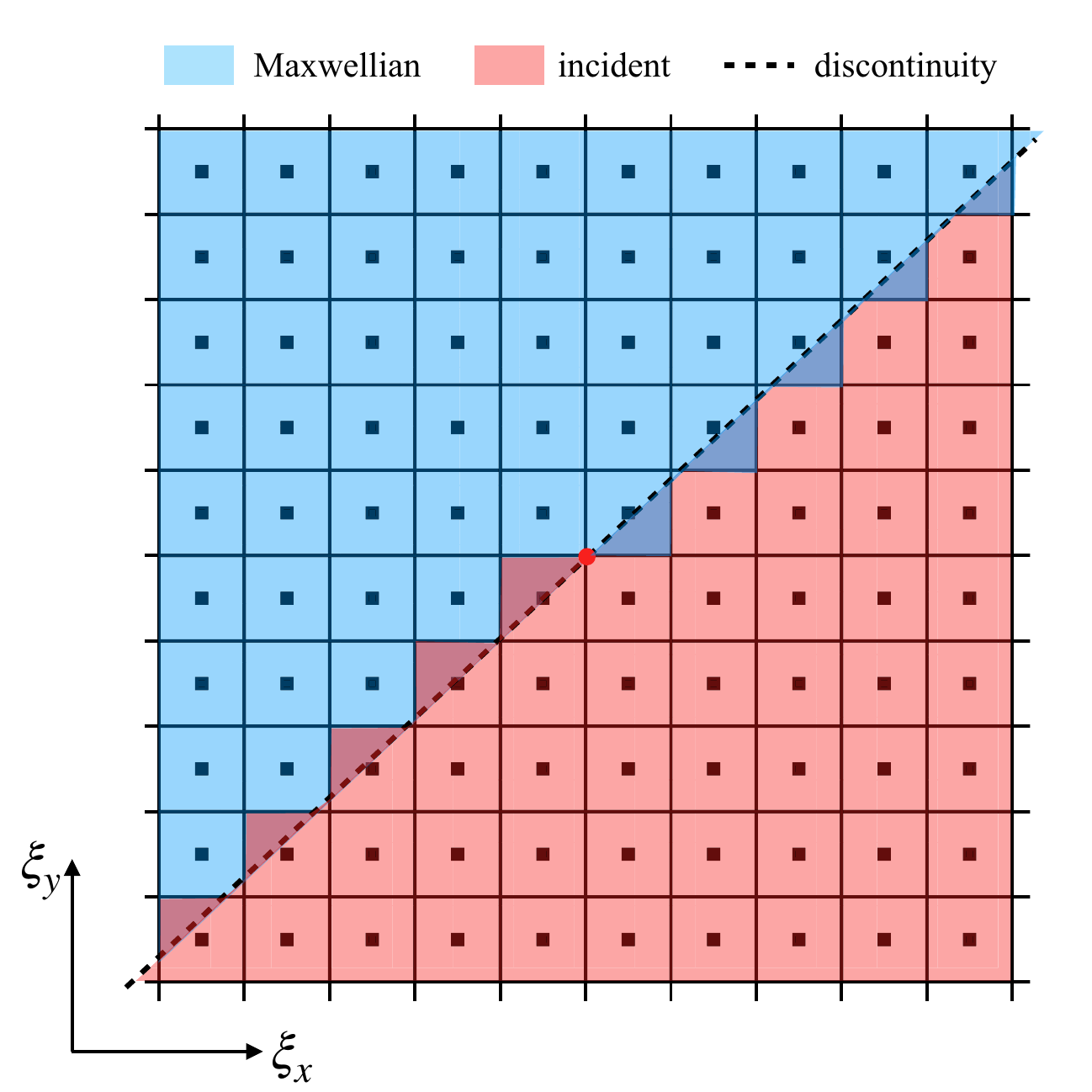}
        \subcaption{Discontinuity in the distribution function in the velocity space induced by the wall surface}\label{fig:big}
        \label{fig:VS}
    \end{subfigure}
    \hfil
    \begin{subfigure}[c]{0.4\textwidth}
        \centering
        \includegraphics[width=\textwidth]{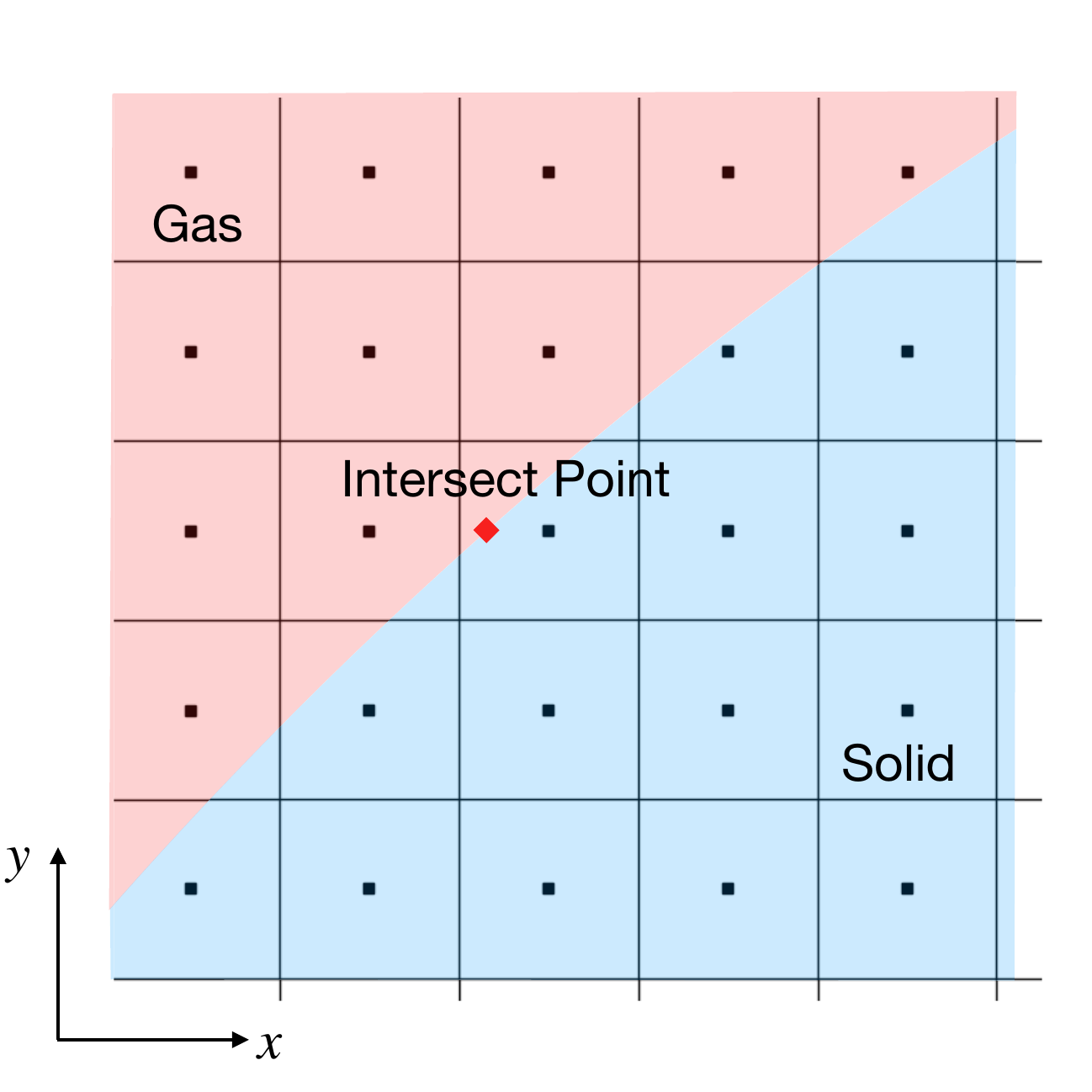}
        \subcaption{The surface and intersect point in physical space}
        \label{fig:VS-PS}
    \end{subfigure}
    \caption{Schematic of the discontinuity in velocity space at the intersect point.}
    \label{fig:discontinuity VS schematic}
\end{figure}
\Cref{fig:VS} shows an example of the distribution function in the two-dimensional velocity space at the intersection point of the wall surface depicted in \Cref{fig:VS-PS}.
As illustrated, \Cref{eq:general bc}  introduces a discontinuity in the distribution function along the tangent to the curved surface (indicated by the dashed line in \Cref{fig:VS}).
As can be seen, the dashed line passes through some velocity cells, which we refer to as cut cells.
Ideally, these cells should capture the combined influence of both incident and reflected Maxwellian distributions. 
However, the piecewise formulation in \Cref{eq:general bc} relying solely on computing $\xi_{n,k}=\bm\xi_k \cdot \bm n_{\rm w}$ at the velocity cell center is clearly insufficient for this purpose.
As a result, portions of cut cells have been erroneously attributed to the opposite distribution (indicated by semi-transparent shades in \Cref{fig:VS}).

\begin{figure}[htbp!]
    \centering
    \includegraphics[width=0.35\textwidth]{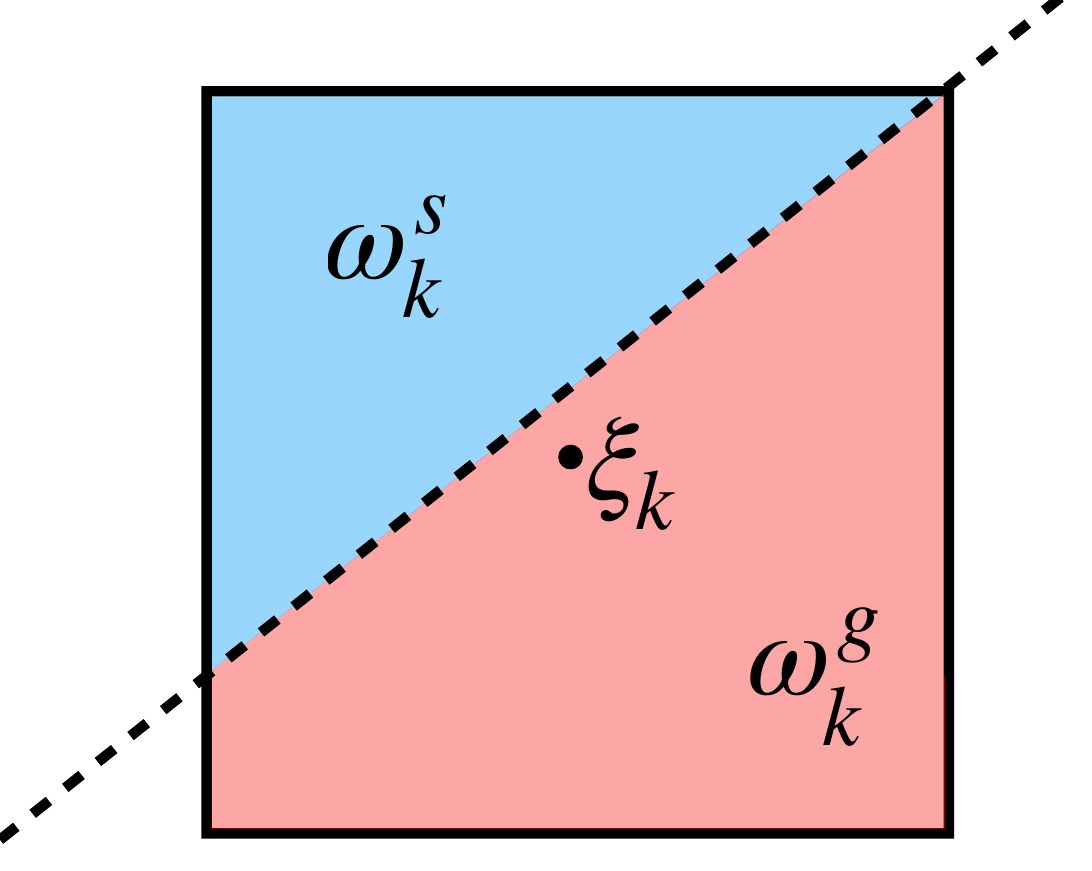}
    \caption{A single cut cell in velocity space.}
    \label{fig:CVC}
\end{figure}

We employ the following equation to correct the distribution function in a cut cell,
\begin{equation}
    \overline{f_{{\rm w},k}^*}=\frac{1}{\omega_k}(\omega_k^g f_{{\rm w},k}^i+\omega_k^s g_{{\rm w},k}),
    \label{eq:fwc}
\end{equation}
where $\omega_k^g=\Delta\bm\xi_{{\rm gas},k}$ and $\omega_k^s=\Delta\bm\xi_{{\rm solid},k}$ denote the respective areas occupied by the gaseous and solid phases, as illustrated in \Cref{fig:CVC}. 
The corresponding no penetration condition for determining the wall density in \Cref{eq:dnpc} becomes
\begin{equation}
    \sum_{k\in \rm solid}\omega_k\xi_{n,k}g_{{\rm w},k}+\sum_{k\in \mathcal C}\omega_k^s\xi_{n,k}g_{{\rm w},k}+\sum_{k\in \rm gas}\omega_k\xi_{n,k}f_{{\rm w},k}^i+\sum_{k\in \mathcal C}\omega_k^g\xi_{n,k}f_{{\rm w},k}^i=0,
    \label{eq:cc_np}
\end{equation}
which yields
\begin{equation}
    \rho_w=-\frac{\sum_{k\in \mathrm{gas}}\omega_k\xi_{n,k}f_{w,k}+\sum_{k\in\mathcal C}\omega_{k}^g\xi_{n,k}f_{w,k}}{\sum_{k\in\mathrm{solid}}\omega_k\xi_{n,k}G_{w,k}+\sum_{k\in\mathcal C}\omega_k^s\xi_{n,k}G_{w,k}},
    \label{eq:rhowc}
\end{equation}
where $G_{{\rm w},k}=\xi_{n,k}(\frac{1}{2\pi RT_{\rm w}})^{3/2}e^{-\frac{1}{2RT_{\rm w}}(\bm\xi_k-\bm U_{\rm w})^2}$, and c.c. denotes the abbreviation of cut cell. 
The subscripts \textit{gas} and \textit{solid} in \Cref{eq:cc_np} and \Cref{eq:rhowc} denote the index sets where the velocity cells fall entirely within the incident and the reflected Maxwellian regions, respectively, and $\mathcal C$ is the index sets of cut cells as shown in \Cref{fig:VS}.
Detailed analysis of numerical accuracy of the cut-cell correction will be presented in \Cref{sec:quadrature accuracy}, whereas \Cref{sec:convergence} provides the numerical verification.

It is known that the terminology “cut cell” typically refers to a sharp-interface fluid-structure interaction model in physical space.
We wish to emphasize that the cut-cell correction proposed in this subsection operates in particle velocity space, representing a fundamental difference from the physical-space counterpart.
The discontinuity induced by the wall surface exhibits elementary topological structures (straight lines in two-dimensional and planar surfaces in three-dimensional configurations, as shown in \Cref{fig:VS}).
Since \Cref{eq:shakhov} does not include a convection operator in velocity space, the small cells that are cut out do not impose strict time-step constraints.
These intrinsic properties enable effective optimization of algorithmic complexity and efficient numerical implementation.


\subsubsection{Summary}

\begin{figure}[htbp!]
    \centering
    \includegraphics[width=1.0\linewidth]{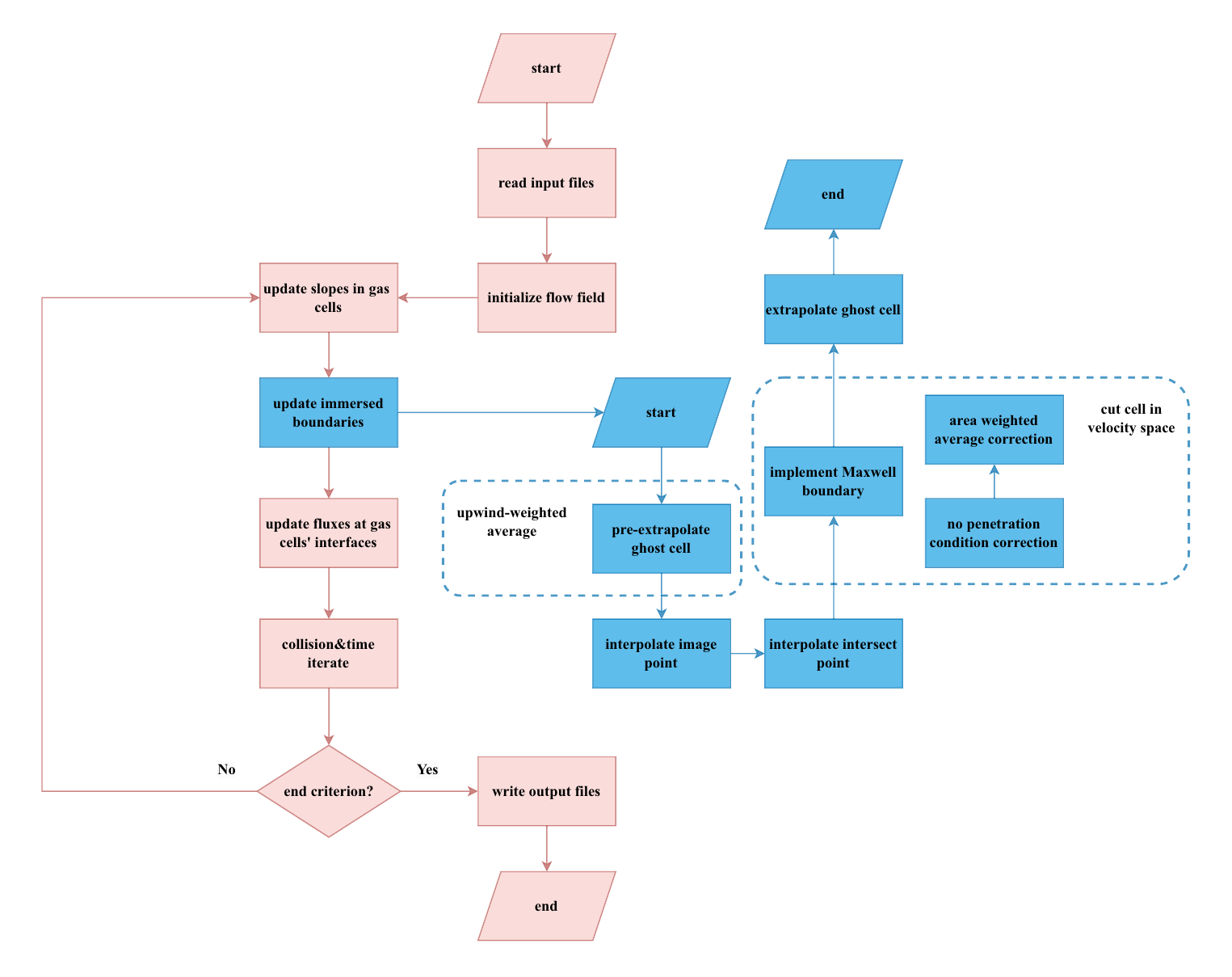}
    \caption{Flow chart of the solution. The boxes labeled with blue refer to the immersed boundary implementation, while those in red represent the outside solution procedure.}
    \label{fig:flowchart}
\end{figure}

The algorithmic flow of the GCIBM from \Cref{sec:gc} to \Cref{sec:CVC} is summarized in \Cref{fig:flowchart}.

\subsection{Numerical analysis}
\label{sec:analysis}
\subsubsection{Surface discontinuity}
\label{sec:surface}
\begin{figure}[htbp]
    \centering
    \begin{subfigure}[c]{0.4\textwidth}
        \includegraphics[width=\textwidth]{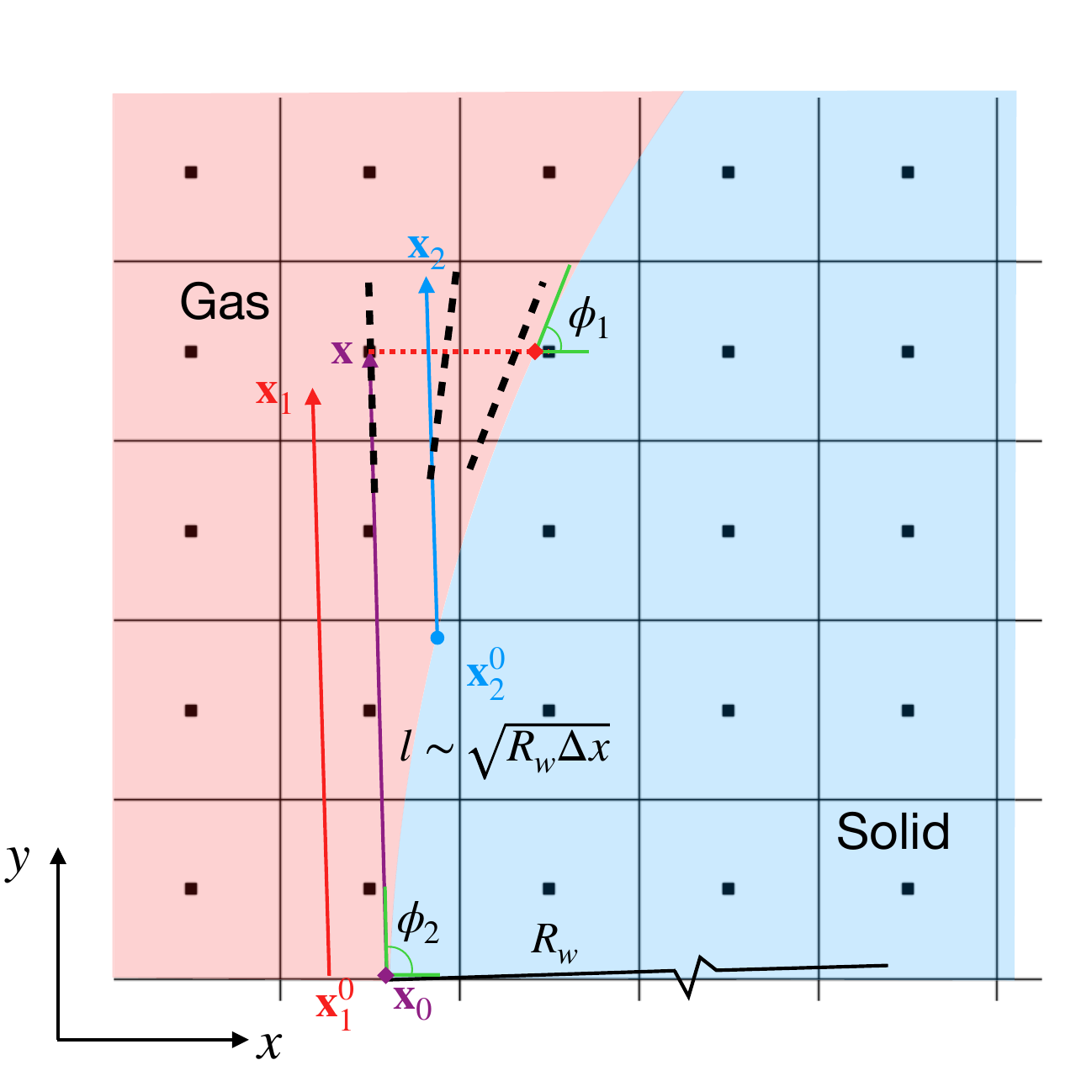}
        \caption{$\Delta x<c\lambda^2/R_w$}
        \label{fig:discontinuity_PS1}
    \end{subfigure}
    \hfil
    \begin{subfigure}[c]{0.4\textwidth}
        \includegraphics[width=\textwidth]{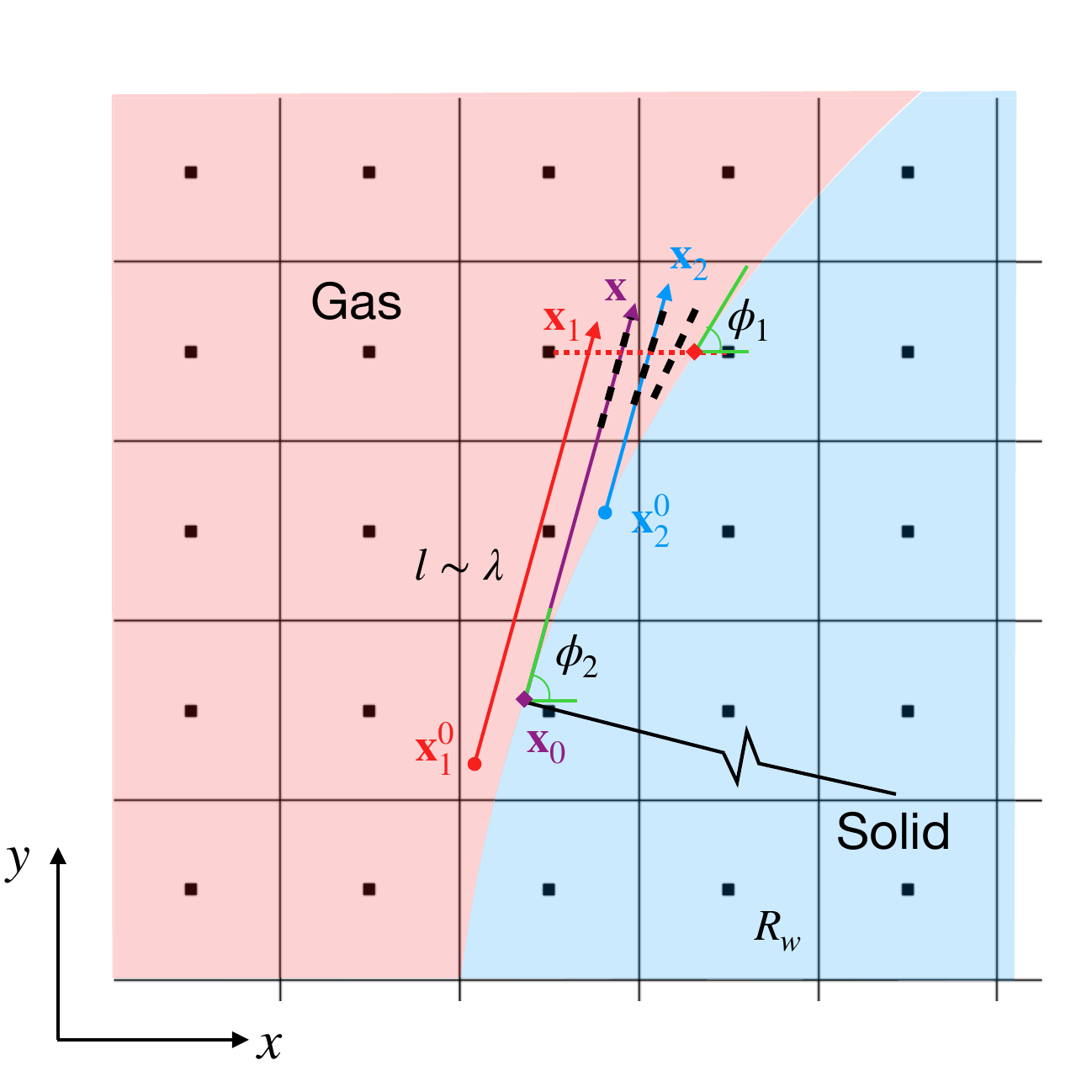}
        \caption{$\Delta x>c\lambda^2/R_w$}
        \label{fig:discontinuity_PS2}
    \end{subfigure}
    \caption{Schematic of the discontinuity induced by convex boundaries. The arrowed lines denote the characteristic lines at the same discrete velocity emitted from different physical points. The red line denotes the characteristic line that is dominated by the gas, the blue denotes the one dominated by the solid boundary, and the purple is tangent to the boundary at $\bm x_0$, along which the discontinuity propagates. The black zigzag line measures the curvature radius of the wall boundary. The black dashed lines denote the approximate location of discontinuities in the distribution function and the direction of discrete velocities. $\phi_1$ and $\phi_2$ are the angles of tangents at the intersect point (red diamond) of the immersed boundary stencil and the tangency point (purple diamond) between the discontinuous characteristic line and the solid boundary.}
    \label{fig:discontinuity}
\end{figure}


Two types of discontinuity phenomena can emerge within gaseous flows.
The first type refers to the discontinuity between adjacent gas and predefined wall boundary conditions, caused by the piecewise distribution function in \Cref{eq:general bc}.
At the macroscopic level, this manifests as the effects of velocity slip and temperature jump.
This phenomenon is not directly related to strong discontinuities within the physical space of the gas region.
Solvers based on body-conformal meshes as well as the method introduced in \Cref{sec:gc} should be able to resolve the Knudsen layer without requiring special discontinuity handling strategies.
The second type of discontinuity is associated with characteristic lines of the kinetic equation, where the distribution functions at the same velocity position exhibit discontinuities in the physical space.
This subsection will specifically address the feasibility of the proposed method under this issue.



The formation of discontinuities in the particle distribution function at the surface of convex objects and their propagation in the physical domain have been confirmed through theoretical analysis \cite{kim_formation_2011} and numerical experiments \cite{sone_discontinuity_1992}.
\Cref{fig:discontinuity} illustrates the origin of discontinuities from the perspective of characteristic lines.
In the two subfigures, the purple line represents the characteristics of the kinetic equation, where the particle velocity falls into the discontinuity region within the velocity space (denoted by the dashed line in \Cref{fig:VS}).
The characteristic line is tangent to the wall boundary.
On either side of a characteristic line, the particle distribution function is determined by the gas (red line) or the solid wall (blue line).
Assuming the particle velocity is $\bm\xi_k$, the characteristic solutions of \Cref{eq:shakhov} along the red and blue lines write
\begin{equation}
\begin{aligned}
    & f_k(\bm x_1,t)=f_k(\bm x^0_1,t_1)+\int_{t_1}^t\frac{g^S_k(\bm x_1-\bm\xi_k(t'-t_1),t')-f_k(\bm x_1-\bm\xi_k(t'-t_1),t')}{\tau(\bm x_1-\bm\xi_k(t'-t_1),t')}dt',\\
    & f_k(\bm x_2,t)=g_{\mathrm{w},k}(\bm x^0_2,t_2)+\int_{t_2}^t\frac{g^S_k(\bm x_2-\bm\xi_k(t'-t_2),t')-f_k(\bm x_2-\bm\xi_k(t'-t_2),t')}{\tau(\bm x_2-\bm\xi_k(t'-t_2),t')}dt',
\end{aligned}
\end{equation}
Here, $\bm x_1$ and $\bm x_2$ denote points on either side of an arbitrary point $\bm x$ on the characteristic line, influenced by gas and solid, respectively, while $\bm x_1^0$ and $\bm x_2^0$ represent their upstream starting points with $t_1=|\bm x_1-\bm x^0_1|/|\bm\xi_k|$ and $t_2=|\bm x_2-\bm x^0_2|/|\bm\xi_k|$. 
Given that $\bm x_1^0\rightarrow \bm x_0$, $\bm x_2^0\rightarrow\bm x_0$ and $t_1=t_2=t_*$, we have $\bm x_1, \bm x_2\rightarrow \bm x$. 
Note that $f_k(\bm x_0,t_*)$ and $g_{w,k}(\bm x_0,t_*)$ can exhibit significant differences.
The discontinuity in the physical space thus arises and propagates further along the characteristic line into the gas field.
If $\Delta t=t-t_*$ is short enough, the effects of intermolecular collisions can be neglected, and the jump condition of the distribution function at $(t,\bm x,\bm \xi_k)$ can be written as
\begin{equation}
    \left[f_k(\bm x,t)\right]\sim \left|g_{\mathrm{w},k}(\bm x_0,t_*)-f_k(\bm x_0,t_*)\right|.
\end{equation}
This discontinuity effect decays with intermolecular collisions and is only clearly appreciable within distances on the order of the mean free path \cite{sone_discontinuity_1992}. 
Thus, it appears only at the bottom of the Knudsen layer. 
Such region containing discontinuities is termed the S layer \cite{sone_new_1973}, with a thickness of $O(\lambda\mathrm{Kn_w})$ magnitude, where $\lambda$ is the mean free path, and $\mathrm{Kn_w}$ denotes the Knudsen number based on the boundary curvature radius $R_\mathrm{w}$, i.e.,
\begin{equation}
    \mathrm{Kn_w} = \frac{\lambda}{R_\mathrm{w}}.
\end{equation}


Reviewing the origins of the aforementioned discontinuities, the premise that $\bm\xi_k$ precisely tangents the boundary, coupled with the discontinuity attenuation with collisions, restricts the influence to an exceedingly limited subset of discrete particle velocities.
Depending on the relative magnitudes of the mean free path, the curvature radius, the mesh size $\Delta x$, and a constant $c$ of $O(1)$ that depends on the geometry of the boundary and the distance between the center of the donor cell and the intersect point, the angle interval of the particle velocity that exhibits discontinuities $\theta$ can be estimated as
\begin{itemize}
    \item[1.] $\Delta x<c\lambda^2/R_\mathrm{w}$: $\theta\sim\sqrt{\Delta x/R_\mathrm{w}}$. In this case, the mesh resolution is sufficiently high, and thus the discontinuities at different particle velocities can fully occupy the interval between the donor cell and the intersect point, as shown in \Cref{fig:discontinuity_PS1}. 
    Note that discontinuities extending beyond the donor cell belong to internal flow field and can be solved using standard slope limiters; they are not included in the boundary module.
    Thus, $\bm x_0$ is determined by constructing the tangent to boundary passing through the center of the donor cell, and the angle between $\phi_1$ and $\phi_2$ is determined by $\Delta x$ and $R_{\rm w}$.
    \item[2.] $\Delta x\ge c\lambda^2/R_\mathrm{w}$: $\theta\sim \lambda/R_\mathrm{w}$. In this case, the mesh cannot fully resolve the S layer, and discontinuities only exist at the bottom of the interval between the donor cell and intersect point, as shown in \Cref{fig:discontinuity_PS2}. The farthest propagated (with the distance of $\lambda$) discontinuity determines the angle between $\phi_1$ and $\phi_2$.
\end{itemize}

\begin{figure}[htbp!]
    \centering
    \begin{subfigure}[t]{0.49\textwidth}
        \centering
        \includegraphics[height=0.8\textwidth]{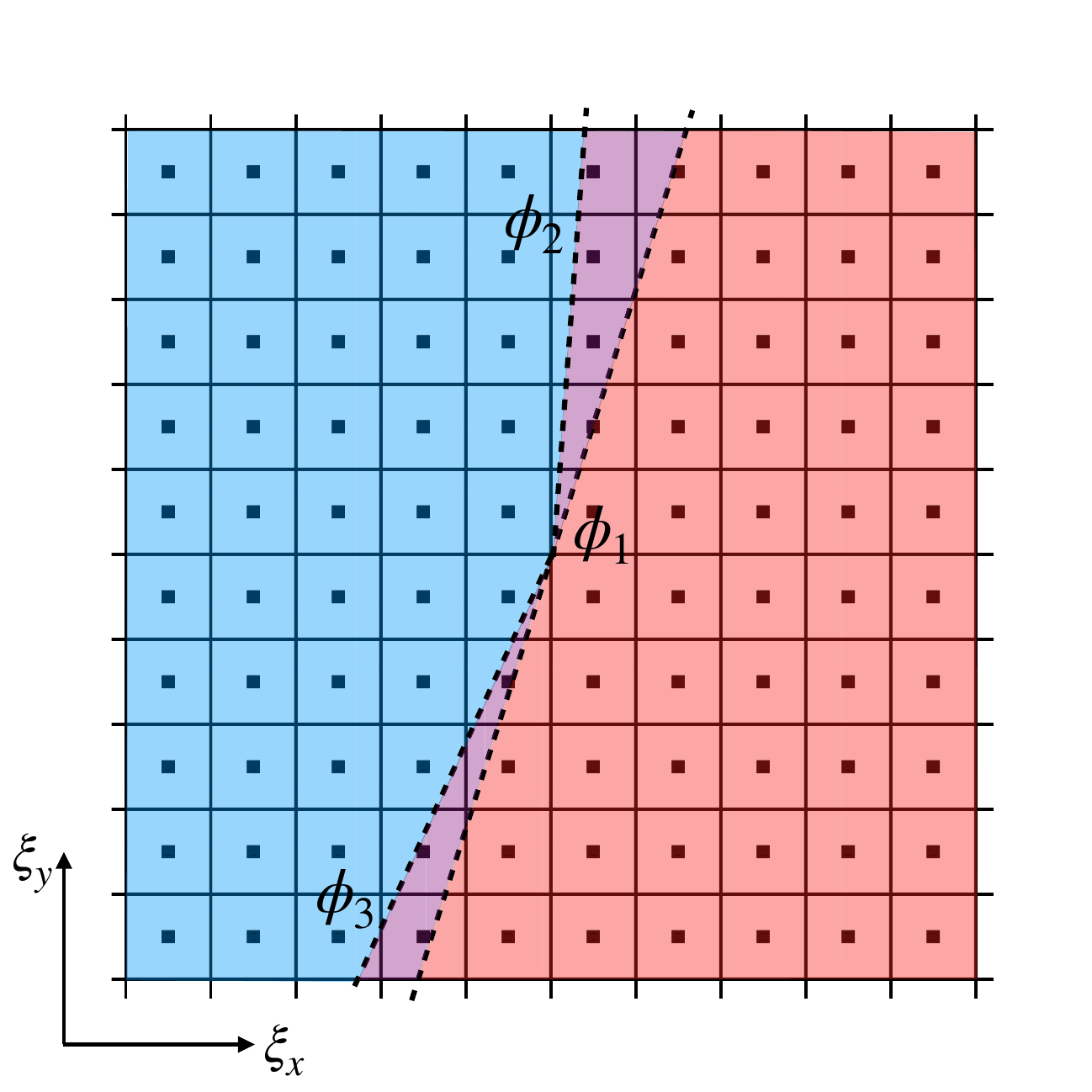}    
        \caption{Discontinuity proportion in velocity space: the discrete velocities in the purple region exhibit discontinuities within physical space, where $\phi_3$ denotes the direction of the tangent from the opposite side}
        \label{fig:discontinuity_VS}
    \end{subfigure}
    \hfil
    \begin{subfigure}[t]{0.49\textwidth}
        \centering
        \includegraphics[height=0.8\textwidth]{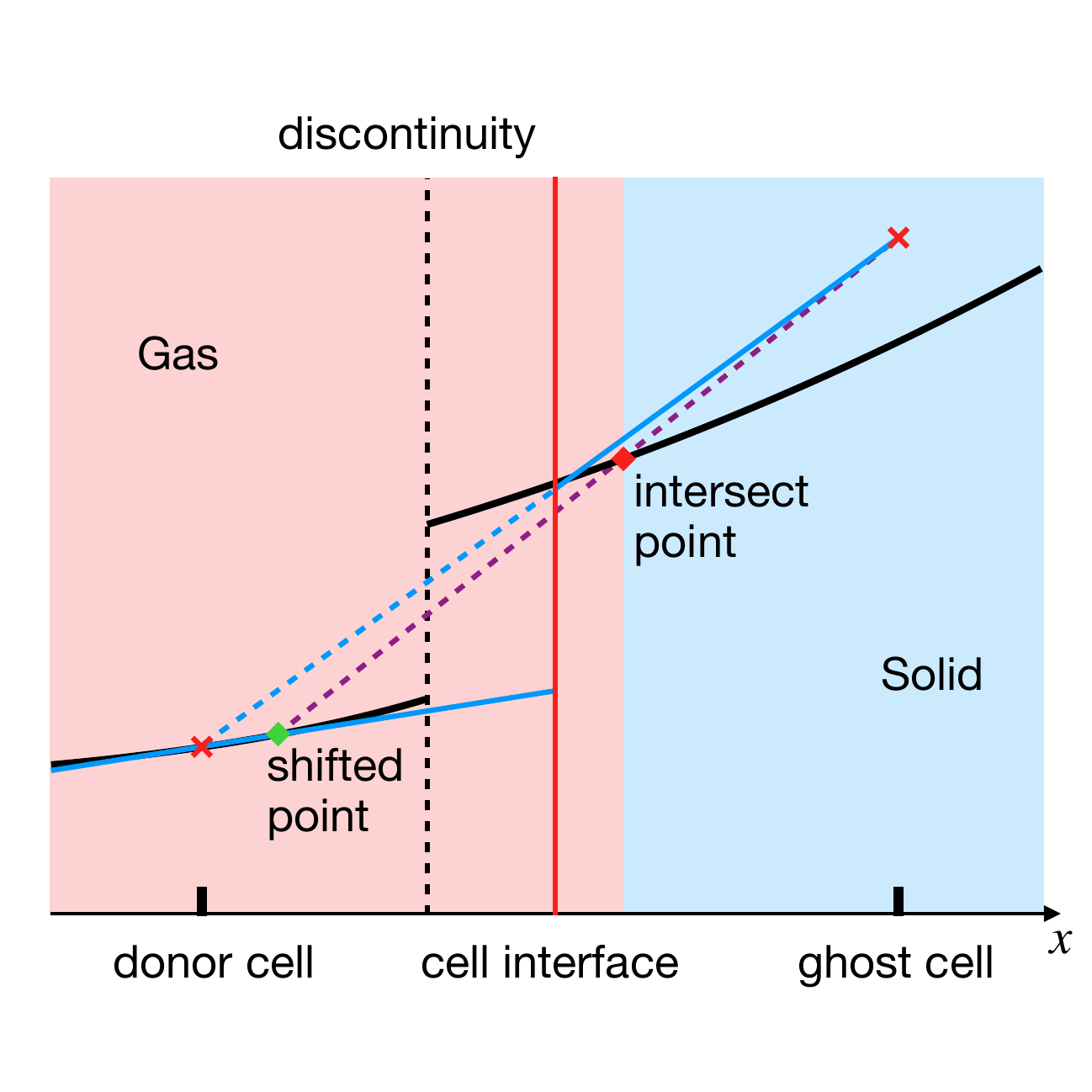}
        \caption{Reconstruction of discontinuous velocities: the black line represents the exact solution, the solid blue line denotes the reconstructed distribution, and the purple line indicates the construction process of the ghost cell}
        \label{fig:discontinuous reconstruction}
    \end{subfigure}
    \caption{Schematic of the proportion of discontinuous velocities and the approximation of the discontinuity in physical space.}
    \label{fig:discontinuity approximation}
\end{figure}

The proportion of discontinuous velocities has been estimated in \Cref{fig:discontinuity_VS}.
For the vast majority of discrete velocity $\bm\xi_k$, the discrete distribution function $f_k$ remains continuous in physical space.
For discontinuous velocities, the approximation process near the boundary is briefly summarized in \Cref{fig:discontinuous reconstruction}.
The discontinuous solution (black curve) is linearly reconstructed (blue line).
With different slopes reconstructed in the donor cell and the ghost cell, the discontinuity effect is modeled at the face.
Comparing the blue line and black curve, the error introduced by this process can be estimated as $O([f]\Delta x)$, where $[f]\sim O(1)$ denotes the jump of the discontinuity, and the approximation error is of first order.

Furthermore, in both of the above cases in \Cref{fig:discontinuity}, the characteristic lines of discontinuous velocities graze and point away from the intersect point. This indicates that during the construction of ghost-cell values, these collocation points lie within the region of the reflected velocity space (where the Maxwellian distribution will subsequently be constructed), thus exerting no direct influence on the formulation of wall distribution function.
Thus, the design accuracy of $f_{\mathrm{w},k}^i$ and $f_\mathrm{w}^*$ should be minimally affected, which further guarantees the validity of the approximation to the distribution in ghost cells. 
In engineering simulations, one may be more concerned with macroscopic moments of distribution functions.
Considering the application of the quadrature rule, \Cref{eq:quadrature}, on \Cref{eq:moments1}, \Cref{eq:moments2}, and \Cref{eq:moments3}, the influence of discontinuities is determined by two aspects.
\begin{itemize}
    \item[1.] The error introduced by discontinuities into the distribution function on each velocity collocation point (estimated as $O(\Delta x)$).
    \item[2.] The proportion of discontinuous velocities in the velocity space. For an infinite velocity space, this proportion can be estimated as $\theta/2\pi$.
\end{itemize}
We confine our discussion to boundaries with moderate curvature, i.e., $R_w\sim L$. 
In the case of $\Delta x\ge c\lambda^2/R_\mathrm{w}$ and $\theta\sim \lambda/R_\mathrm{w}$, the error induced by the discontinuity to macroscopic variables is thus of $O(\Delta x)\cdot \theta/2\pi$.
A suggested height of the first layer mesh is $\Delta x=0.2\lambda$ \cite{luo_multiscale_2024,yang_efficient_2024,chen_global_2024}.
Following this requirement, the error can be written as $O(\Delta x)\cdot O(\lambda/R_{\rm w})\sim O(\Delta x^2)$, given $\Delta x\sim O(\lambda)$ and $R_{\rm w}\sim O(1)$.
On the other hand, when the mesh are sufficiently refined, i.e., $\Delta x<c\lambda^2/R_\mathrm{w}$ and  $\theta\sim\sqrt{\Delta x/R_\mathrm{w}}$, the error will converge with the order of $O(\Delta x)\cdot O(\sqrt{\Delta x/R_{\rm w}})\sim O(\Delta x^{3/2})$.

In summary, the effect of discontinuities at the wall surface is moderate, and the solution algorithm developed in \Cref{sec:gc} is effective.
Note that when the solid boundary contains both convex and concave segments with small characteristic scales of variation, the discontinuity effects can be more complicated, and we do not consider this scenario in the current study.

\subsubsection{Spatial accuracy}
\label{sec:accuracy}
This subsection proves the developed solution algorithm achieves second-order accuracy for particle velocities that are continuous in the physical space.
We firstly introduce a test function of the form
\begin{equation}
    f(x,y,\bm{\xi})=a(\bm\xi)x+b(\bm\xi)y+c(\bm\xi),
\end{equation}
where $a$, $b$, and $c$ are functions solely of the particle velocity. 
Without loss of generality, we take the case in \Cref{fig:ghost cell} as an example. The slope $\bm{\sigma}(\bm\xi)=\left[a(\bm\xi),b(\bm\xi)\right]^{\mathrm T}$ applies to all positions in the stencil. Substituting it into \Cref{eq:extrapolation 1} and \Cref{eq:upwind} yields
\begin{equation}
    f_{i}(\bm\xi)-\bm\sigma_{i}(\bm\xi)\cdot\bm r_i=a(\bm\xi)x_{\rm gc}+b(\bm\xi)y_{\rm gc}+c(\bm\xi).
\end{equation}
Given that the upwind weight $w_{i}(\bm\xi)$ satisfies $\sum_i^{N_S} w_{i}(\bm\xi)=1$, we have
\begin{equation}
    f_{\rm gc}(\bm\xi)=\left[a(\bm\xi)x_{\rm gc}+b(\bm\xi)y_{\rm gc}+c(\bm\xi)\right]\sum_{i}^{N_S}w_{i}(\bm\xi)=a(\bm\xi)x_{\rm gc}+b(\bm\xi)y_{\rm gc}+c(\bm\xi).
\end{equation}
Similarly, one-sided difference yields $f_{\rm sp}(\bm\xi)=a(\bm\xi)x_{\rm sp}+b(\bm\xi)y_{\rm sp}+c(\bm\xi)$. Substituting it into \Cref{eq:fw} and noticing that $x_{\rm sp}=x_{\rm gc}=x_{\rm w}$, we have
\begin{equation}
    f_{\rm w}(\bm\xi)=a(\bm\xi)x_{\rm w}+b(\bm\xi)y_{\rm w}+c(\bm\xi),
\end{equation}
which is in perfect agreement with the exact distribution function at the intersect point.
After determining $\rho_{\rm w}$ with the no penetration condition in \Cref{eq:npc},
\begin{equation}
    \rho_{\rm w}=-\frac{\int_{\xi_n >0}\xi_{n}f_{\rm w}d\bm\xi}{\int_{\xi_n <0}\xi_{n}\left( \frac{1}{2 \pi R T_{\rm w}} \right)^{3/2} e^{-\frac{1}{2 R T_{\rm w}} ((\bm{\xi} - \bm{U}_{\rm w})^2)}d\bm\xi},
    \label{eq:rhow}
\end{equation}
the Maxwell gas-surface interaction in \Cref{eq:general bc} is obtained without spatial discretization error. Subsequently, the accuracy of the boundary condition depends on the extrapolation used to construct $f^*_{\rm gc}(\bm\xi)$.

After \Cref{eq:general bc}, the independence of spatial coordinates of $a$, $b$, and $c$ is compromised.
The analysis based on the test function above no longer holds, and we continue the proof based on the Taylor expansion.
Based on the image point, the expansion takes the form
\begin{equation}
    f_{\rm w}^*(\bm\xi)=f_{\rm sp}(\bm\xi)+\left.\frac{\partial f(x_{\rm dc},y,\bm\xi)}{\partial y}\right|_{y=y_{\rm sp}}(y_{\rm w}-y_{\rm sp})+O(\Delta y^2),
\end{equation}
which yields
\begin{equation}
    \left.\frac{\partial f(x_{\rm dc},y,\bm\xi)}{\partial y}\right|_{y=y_{\rm sp}}=\frac{f_{\rm w}^*(\bm\xi)-f_{\rm sp}(\bm\xi)}{y_{\rm w}-y_{\rm sp}}+\frac{O(\Delta y^2)}{y_{\rm w}-y_{\rm sp}}.
    \label{eq:slope}
\end{equation}
The Taylor expansion also gives
\begin{equation}
    f_{\rm gc}(\bm\xi)=f_{\rm sp}(\bm\xi)+\left.\frac{\partial f(x_{\rm dc},y,\bm\xi)}{\partial y}\right|_{y=y_{\rm sp}}(y_{\rm gc}-y_{\rm sp})+O(\Delta y^2).
\end{equation}
Substituting \Cref{eq:slope} into the right hand side, we have
\begin{equation}
    f_{\rm gc}(\bm\xi) = f_{\rm sp}(\bm\xi)+(\frac{f_{\rm w}^{*}(\bm\xi)-f_{\rm sp}(\bm\xi)}{y_{\rm w}-y_{\rm sp}}+\frac{O(\Delta y^2)}{y_{\rm w}-y_{\rm sp}})(y_{\rm gc}-y_{\rm sp})+O(\Delta y^2).
    \label{eq:fg error}
\end{equation}
The order of the truncation error depends on $y_{\rm w}-y_{\rm sp}$, i.e., the distance between the image point and the intersect point. When $({y_{\rm gc}-y_{\rm sp}})/({y_{\rm w}-y_{\rm sp}})\sim 1$, the truncation error terms in \Cref{eq:fg error} are consistently of second order. Based on the identification criterion provided in \Cref{sec:gc}, the value of $|y_{\rm w}-y_{\rm gc}|$ lies within the interval $\left(0,\Delta y\right ]$. We suggest the set as \Cref{eq:sp}
for the balance of the accuracy and the physical locality. With the shifted point determined by \Cref{eq:sp}, we obtain
\begin{equation}
    f_{\rm gc}(\bm\xi)=f_{\rm sp}(\bm\xi)+\frac{f_{\rm w}^{*}(\bm\xi)-f_{\rm sp}(\bm\xi)}{y_{\rm w}-y_{\rm sp}}(y_{\rm gc}-y_{\rm sp})+O(\Delta y^2),
\end{equation}
which indicates the reconstruction of ghost-cell values following \Cref{eq:extrapolation 2} exhibits a second-order spatial accuracy.

\subsubsection{Quadrature accuracy}
\label{sec:quadrature accuracy}

This subsection demonstrates the effectiveness of the cut-cell correction proposed in \Cref{sec:CVC}.
As analyzed in \Cref{sec:surface}, at the intersect point, the incident distribution $f^i_{\rm w}$ typically differs from the reflected Maxwellian distribution $g_{\rm w}$, which forms a discontinuity in the velocity space through \Cref{eq:general bc}.
We assume $\Delta\xi_x=\Delta\xi_y=\Delta\xi$, and the quadrature weights in \Cref{eq:quadrature} takes $\omega_k=\omega=\Delta\xi^2$.
We denote the quadrature error by $E^*$, and the integral over the cut cells writes
\begin{equation}
        \int_{\mathrm c.c.}\phi fd\bm\xi=\sum_{k\in \mathcal C}\omega_k\phi_kf_k+E^*=\omega\left(\sum_{k\in \mathcal G}\phi_kf^i_{{\rm w},k}+\sum_{k\in \mathcal S}\phi_k g_{{\rm w},k}\right)+E^*,
        \label{eq:cc quadrature}
\end{equation}
where $\mathcal C$ represents the set of the cut cells' indices, while $\mathcal G$ and $\mathcal S$ denote the subsets of $\mathcal C$ whose elements satisfy $\xi_n=\bm\xi_k\cdot \bm n_{\rm w}\le0$ and $\xi_n=\bm\xi_k\cdot \bm n_{\rm w}>0$, respectively. 
Considering the individual continuity of $g_{{\rm w}}$, $f_{\rm w}$, and $\phi$ respectively, the Taylor expansion in the cut cells leads to
\begin{equation}
    \begin{aligned}
        &\phi g_{\rm w}=\phi_kg_{{\rm w},k}+(\bm\xi-\bm\xi_k)\cdot\frac{\partial \phi g_{\rm w}}{\partial\bm\xi}+O(\Delta\xi^2),\\
        &\phi f^i_{\rm w} = \phi_kf^i_{{\rm w},k}+(\bm\xi-\bm\xi_k)\cdot\frac{\partial \phi f^i_{\rm w}}{\partial\bm\xi}+O(\Delta\xi^2).
    \end{aligned}
    \label{eq:CVC Taylor}
\end{equation}
Applying \Cref{eq:CVC Taylor} to cut cells yields
\begin{equation}
    \begin{split}
        \iint_\mathcal C\phi fd\bm\xi=&\sum_{k\in \mathcal C}\left[\iint_{\Delta\bm\xi_{{\rm gas},k}}\left(\phi_k f^i_{{\rm w},k}+O(\Delta\xi)\right)d\bm\xi+\iint_{\Delta\bm\xi_{{\rm solid},k}}\left(\phi_k g_{{\rm w},k}+O(\Delta\xi)\right)d\bm\xi\right]\\
            =&\sum_{k\in \mathcal G}\left[\iint_{\Delta\bm\xi_k}\phi_k f^i_{{\rm w},k}d\bm\xi+\iint_{\Delta\bm\xi_{{\rm solid},k}}\phi_k(g_{{\rm w},k}-f^i_{{\rm w},k})d\bm\xi+O(\Delta\xi^3)\right]+\\
            &\sum_{k\in \mathcal S}\left[\iint_{\Delta\bm\xi_k}\phi_k g_{{\rm w},k}d\bm\xi+\iint_{\Delta\bm\xi_{{\rm gas},k}}\phi_k(f^i_{{\rm w},k}-g_{{\rm w},k})d\bm\xi+O(\Delta\xi^3)\right]\\
            =&\omega\left(\sum_{k\in \mathcal G}\phi_kf^i_{{\rm w},k}+\sum_{k\in \mathcal S}\phi_k g_{{\rm w},k}\right)+\sum_{k\in \mathcal G}\iint_{\Delta\bm\xi_{{\rm solid},k}}\phi_k(g_{{\rm w},k}-f^i_{{\rm w},k})d\bm\xi+\\
            &\sum_{k\in \mathcal S}\iint_{\Delta\bm\xi_{{\rm gas},k}}\phi_k(f^i_{{\rm w},k}-g_{{\rm w},k})d\bm\xi+O(\Delta\xi^2)
    \end{split}.
    \label{eq:ncvc_error 1}
\end{equation}
Here, $\Delta\bm\xi_{{\rm gas},k}$ and $\Delta\bm\xi_{{\rm solid},k}$ represent the partition of the two distribution function to the $k$-th cut cell. The last equality is obtained with the estimation that the number of the cells that the discontinuity goes through is of comparable magnitude to $\sqrt{N_V}\sim\sqrt{RT_0}/\Delta\xi$, where $R$ is the gas constant and $T_0$ is the reference temperature.
Comparing \Cref{eq:ncvc_error 1} with \Cref{eq:cc quadrature}, we obtain
\begin{equation}
    E^*=\sum_{k\in \mathcal G}\iint_{\Delta\bm\xi_{{\rm solid},k}}\phi_k(g_{{\rm w},k}-f^i_{{\rm w},k})d\bm\xi+
            \sum_{k\in \mathcal S}\iint_{\Delta\bm\xi_{{\rm gas},k}}\phi_k(f^i_{{\rm w},k}-g_{{\rm w},k})d\bm\xi+O(\Delta\xi^2).
    \label{eq:ncvc_error 2}
\end{equation}
In extreme cases (depending on the normal vector of the boundary), we have $\Delta\bm\xi_{{\rm solid},k}\sim\Delta\bm\xi_{{\rm gas},k}\sim0.5\Delta\bm\xi_k\sim\Delta\xi^2$, and the error can be estimated as
\begin{equation}
    E^*\sim \sqrt{N_V}\omega\overline{|\phi(g_{{\rm w}}-f^i_{\rm w})|}\sim\frac{1}{\Delta\xi}\Delta\xi^2\overline{|\phi(g_{\rm w}-f^i_{\rm w})|}=\Delta\xi\overline{|\phi(g_{\rm w}-f^i_{\rm w})|}\sim O(\Delta\xi),
    \label{eq:ncvc_error 3}
\end{equation}
where $\overline{|\phi(g_{\rm w}-f_{\rm w})|}$ represents the average difference between the Maxwellian and extrapolated non-equilibrium distribution over the cut cells. 
The conclusion depends on the premise that $\Delta\xi<<|\phi(g_{\rm w}-f_{\rm w})|$, i.e., the resolution is fine enough to resolve the variation of $\phi(\bm\xi)$.

We wish to point out the above issues also exist for higher-order quadrature.
For the closed curve boundary, the discontinuities at different boundary points will sweep over the whole velocity space, which makes the correction by adopting the specific quadrature points distribution invalid.
Employing different collocation points at different locations can be helpful.
However, this requires mappings between different collocation points, thereby complicating the computational process and introducing additional aliasing errors. 
The cut-cell method in velocity space represents a more readily implementable approach.
We return to the second equality in \Cref{eq:ncvc_error 1}. If the quadrature on $\Delta\bm\xi_{{\rm gas},k}$ and $\Delta\bm\xi_{{\rm solid},k}$ is handled separately, we arrive at
\begin{equation}
    \begin{split}
        \iint_{\mathrm c.c.}\phi fd\bm\xi=&\sum_{k\in \mathcal C}\left(\iint_{\Delta\bm\xi_{{\rm gas},k}}\phi f^i_{\rm w}d\bm\xi+\iint_{\Delta\bm\xi_{{\rm solid},k}}\phi g_{\rm w}d\bm\xi\right)\\
        =&\sum_{k\in \mathcal C}\left[\iint_{\Delta\bm\xi_{{\rm gas},k}}\left(\phi_k f^i_{{\rm w},k}+O(\Delta\xi)\right)d\bm\xi+\iint_{\Delta\bm\xi_{{\rm solid},k}}\left(\phi_k g_{{\rm w},k}+O(\Delta\xi)\right)d\bm\xi\right]\\
        \sim&\sum_{k\in \mathcal C}\left(\phi_k\left(\Delta\bm\xi_{{\rm gas},k} f^i_{{\rm w},k}+\Delta\bm\xi_{{\rm w},k}g_{{\rm w},k}\right)+O(\Delta\xi)\Delta\xi^2\right)\\
        \sim&\sum_{k\in \mathcal C}\Delta\bm\xi_k\phi_k\overline {f^*_{{\rm w},k}}+\sqrt{N_V}\Delta\xi^2O(\Delta\xi)\\
        \sim&\sum_{k\in \mathcal C}\Delta\bm\xi_k\phi_k\overline {f^*_{{\rm w},k}}+O(\Delta\xi^2)
    \end{split},
    \label{eq:VSerror}
\end{equation}
where
\begin{equation}
    \overline{f^*_{w,k}}=\frac{\Delta\bm\xi_{{\rm{gas}},k}}{\Delta\bm\xi_k}f_{{\rm w},k}^i+\frac{\Delta\bm\xi_{{\rm w},k}}{\Delta\bm\xi} g_{{\rm w},k}=\frac{1}{\omega_k}(\omega_k^g f_{{\rm w},k}^i+\omega_k^s g_{{\rm w},k}),
\end{equation}
which is exactly \Cref{eq:fwc}. Hence, the quadrature accuracy has been improved to the the second order.
A numerical validation will be presented in \Cref{fig:cylinder velocity space}.

\subsubsection{Stability and well-posedness}
\label{sec:stability}

Finally, we analyze the necessity of the upwind interpolation approach.
To elucidate the motivation and effectiveness of the approach, we firstly identify the source of numerical instability in the IBM based on the direction of the characteristic lines of the kinetic equation.
Due to the introduction of immersed boundaries, the mesh face may not coincide with the solid boundary.
Consequently, the advection operator of the kinetic equation at some discrete velocities (belonging to the incident portion) can be interpreted as beling solved with a fully downwind scheme, which is known for its numerical instability.

\begin{figure}[htbp!]
    \centering
    \begin{subfigure}[b]{0.49\textwidth}
        \centering
        \includegraphics[height=0.7\textwidth]{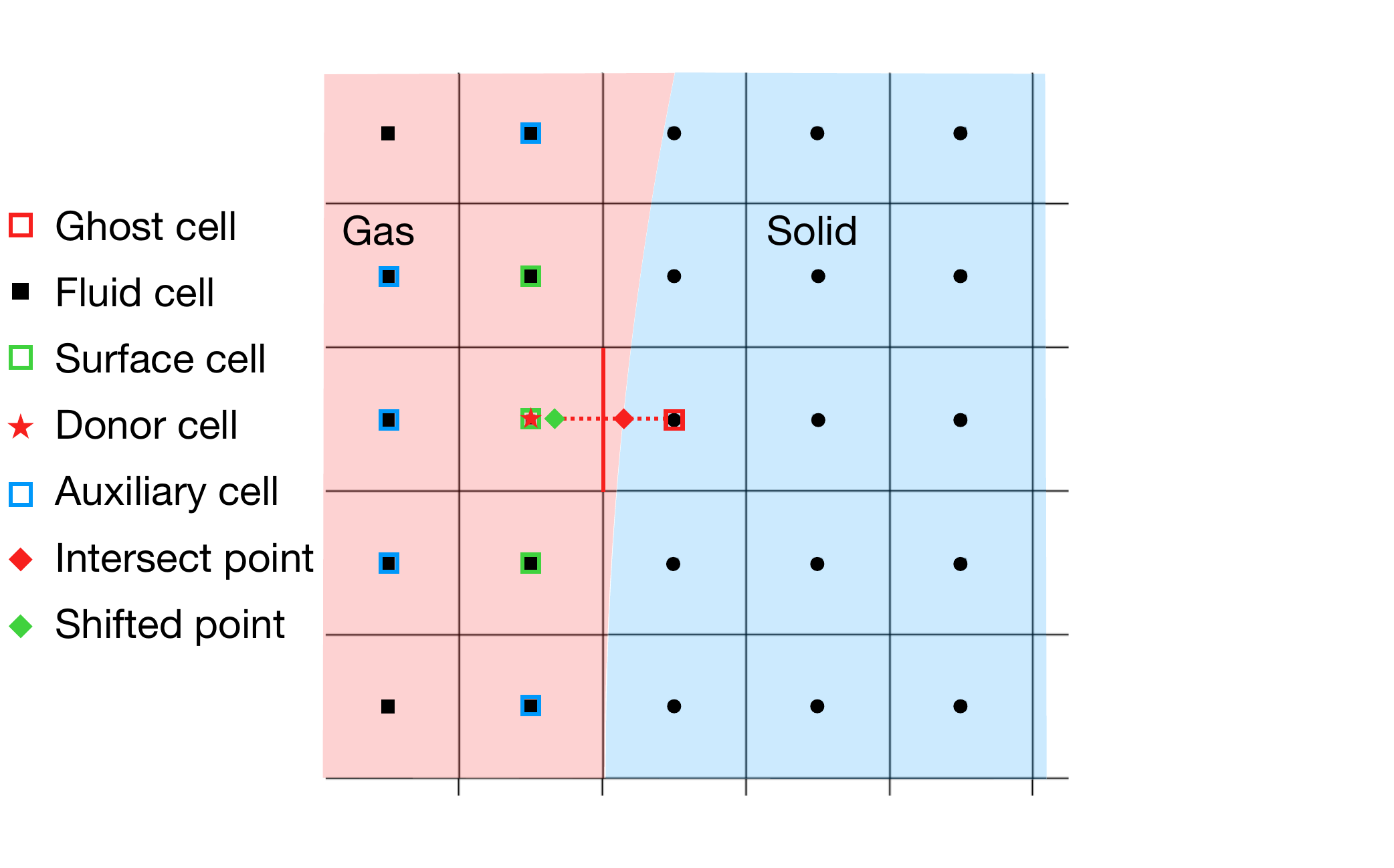}
        \caption{Physical space}
        \label{fig:downwind_ps_1}
    \end{subfigure}
    \hfil
    \begin{subfigure}[b]{0.49\textwidth}
        \centering
        \includegraphics[height=0.7\textwidth]{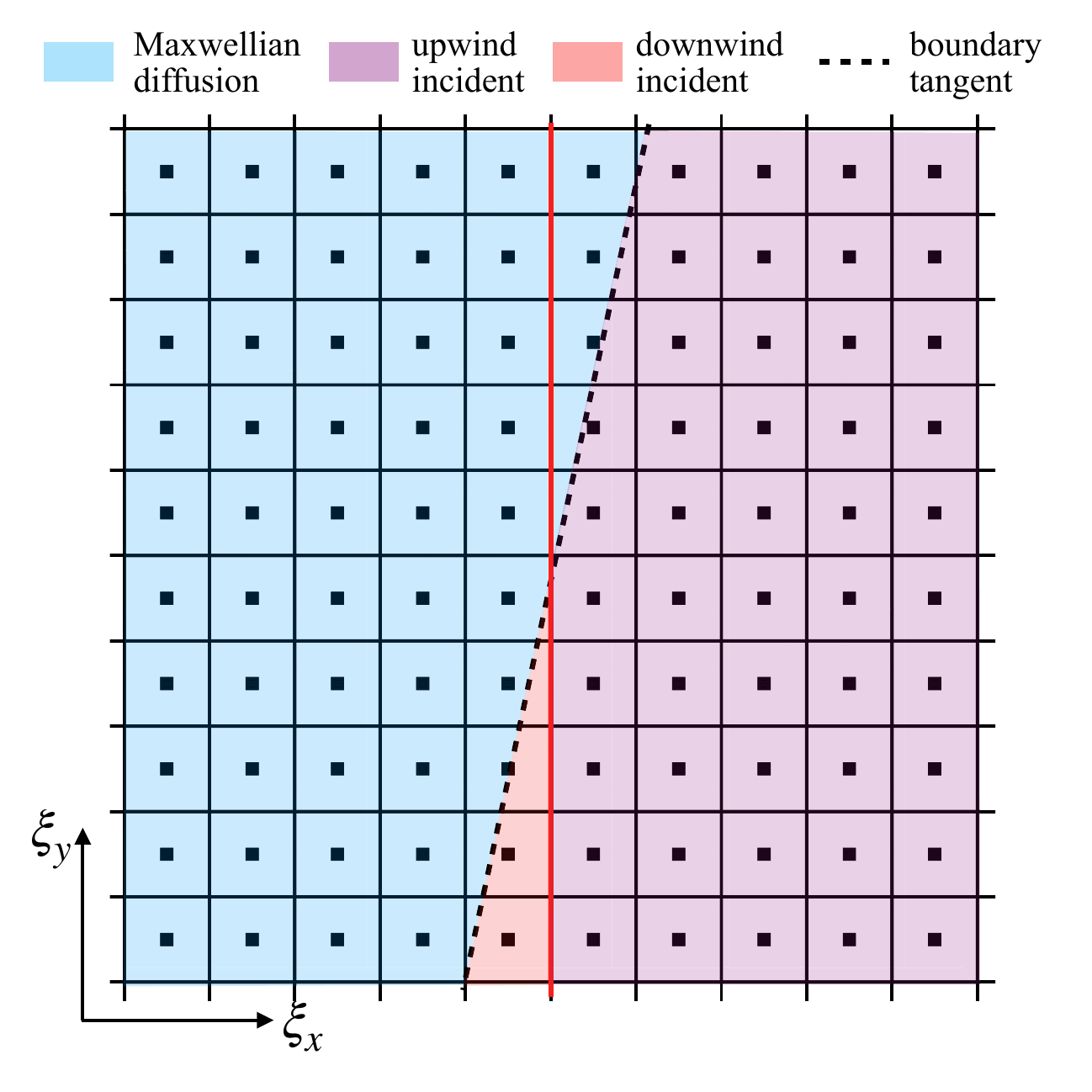}
        \caption{Velocity space}
        \label{fig:downwind_vs_1}
    \end{subfigure}
    \caption{Schematic showing the origin of numerical instability. In this case, the downwind collocation points are of small proportion and the stability can be maintained with sufficient collisions.}
    \label{fig:downwind1}
\end{figure}
\begin{figure}[htbp!]
    \centering
    \begin{subfigure}[b]{0.49\textwidth}
        \centering
        \includegraphics[height=0.7\textwidth]{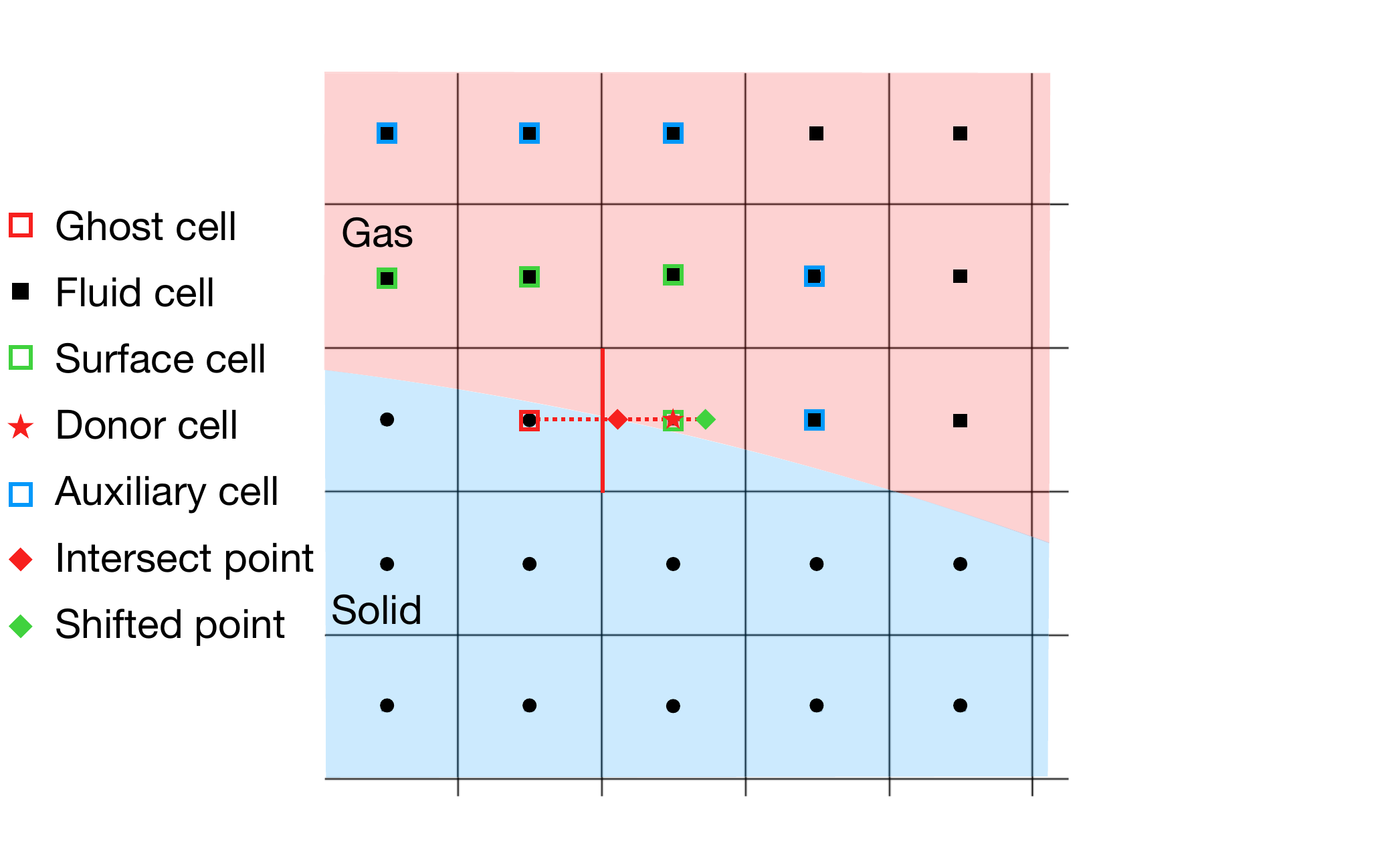}
        \caption{Physical space}
        \label{fig:downwind_ps_2}
    \end{subfigure}
    \hfil
    \begin{subfigure}[b]{0.49\textwidth}
        \centering
        \includegraphics[height=0.7\textwidth]{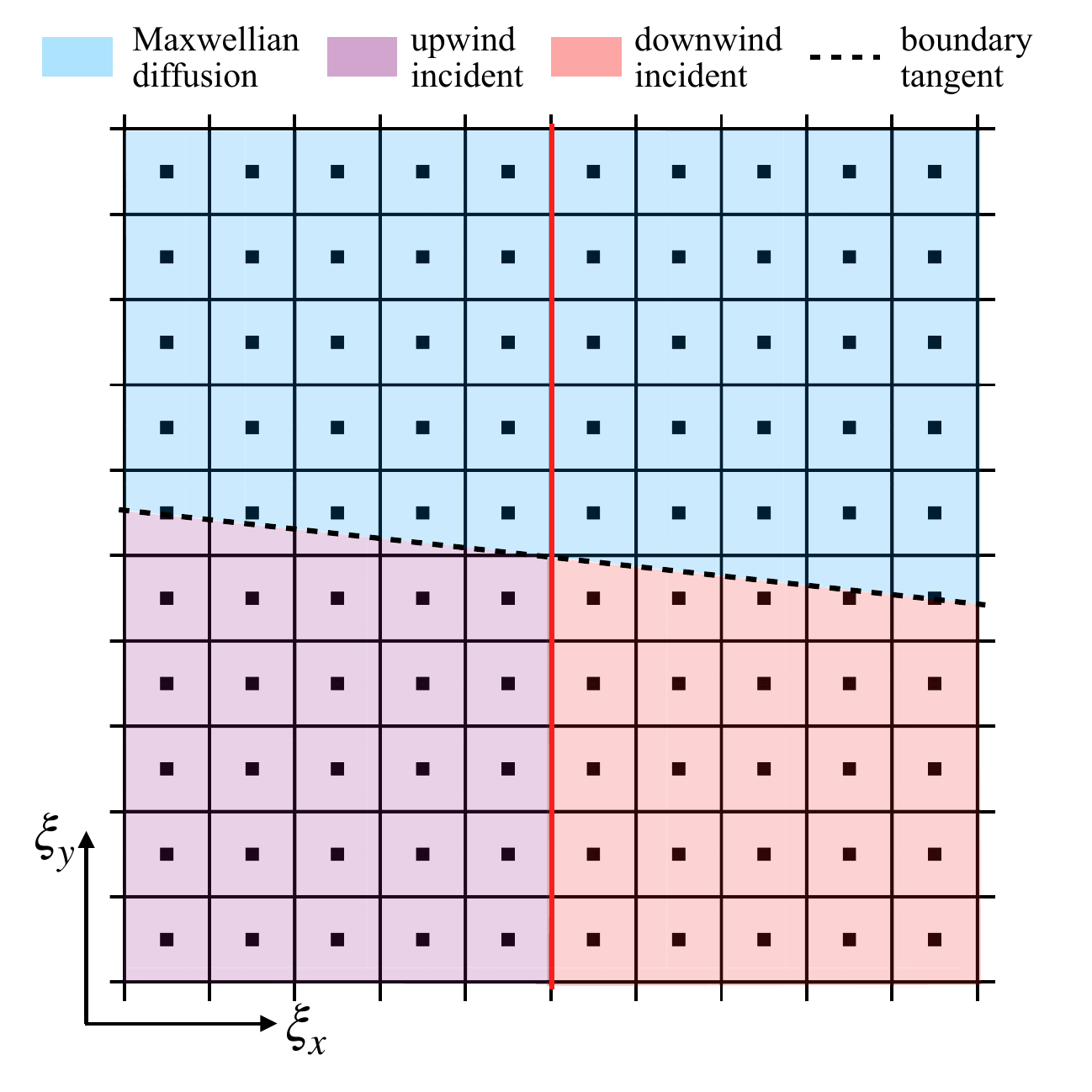}
        \caption{Velocity space}
        \label{fig:downwind_vs_2}
    \end{subfigure}
    \caption{Schematic showing the origin of numerical instability. In this case, the downwind collocation points are of large proportion and the solution tends to be instable from the donor cell.}
    \label{fig:downwind2}
\end{figure}

Referring to \Cref{fig:downwind1}, we elaborate on the root cause of the issue.
In \Cref{fig:downwind_vs_1}, the red and purple parts constitute the incident distribution function, where the particle velocity satisfies $\bm\xi_{n,k}<0$.
As introduced in \Cref{sec:gc}, the distribution function of the ghost cell is extrapolated along $x$-direction. Using it to compute the numerical fluxes at the face labeled with red line in \Cref{fig:downwind_ps_1}, we can obtain the semi-discrete equation for the donor cell $(\bm x_i,\bm y_j)$ at $\bm \xi_k$, i.e.,
\begin{equation}
    \frac{\partial f_{i,j,k}}{\partial t}+\frac{\xi_{x,k}}{\Delta x}\left(f_{i+\frac12,j,k}-f_{i-\frac12,j,k}\right)=\frac{g^S_{i,j,k}-f_{i,j,k}}{\tau_{i,j}}-\frac{\xi_{y,k}}{\Delta y}(f_{i,j+\frac12,k}-f_{i,j-\frac12,k}).
    \label{eq:discretized donor}
\end{equation}
Hereafter we omit the collision term and fluxes along $y$-direction (non-related to the immersed boundary).

The discrete velocities that fall into the incident region satisfy
\begin{equation}
    \xi_{n,k}<0\quad{\rm{and}}\quad \xi_{x,k}<0.
    \label{eq:downwind condition}
\end{equation}
Without the upwind-weighted strategy, the construction of distribution function in the ghost cell can be written as
\begin{equation}
    \begin{split}
        &f_{i+1,j,k}=2f_{i,j,k}-f_{i-1,j,k},\\
        &\sigma_{i+1,j,k}^x=\frac{1}{\Delta x}(f_{i,j,k}-f_{i-1,j,k})
    \end{split}
\end{equation}
Substituting $f_{i\pm\frac12,j,k}$ provided in \Cref{eq:reconstruction} into the left-hand side of  \Cref{eq:discretized donor}, we arrive at
\begin{equation}
    \frac{\partial f_{i,j,k}}{\partial t}+\frac{\xi_{x,k}}{\Delta x}\left(f_{i,j,k}-f_{i-1,j,k}\right)=0.
    \label{eq:downwind}
\end{equation}
Given that $\xi_{x,k}<0$, the advection operator is discretized with a downwind scheme along $x$-direction, which can cause the numerical instability.
Note that the existence of downwind collocations is entirely attributed to the non-coincidence of mesh face with solid boundary. If the two are aligned, we have
\begin{equation}
    \xi_{n,k}=-\xi_{x,k},
\end{equation}
and \Cref{eq:downwind condition} will not hold.

The above analysis does not account for the effect of collision term.
Preliminary studies have been conducted on collision effects in the near-Maxwellian regime \cite{chen_bgk_2025}.
While still in the nascent stages, initial findings suggest that collision terms exert a stabilizing influence upon the numerical system.
For the case of \Cref{fig:downwind1}, the portion of the downwind region is very low.
The instability induced by the downwind scheme is expected to be counteracted by collision effects.
For the case of \Cref{fig:downwind2}, the influence exerted by the downwind portion is stronger, making it more difficult for a naive interpolation method to maintain the stability.
More systematic work on collision term analysis will be carried out in future endeavors.


We wish to point out that, when viewed through an alternative lens, the solution of the advection operator using a downwind scheme should require an additional inflow boundary condition along $x$-direction.
Due to the finite numerical resolution, direct interpolation without the upwind-weighted strategy can fail to meet this requirement, leading to the ill-posedness in the discretized algebraic system.
From the perspective of particle transports, particles with downwind velocities transport tangentially along the wall surface, indicating that additional information should be provided from other dimensions.
This inspires us to develop the upwind-weighted strategy. 
\section{Numerical Experiments}
\label{sec:4}
This section validate the solution algorithm through numerical experiments. 
Both two-dimensional and three-dimensional fluid-structure interactions are considered.
For two-dimensional problems, we employ reduced distribution functions to enhance the computational efficiency \cite{yang_rarefied_1995}, which are defined as
\begin{equation}
    h=\int_{-\infty}^{\infty} fd\xi_z,\quad b=\int_{-\infty}^{\infty}\xi_z^2 f d\xi_z,
\end{equation}
where $\xi_z$ represents the molecular velocity along $z$-direction. The non-dimensionalization is performed as
\begin{equation}
    \begin{split}
        &\hat x_i=\frac{x_i}{L},\quad \hat\rho=\frac{\rho}{\rho_0},\quad \hat T=\frac{T}{T_0},\quad \hat\xi_i=\frac{\xi_i}{\sqrt{2RT_0}},\quad \hat U_i=\frac{U_i}{\sqrt{2RT_0}},\quad \hat\mu=\frac{\mu}{\rho_0L\sqrt{2RT_0}},\\
        &\hat p=\frac{p}{2\rho_0RT_0},\quad \hat q_i=\frac{q_i}{\rho_0(2RT_0)^{3/2}},\quad \hat h = \frac{h}{\rho_0/(2RT_0)},\quad \hat b=\frac{b}{\rho_0},
    \end{split}
\end{equation}
where $L$ represents a macroscopic characteristic length. The subscript $0$ denotes the reference state. For brevity, we omit the hat notation henceforth. With above non-dimensionalization, the reduced Maxwellian and corrected equilibrium distributions write
\begin{equation}
    \begin{split}
        &H_k=\frac{\rho}{\pi T}e^{-\frac{c_k^2}{T}},\quad B_k=\frac{T}{2}H_k\\
        &H_k^S = H_k\left[1+\frac{4(1-Pr)}{5\rho T^2}\bm c_k\cdot\bm q\left(\frac{2c_k^2}{T}-4\right)\right]\\
        &B_k^S=B_k\left[1+\frac{4(1-Pr)}{5\rho T^2}\bm c_k\cdot\bm q\left(\frac{2c_k^2}{T}-2\right)\right]
    \end{split}.
\end{equation}
Argon gas is used in all numerical experiments and variable hard sphere (VHS) model is adopted to determine the reference dynamic viscosity,
\begin{equation}
    \mu_0=\frac{5(\alpha+1)(\alpha+2)\sqrt{\pi}}{4\alpha(5-2\omega)(7-2\omega)}\mathrm{Kn},
\end{equation}
where $\alpha=1.0$ and $\omega=0.81$ for Argon. The local viscosity is determined by
\begin{equation}
    \mu=\mu_0\left(\frac{T}{T_0}\right)^\omega.
    \label{eq:viscosity}
\end{equation}
The convergence criterion is set as
\begin{equation}
    \mathrm{res}=\max_i\frac{\sqrt{\sum_j^N (\bm{W}_{i,j}^{n+1}-\bm{W}_{i,j}^{n})^2}}{\sum_j^N |\bm{W}_{i,j}^{n+1}|}<\mathrm{TOL},
\end{equation}
where $W_{i,j}^n$ represents the $i$-th macroscopic conservative variable in the $j$-th physical cell and $n$-th time step, $N$ denotes the total number of the mesh in physical space. In this paper, we set the tolerance as TOL$=10^{-6}$.
The forward Euler scheme is adopted for time marching in all cases.

\subsection{Convergence test}
\label{sec:convergence}
We first test the convergence of the solution algorithm using a lid-driven cavity flow, and an additional stationary cylinder with diameter $D = 0.25$ is positioned at the center of the cavity. The characteristic length is defined as the cavity length ($L = 1.0$), while the lid velocity is set to $U_{\rm w} = 0.15$. The temperature of all solid walls are set as $T_{\rm w} = 1.0$. 
The Knudsen number in the reference state is $0.075$.

\begin{figure}[h]
    \centering
    \includegraphics[width=0.6\linewidth]{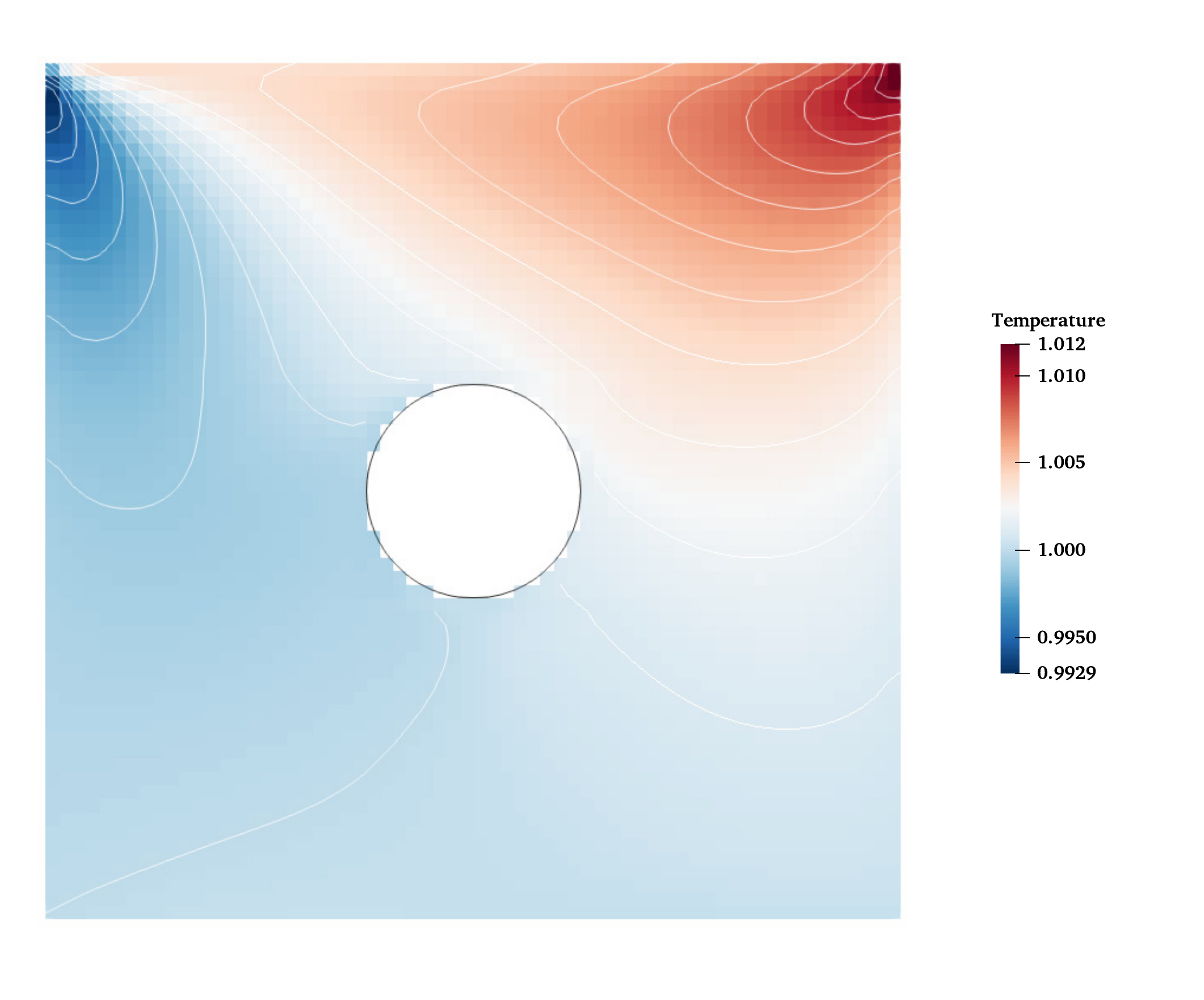}
    \caption{The converged temperature contour with the velocity space mesh of $200\times 200$.}
    \label{fig:converged field}
\end{figure}

We first test the convergence in the velocity space.
A series of uniform collocation points with resolutions of $40\times40$, $60\times60$, $80\times80$, and $100\times100$ are used in the truncated velocity space $[-5,5]^2$. 
The converged solution on the velocity mesh of $200\times 200$ is adopted as the reference, which is shown in \Cref{fig:converged field}.
The Cartesian mesh in the physical space is fixed at $64\times 64$.

\begin{figure}[htbp!]
    \centering
    \begin{subfigure}[t]{0.49\textwidth}
        \includegraphics[width=\textwidth]{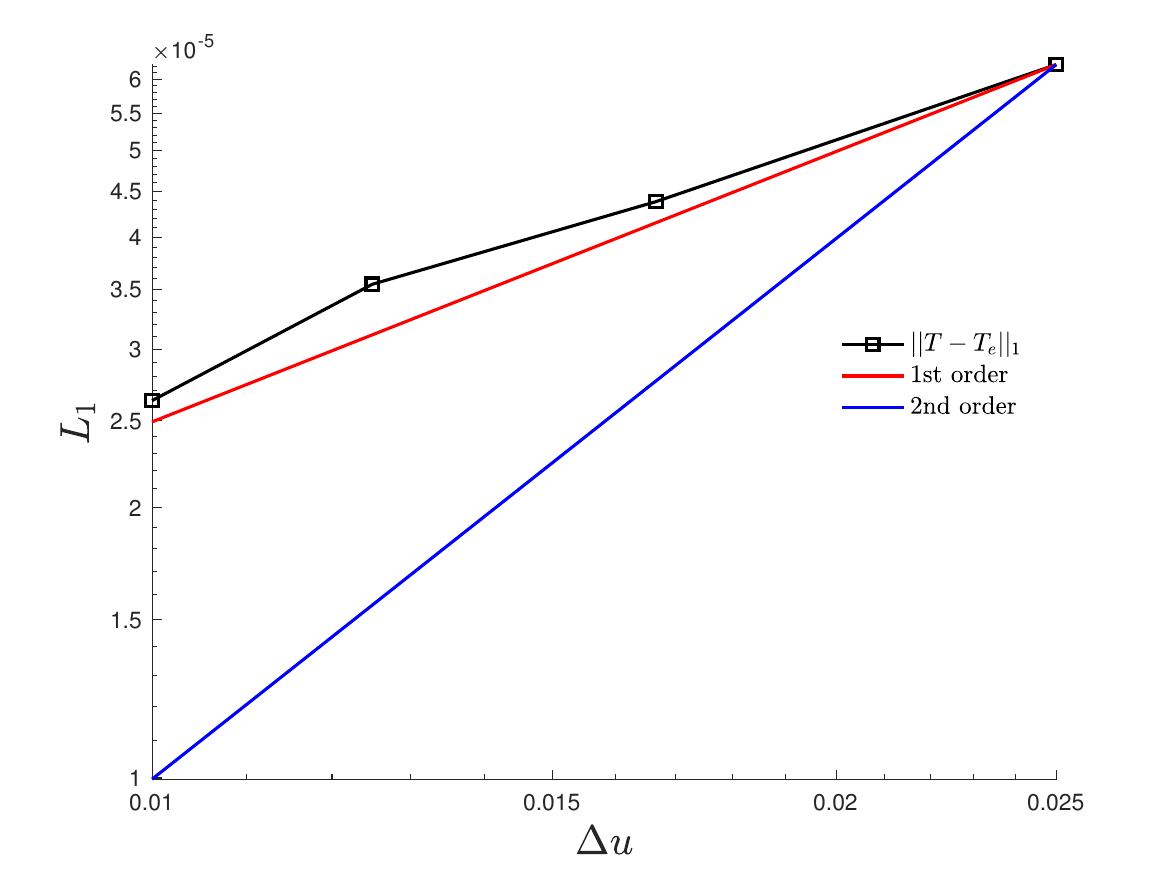}
        \caption{Without cut-cell correction}
        \label{fig:noncvc_order}
    \end{subfigure}
    \hfil
    \begin{subfigure}[t]{0.49\textwidth}
        \includegraphics[width=\textwidth]{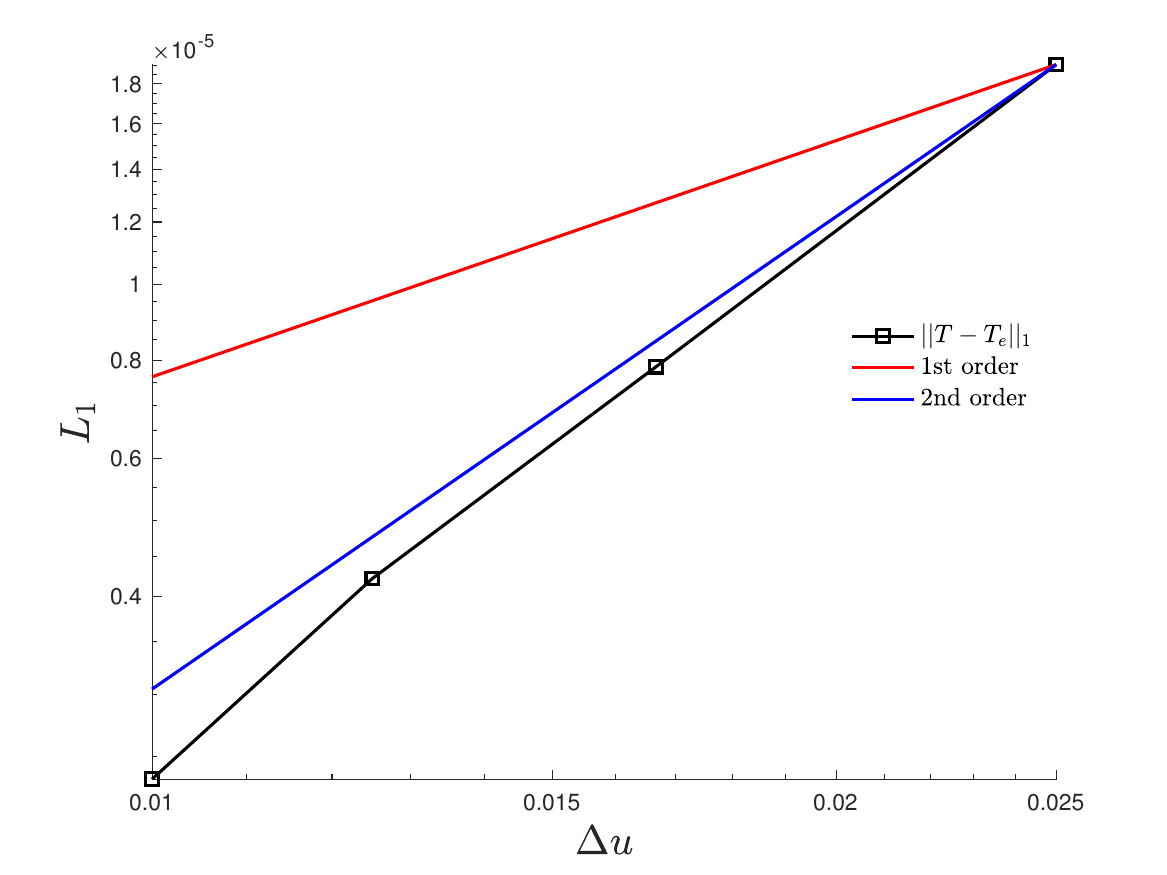}
        \caption{With cut-cell correction}
        \label{fig:cvc_order}
    \end{subfigure}
    \caption{The comparison of the variation of $L_1$ error with velocity space mesh resolution with and without the cut-cell correction in velocity space in the lid-driven cavity flow.}
    \label{fig:convergence comparison}
\end{figure}

\begin{figure}[htbp!]
    \centering
    \begin{subfigure}[t]{0.49\textwidth}
        \includegraphics[width=\textwidth]{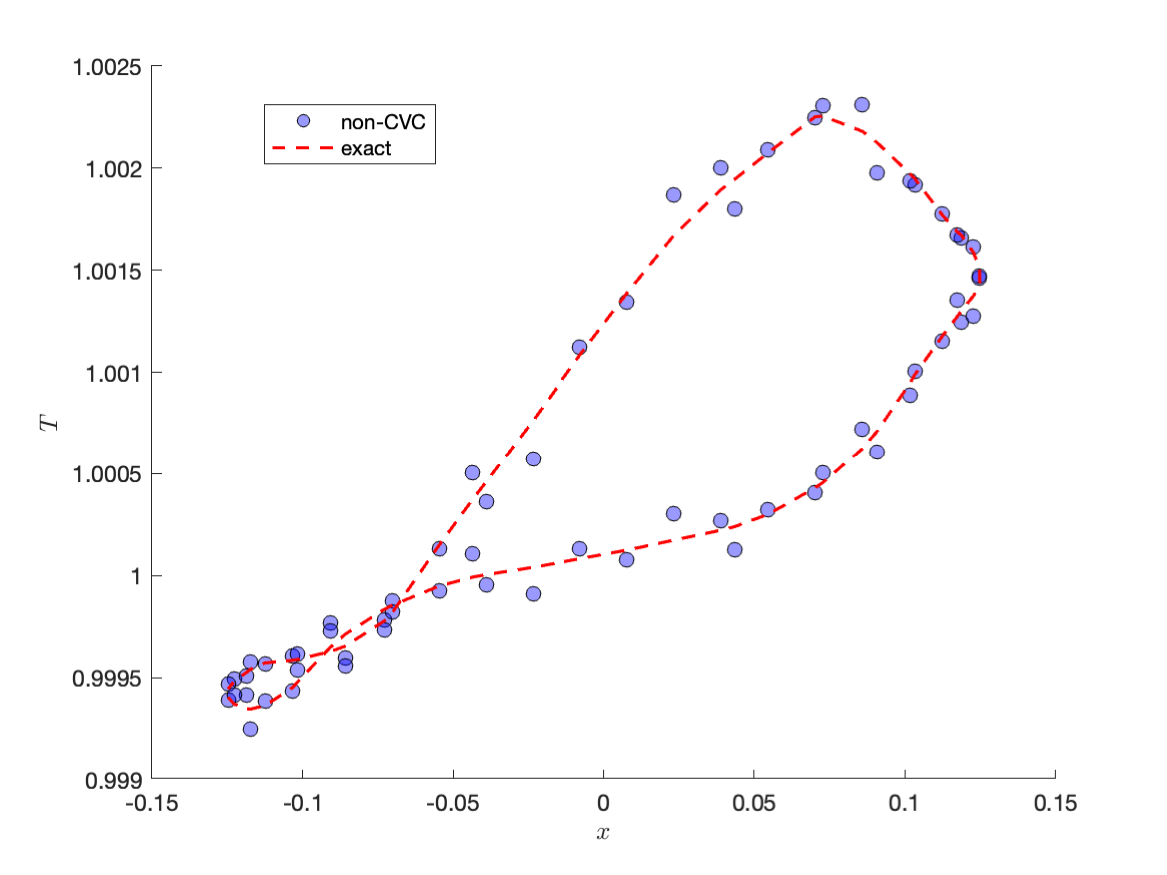}
        \caption{Without cut-cell correction}
        \label{fig:noncvcT}
    \end{subfigure}
    \hfil
    \begin{subfigure}[t]{0.49\textwidth}
        \includegraphics[width=\textwidth]{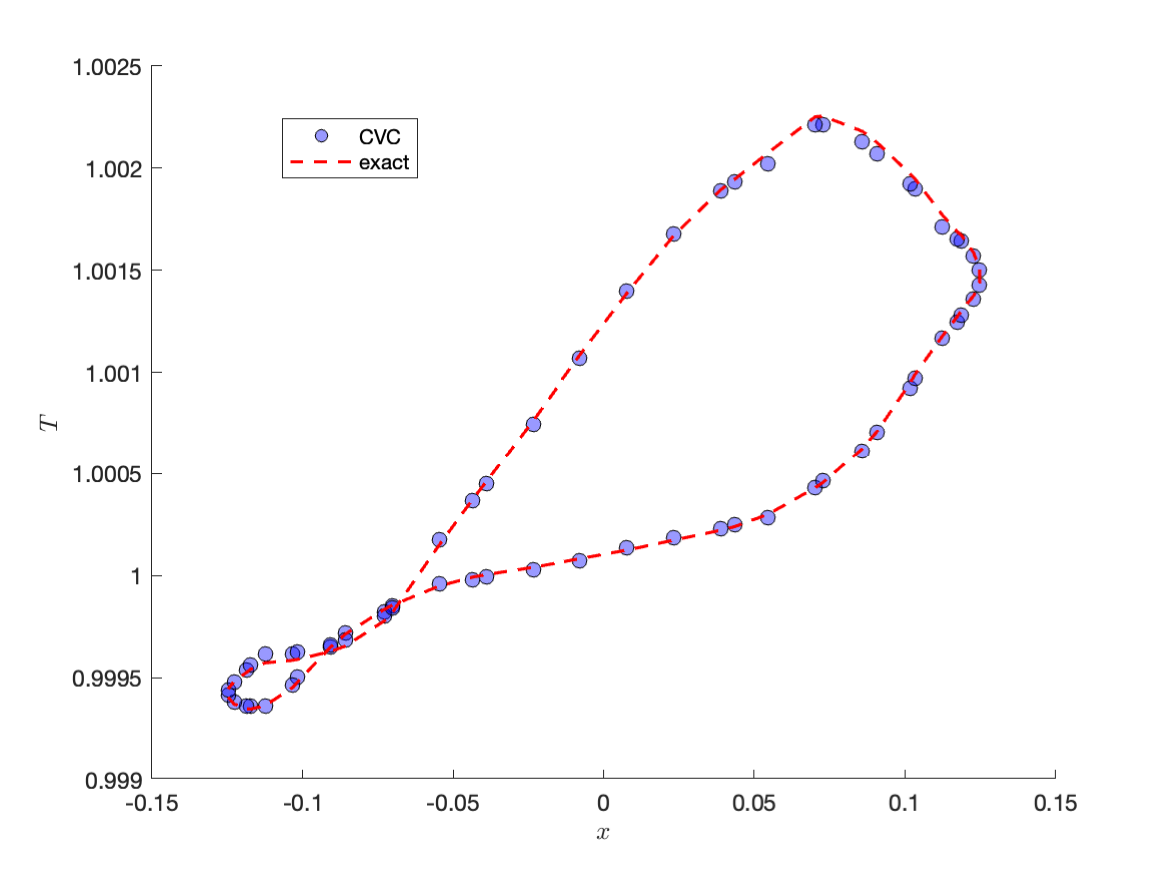}
        \caption{With cut-cell correction}
        \label{fig:cvcT}
    \end{subfigure}
    \caption{The comparison of cylinder surface temperatures with and without the cut-cell method using the same velocity mesh number of $40\times 40$ in the lid-driven cavity flow.}
    \label{fig:temperature comparison}
\end{figure}

The order of convergence is shown in \Cref{fig:convergence comparison}. The $L_1$ error of the temperature is measured at the wall of the cylinder, i.e., the intersect points in \Cref{fig:ghost cell}. The solution algorithm with and without the cut-cell method exhibits second- and first-order accurate convergence, respectively, which is consistent with the analysis in \Cref{sec:CVC}.
\Cref{fig:temperature comparison} shows the comparison of the wall temperature with the same velocity mesh of $40\times 40$. It is clear that the accuracy is remarkably improved and further validates the effectiveness of the proposed cut-cell methodology in velocity space.

\begin{figure}[htbp!]
    \centering
    \begin{subfigure}[t]{0.49\textwidth}
        \includegraphics[width=\textwidth]{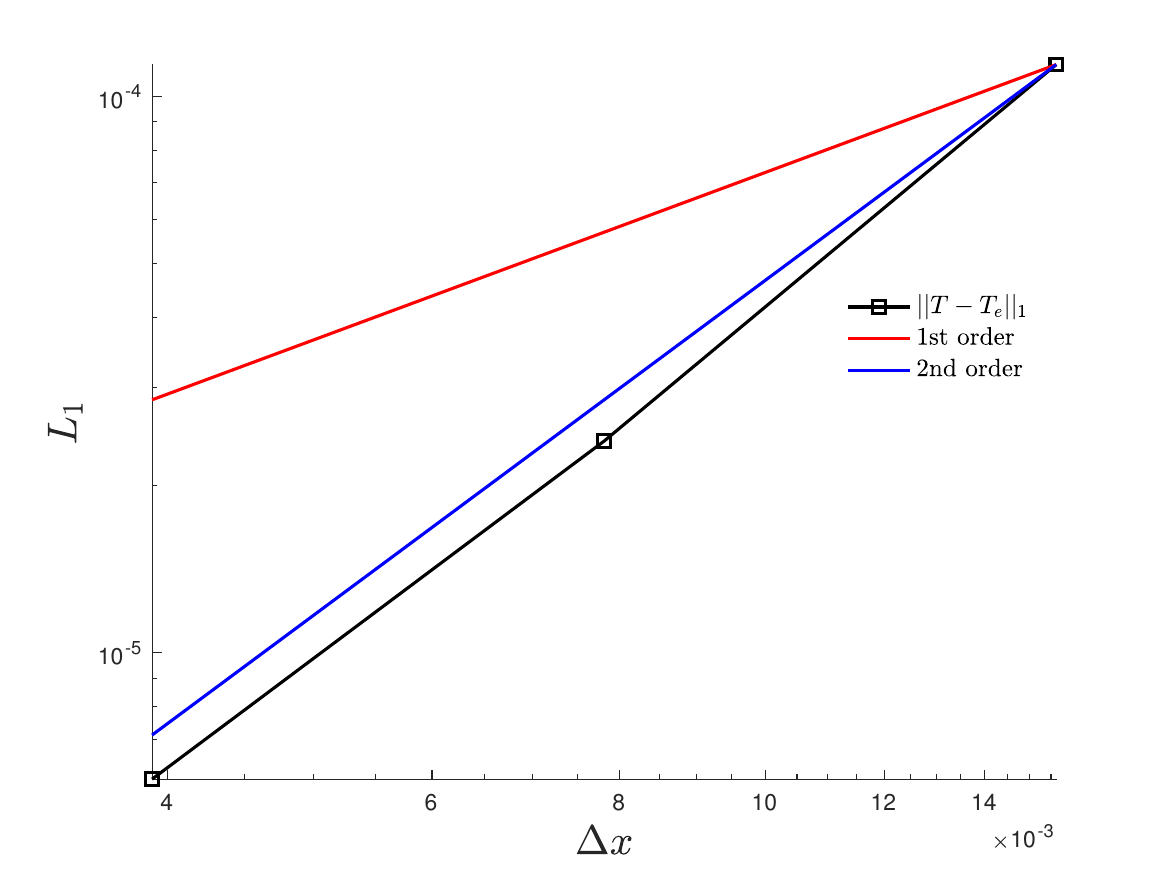}
        \caption{$L_1$ error}
        \label{fig:ps_L1}
    \end{subfigure}
    \hfil
    \begin{subfigure}[t]{0.49\textwidth}
        \includegraphics[width=\textwidth]{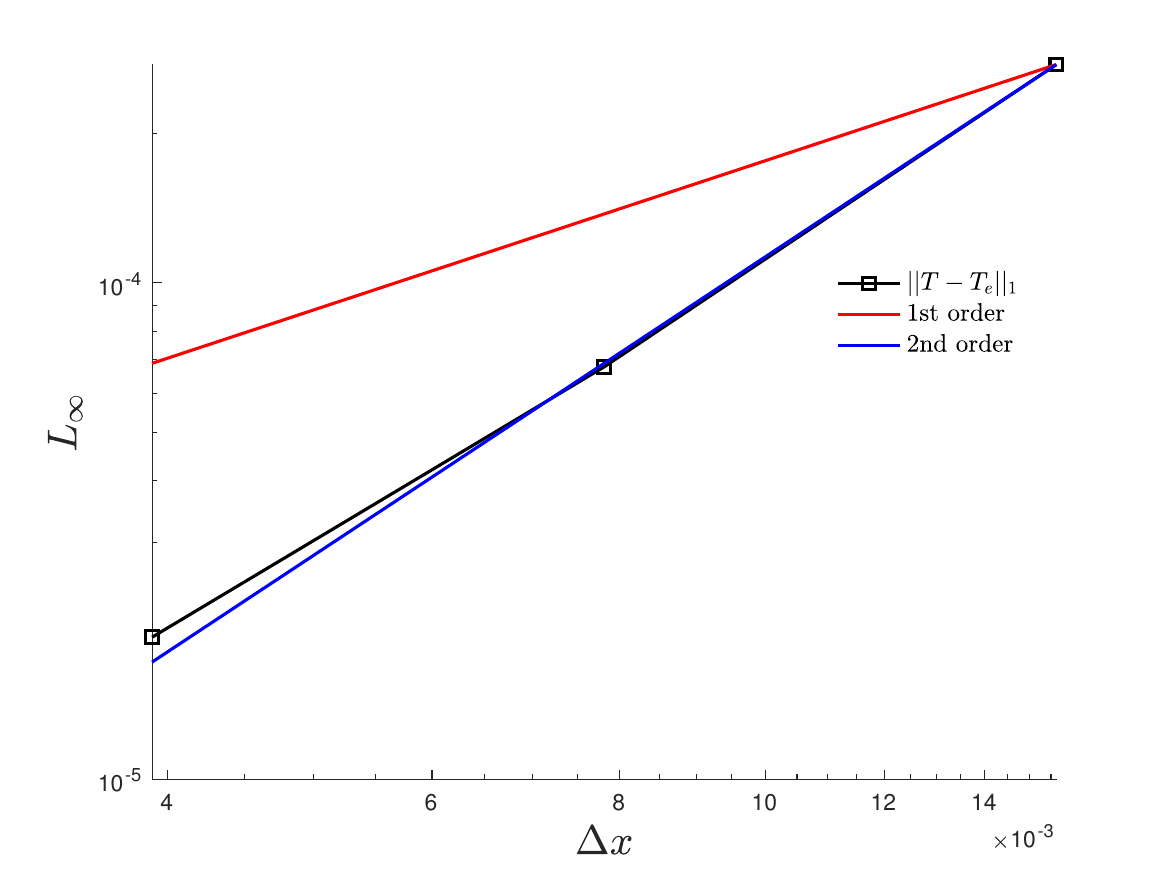}
        \caption{$L_\infty$ error}
        \label{fig:ps_L_infty}
    \end{subfigure}
    \caption{The variation of errors with physical mesh resolution, which validates the second-order convergence of the solution algorithm in the lid-driven cavity flow.}
    \label{fig:ps convergence}
\end{figure}

For convergence testing in physical space, since numerical solution obtained with fine mesh remain inevitably susceptible to the zigzag effect, we construct an exact reference solution of the form
\begin{equation}
    \begin{split}
        &h_k=\frac{\rho^*}{\pi T^*}e^{-\frac{\bm\xi_k-\bm U^*}{T^*}},\quad b_k=\frac{T^*}{2}h_k,\\
        &\rho^*=1+0.1\sin{\pi(x+y)},\quad T^*=1+0.1\cos{\pi(x+y)},\\
        &\bm U^*=0.4[\sin{\pi(x+y)},\cos{\pi(x+y)}]^{\rm T}.
    \end{split}
\end{equation}
%
%
%
For the above solution, we can analytically determine the distribution function at the wall of the cylinder, and use it as the benchmark solutions to measure the numerical error introduced by the numerical method in a time step. 
A series of uniform Cartesian mesh in physical space with the resolution of $64\times64$, $128\times 128$, and $256\times 256$ are employed. 
The velocity collocation points are fixed at $120\times120$.
The $L_1$ and $L_\infty$ errors of the temperature are shown in \Cref{fig:ps convergence}, where the second-order spatial accuracy proved in \Cref{sec:analysis} is validated.

\subsection{Hypersonic flow around circular cylinder}
\label{sec:cylinder}

\begin{figure}[htbp!]
    \centering
    \includegraphics[width=0.6\linewidth]{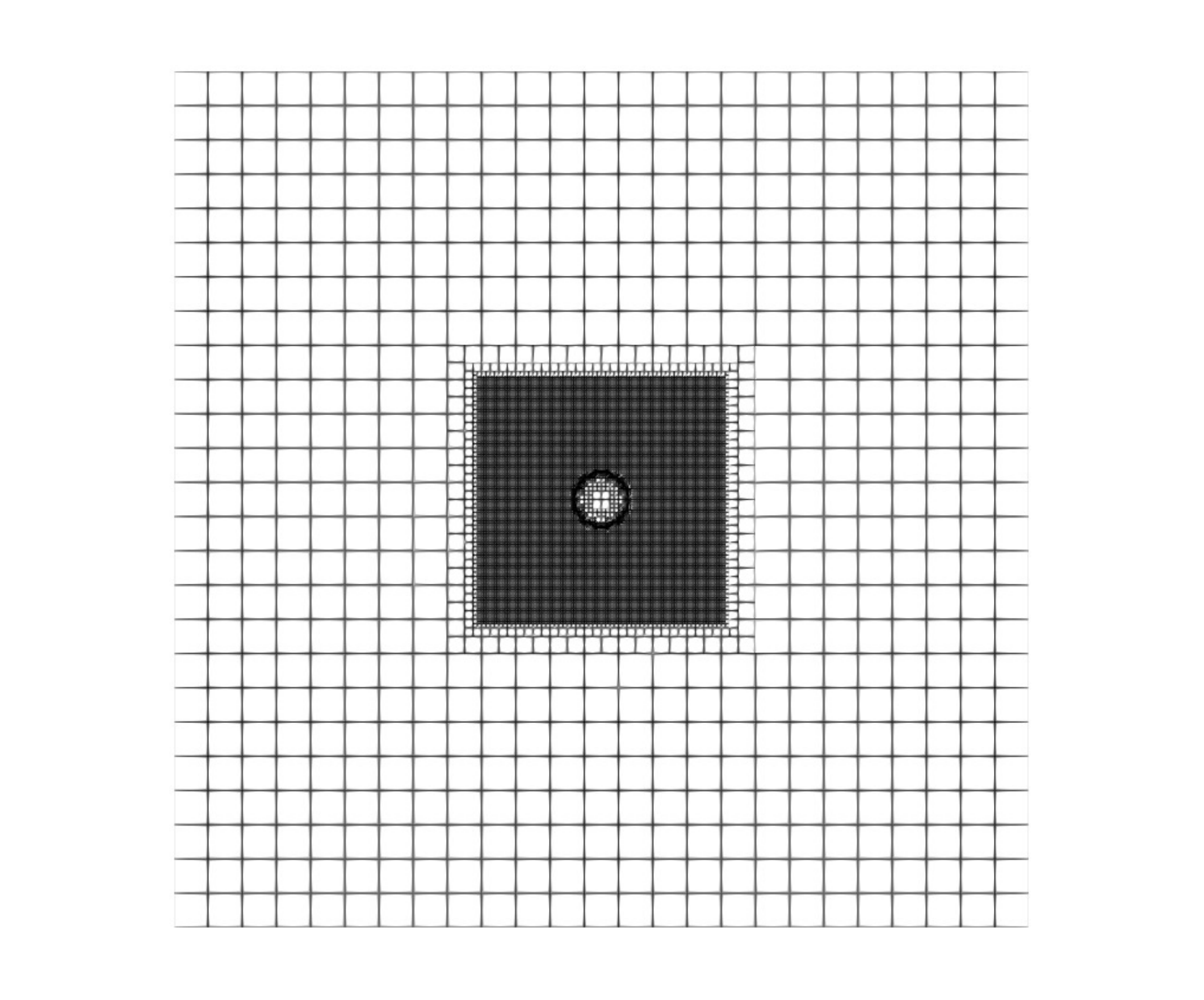}
    \caption{The physical mesh used for simulating the hypersonic flow around the cylinder.}
    \label{fig:cylinder mesh}
\end{figure}

The second numerical experiment is the hypersonic flow around a cylinder. The Mach number is set as $5.0$ and the Knudsen number is $0.1$. The radius of the cylinder serves as the characteristic length, i.e., the dimensionless radius of the cylinder $R_0=1.0$. The state of the incoming flow serves as the reference state, i.e. the dimensionless density $\rho_{\infty}=1.0$, and temperature $T_{\infty}=1.0$. The dimensionless freestream velocity is calculated with
\begin{equation}
    U_{\infty}=\mathrm{Ma}C_{\infty},
\end{equation}
where the dimensionless sound speed $C_{\infty}=\sqrt{\gamma T_{\infty}/2}$. For Argon gas, the specific heat ratio $\gamma=5/3$. The temperature of the cylinder wall is $T_{\rm w}=1.0$. 
The physical domain is set as $[-16,16]^2$, with a stationary cylinder positioned at $(0,0)$.
As shown in \Cref{fig:cylinder mesh}, a tree-based adaptive mesh refinement (AMR) strategy is employed to enhance the numerical efficiency, where the mesh size adjacent to the solid boundary is set as $0.01$ (configured with reference to the body-fitted mesh \cite{yang_efficient_2024}).
The velocity space is truncated as $[-10,10]\times[-10,10]$ and discretized with a uniform Cartesian mesh of $60\times 60$.

\begin{figure}[htbp!]
    \centering
    \begin{subfigure}[t]{0.45\textwidth}
        \includegraphics[width=\textwidth]{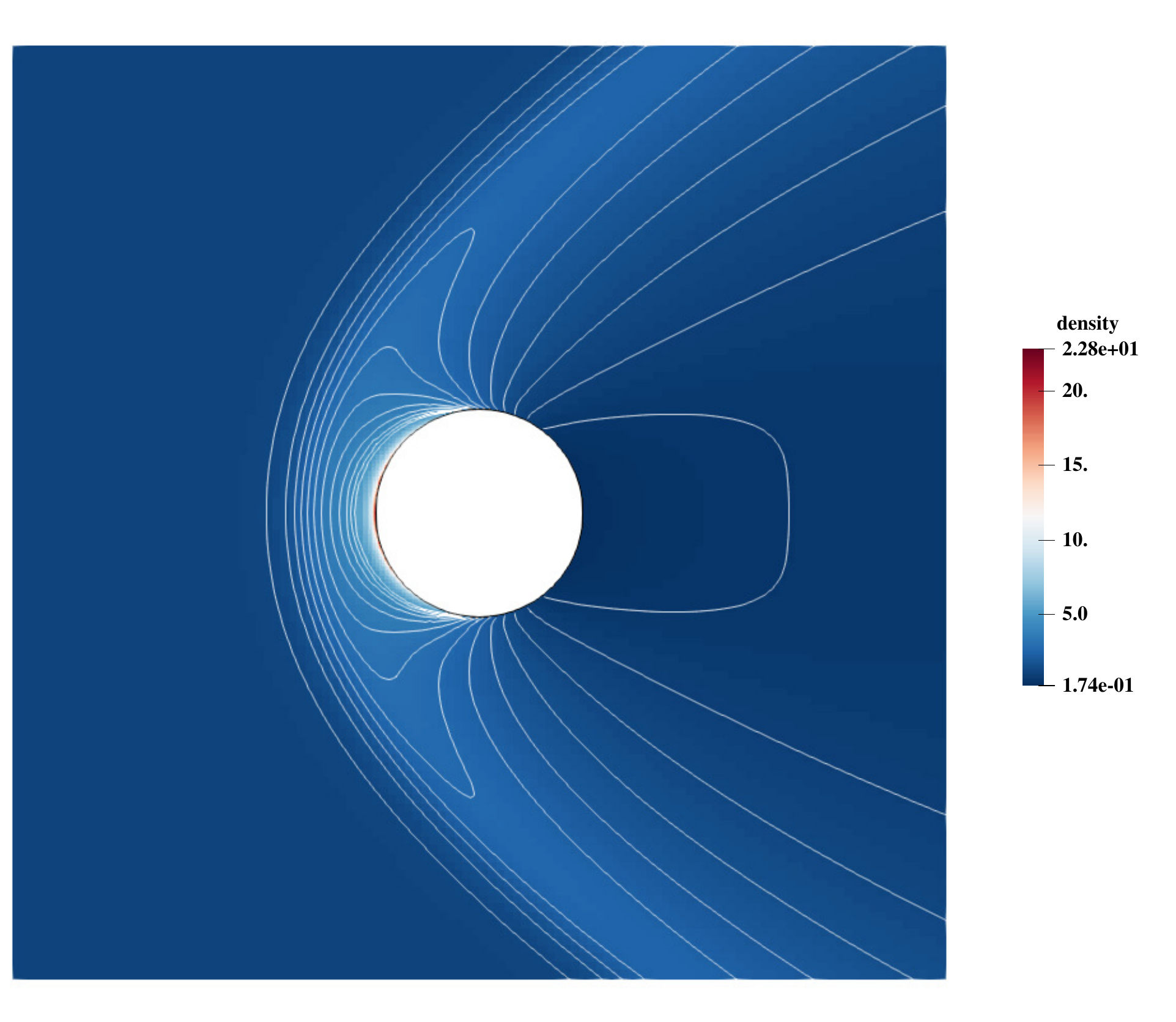}
        \caption{Density}
        \label{fig:cylinder density}
    \end{subfigure}
    \hfil
    \begin{subfigure}[t]{0.45\textwidth}
        \includegraphics[width=\textwidth]{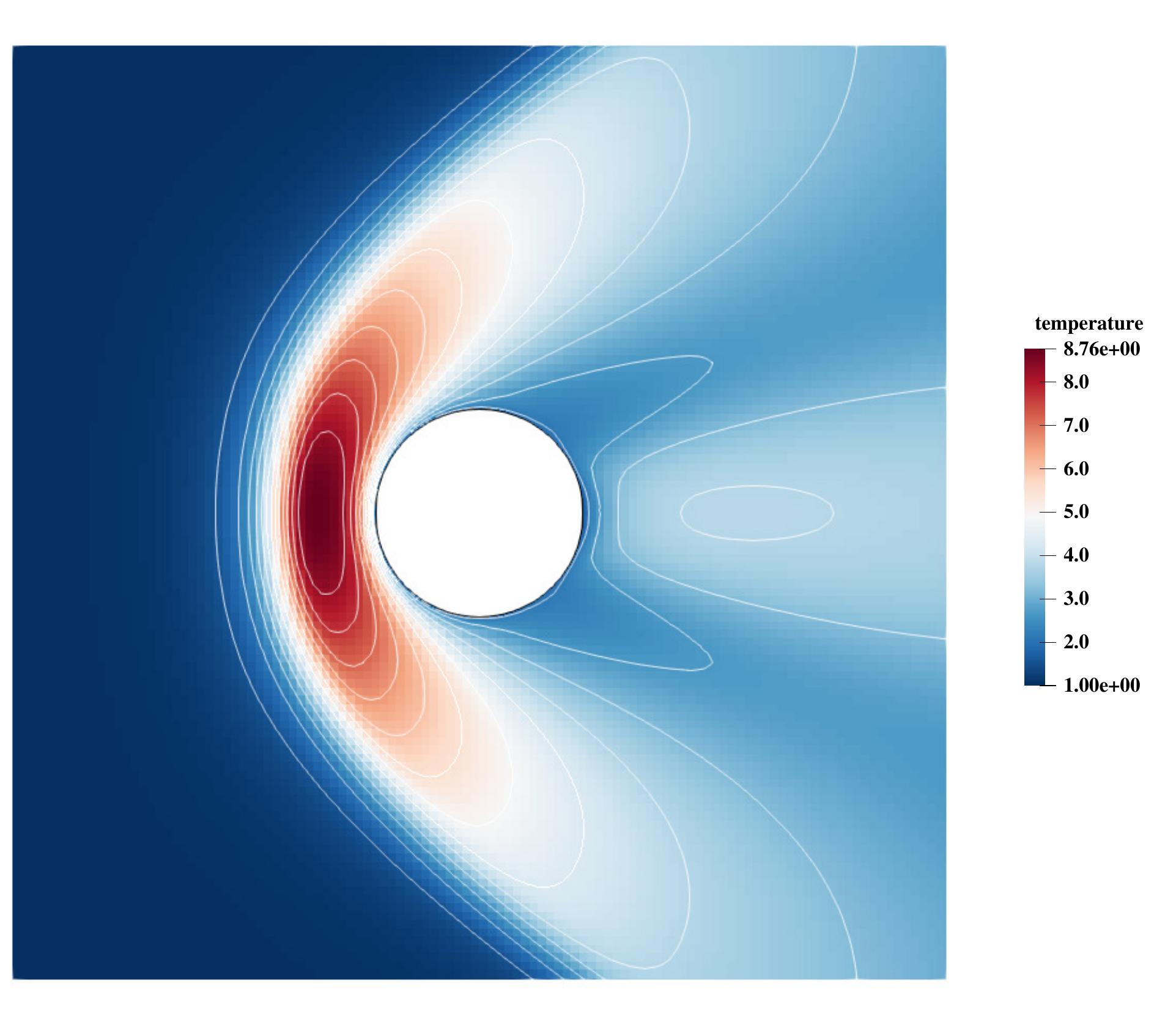}
        \caption{Temperature}
        \label{fig:cylinder temperature}
    \end{subfigure}
    \\
    \begin{subfigure}[t]{0.45\textwidth}
        \includegraphics[width=\textwidth]{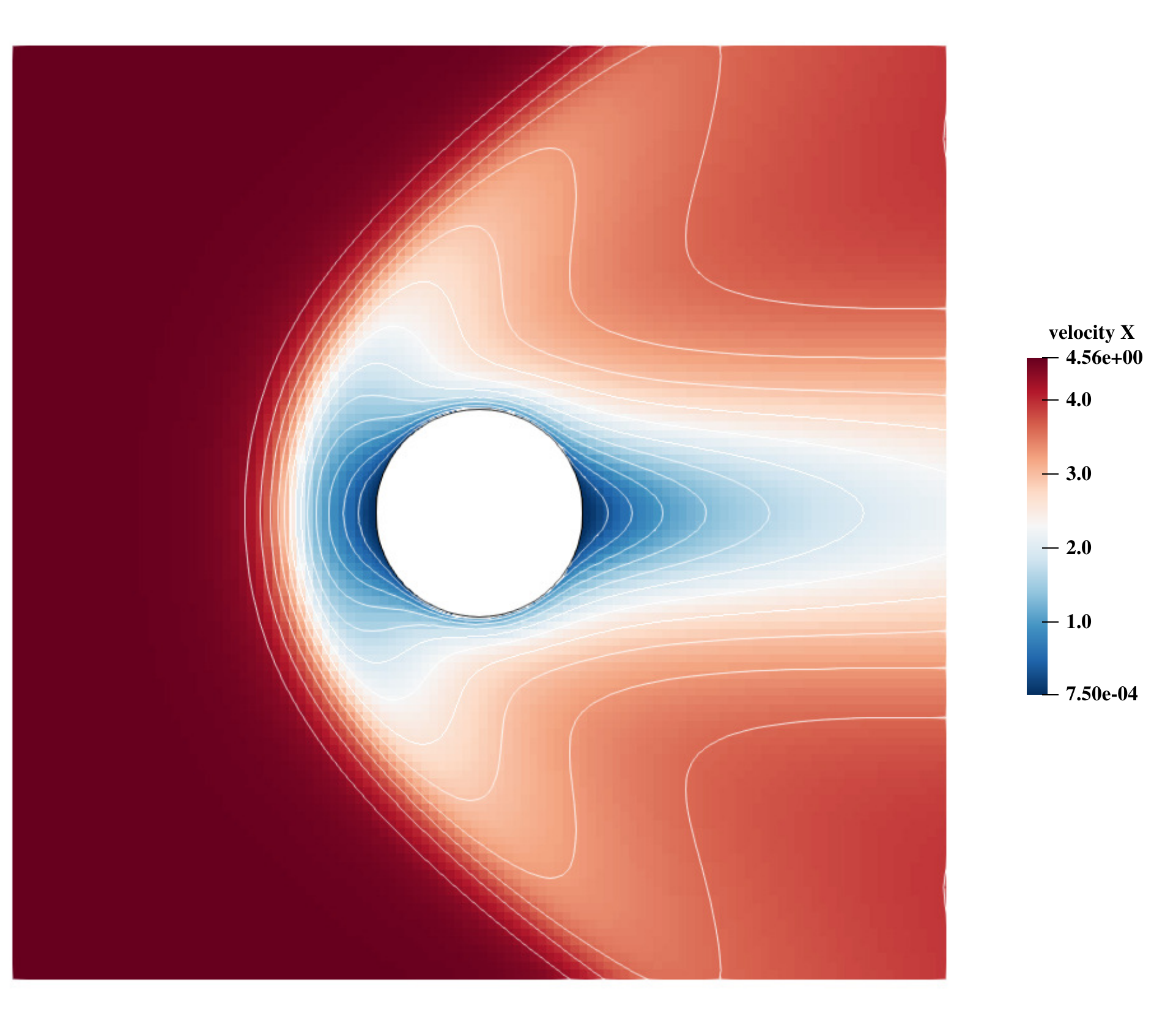}
        \caption{X-component of velocity}
        \label{fig:cylinder qx}
    \end{subfigure}
    \hfil
    \begin{subfigure}[t]{0.45\textwidth}
        \includegraphics[width=\textwidth]{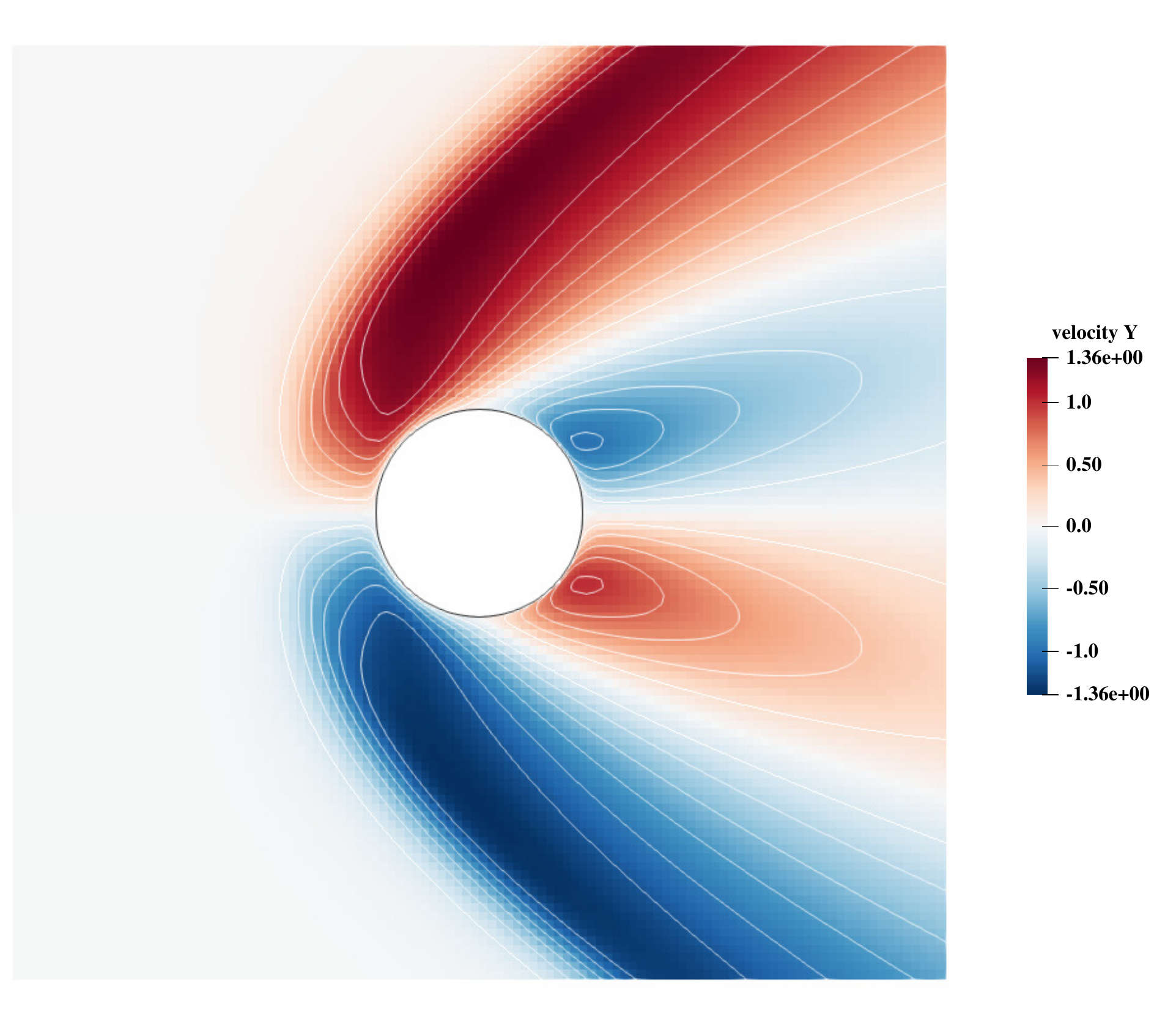}
        \caption{Y-component of velocity}
        \label{fig:cylinder qy}
    \end{subfigure}
    \\
    \begin{subfigure}[t]{0.45\textwidth}
        \includegraphics[width=\textwidth]{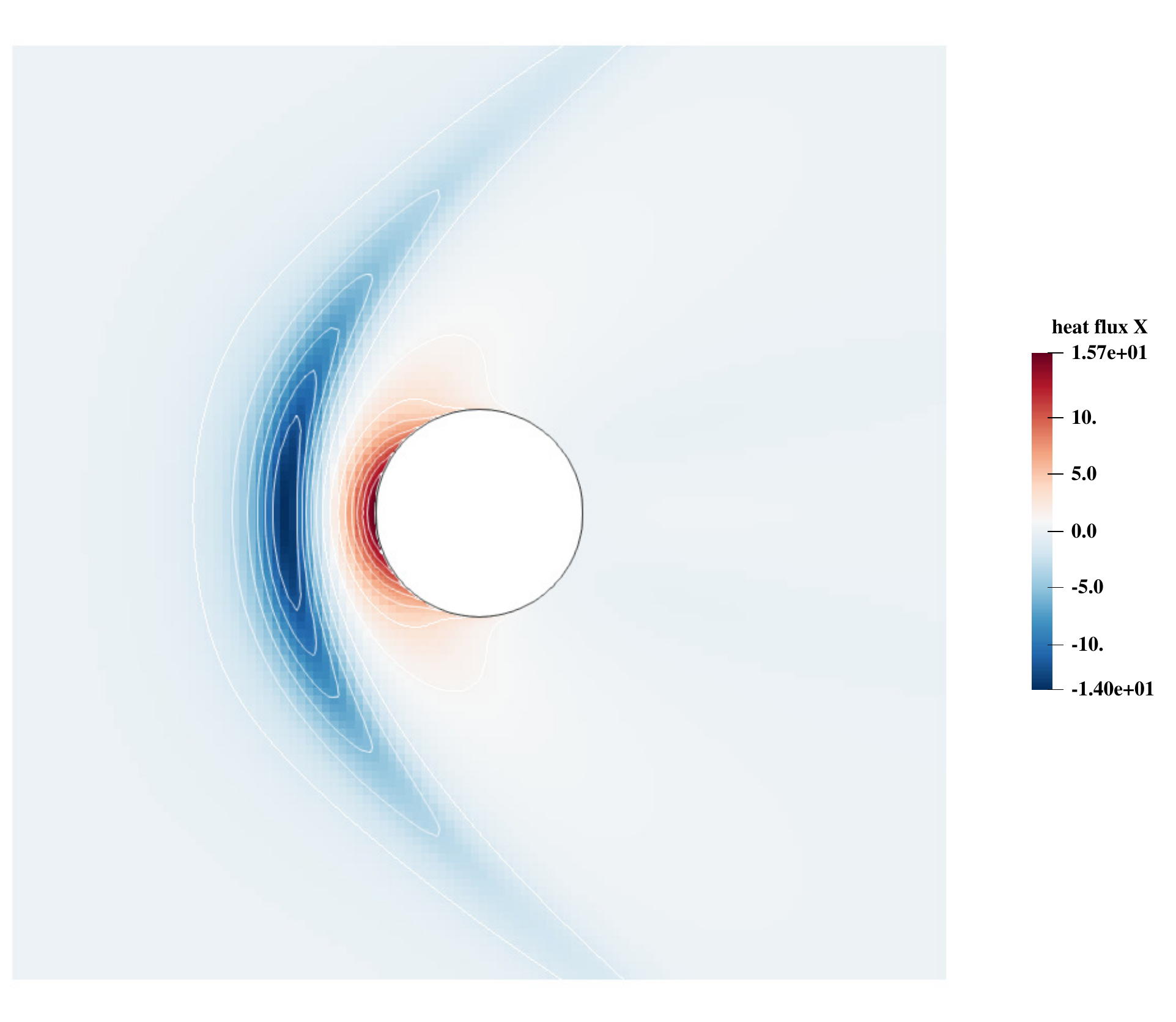}
        \caption{X-component of heat flux}
        \label{fig:cylinder qx}
    \end{subfigure}
    \hfil
    \begin{subfigure}[t]{0.45\textwidth}
        \includegraphics[width=\textwidth]{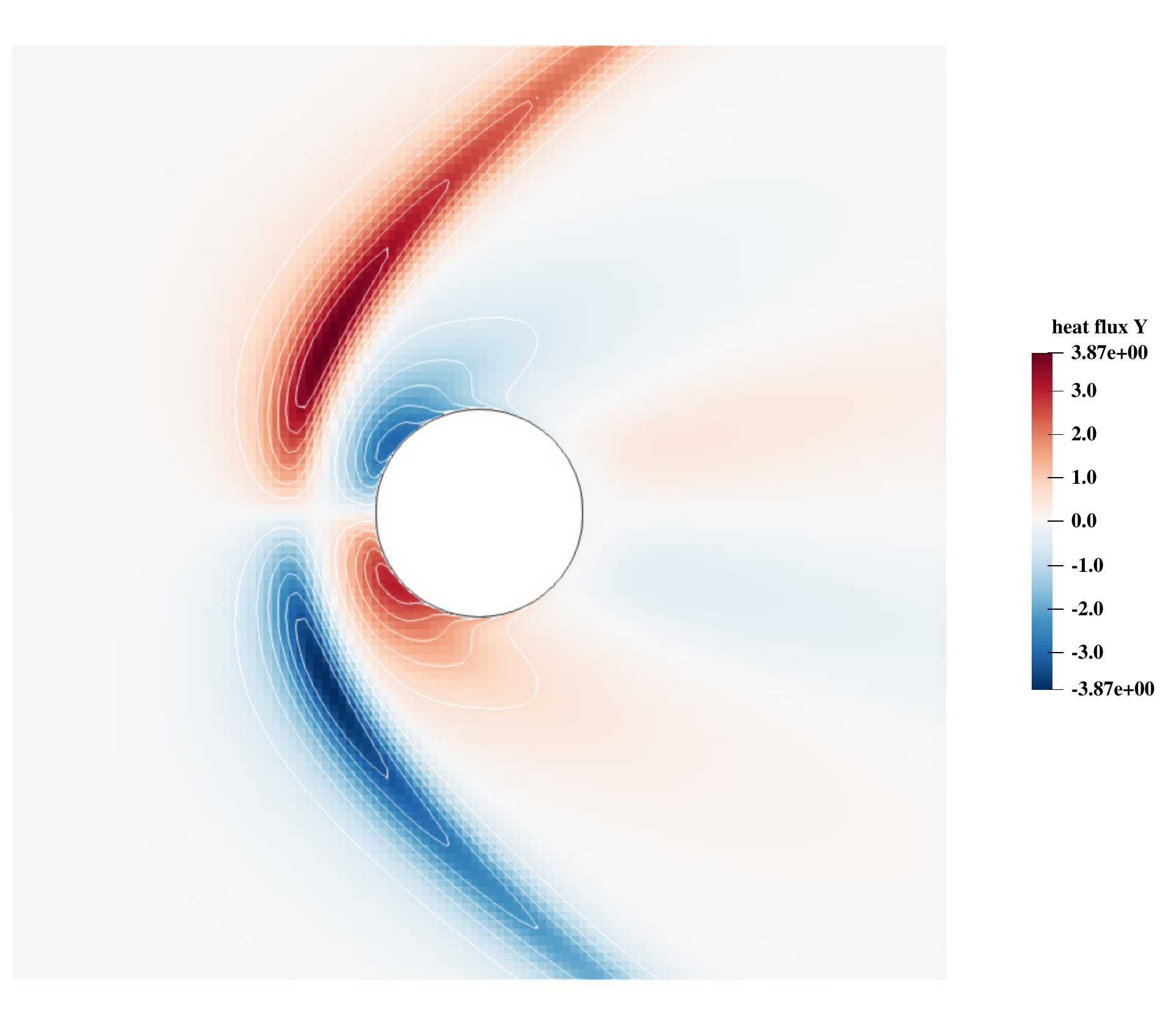}
        \caption{Y-component of heat flux}
        \label{fig:cylinder qy}
    \end{subfigure}
    \caption{The converged contours of macroscopic flow variables in the hypersonic flow around the cylinder.}
    \label{fig:cylinder field}
\end{figure}

\begin{figure}[htbp!]
    \centering
    \begin{subfigure}[t]{0.49\textwidth}
        \includegraphics[width=\textwidth]{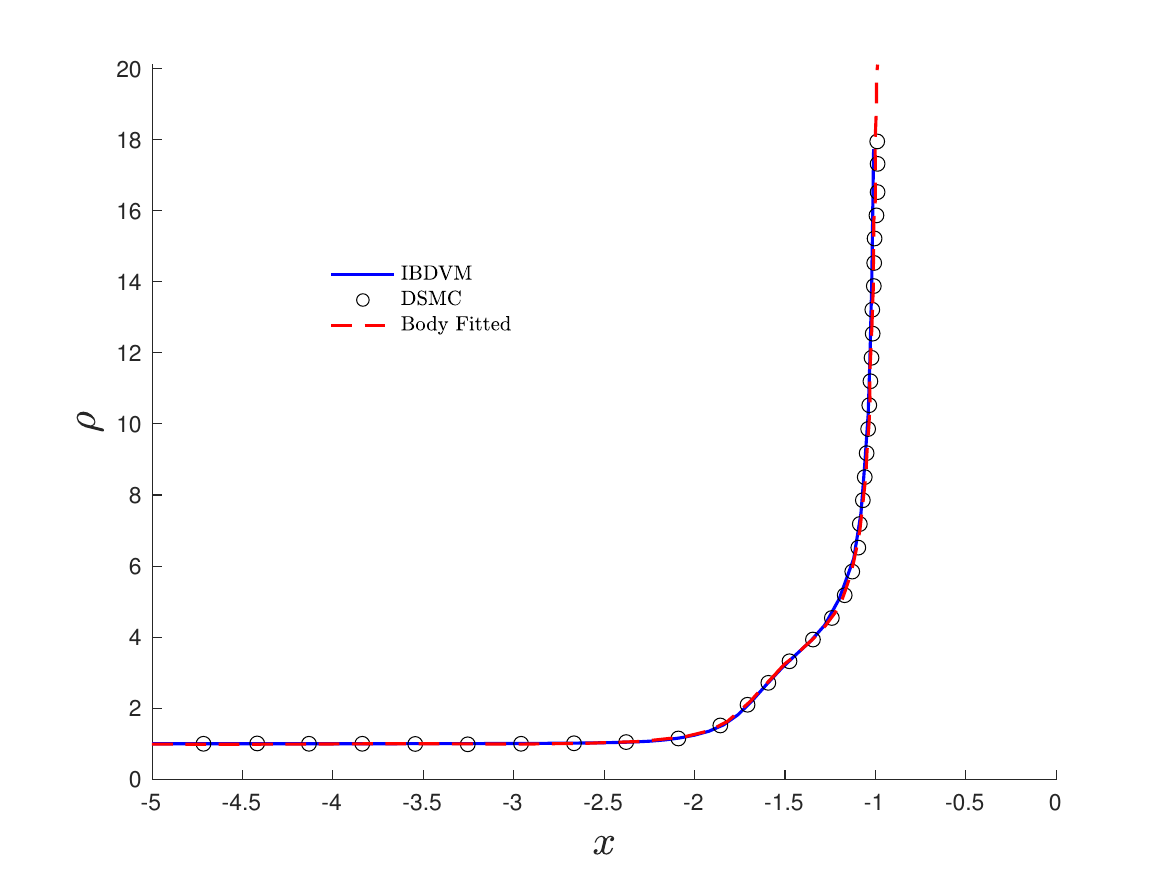}
        \caption{Density along the stagnation line}
        \label{fig:cylinder SL_rho}
    \end{subfigure}
    \hfil
    \begin{subfigure}[t]{0.49\textwidth}
        \includegraphics[width=\textwidth]{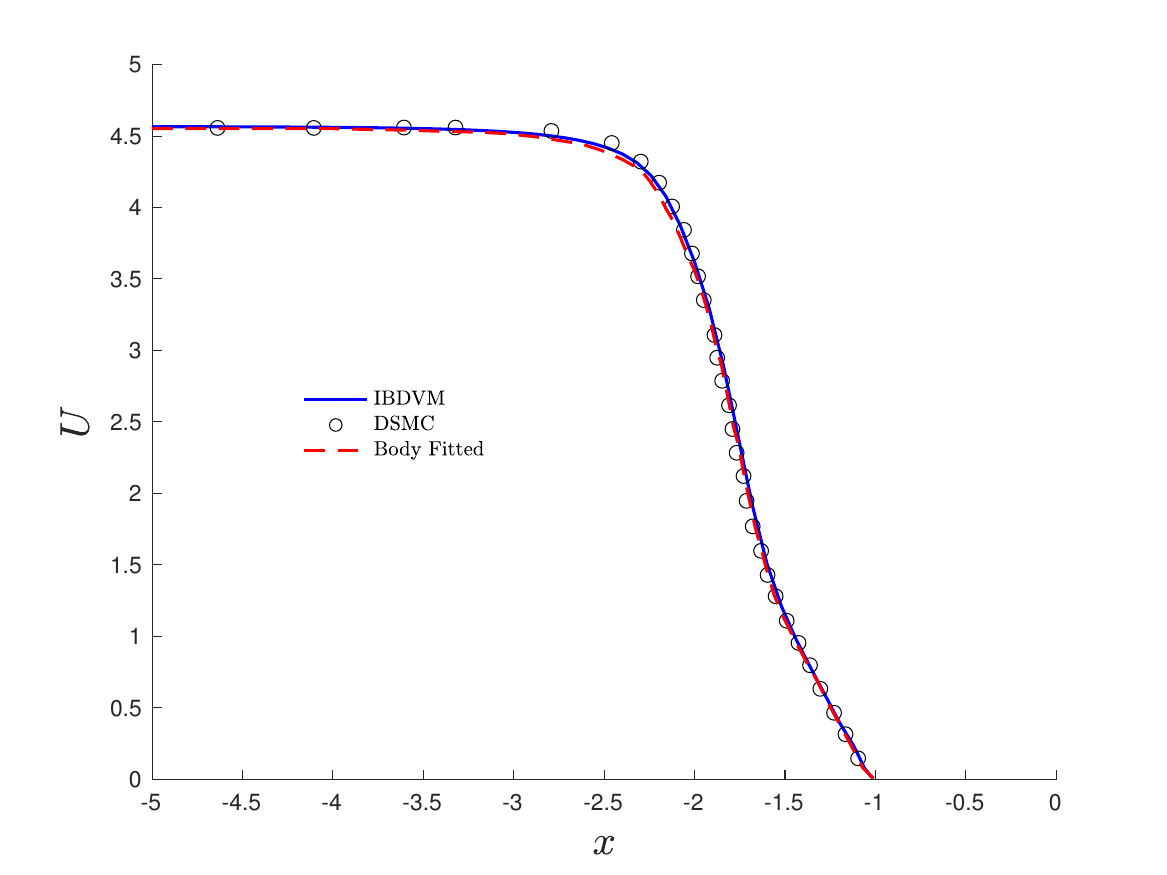}
        \caption{X-component velocity along the stagnation line}
        \label{fig:cylinder SL_U}
    \end{subfigure}
    \\
    \begin{subfigure}[t]{0.49\textwidth}
        \includegraphics[width=\textwidth]{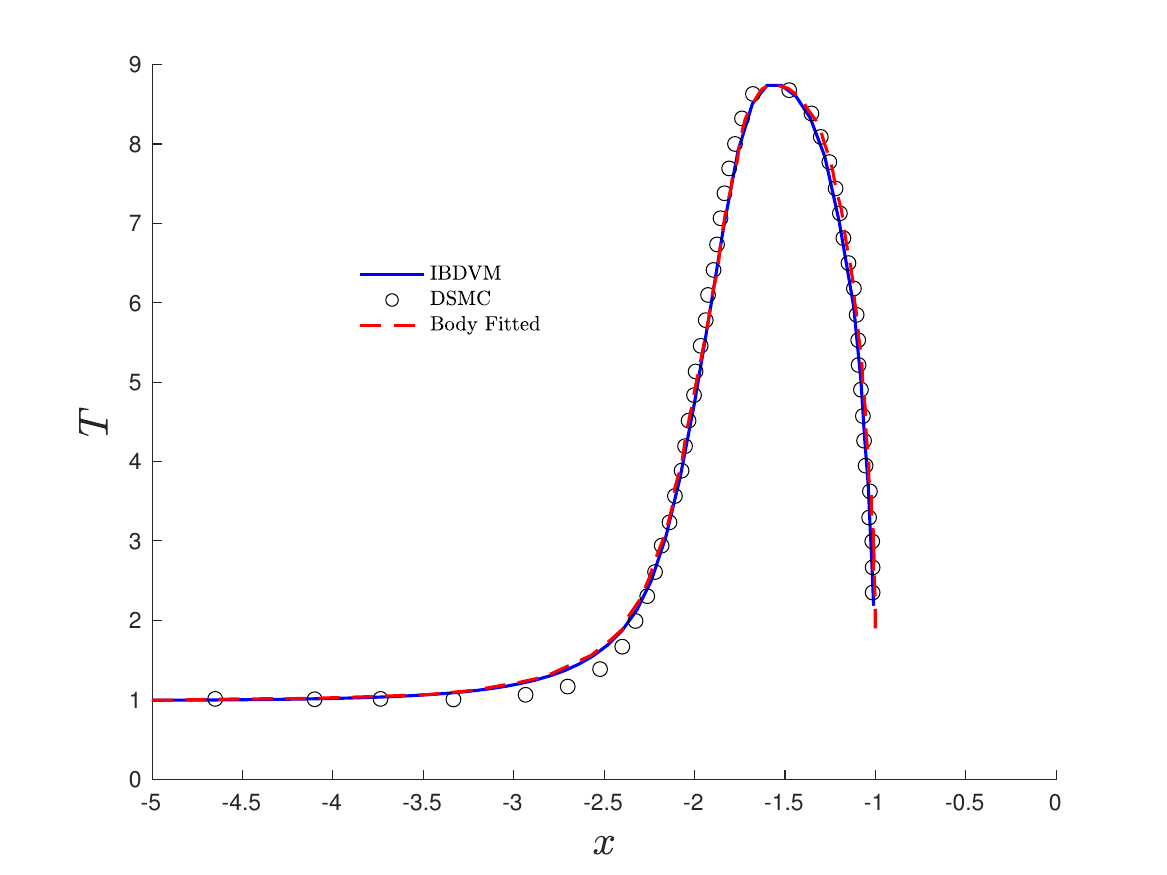}
        \caption{Temperature along the stagnation line}
        \label{fig:cylinder SL_T}
    \end{subfigure}
    \hfil
    \begin{subfigure}[t]{0.49\textwidth}
        \includegraphics[width=\textwidth]{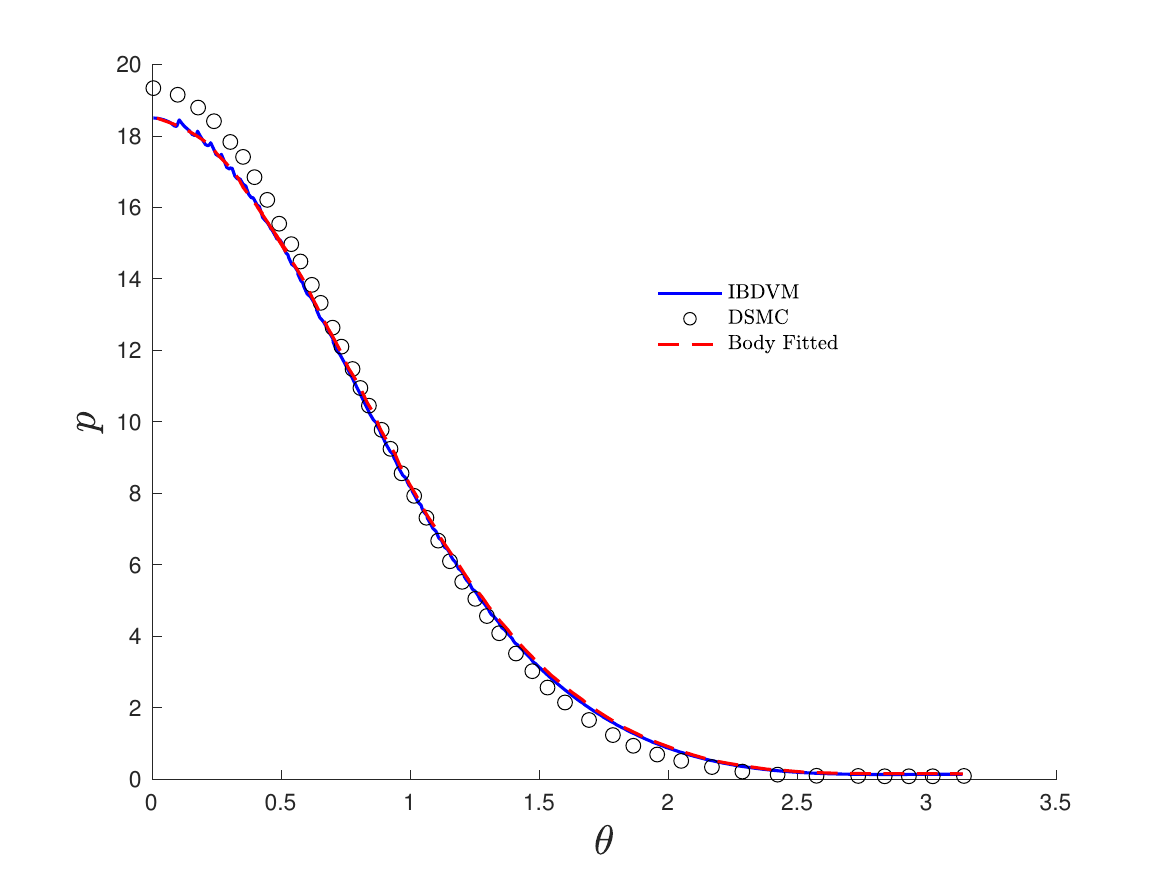}
        \caption{Pressure along the surface of the cylinder}
        \label{fig:cylinder p}
    \end{subfigure}
    \\
    \begin{subfigure}[t]{0.49\textwidth}
        \includegraphics[width=\textwidth]{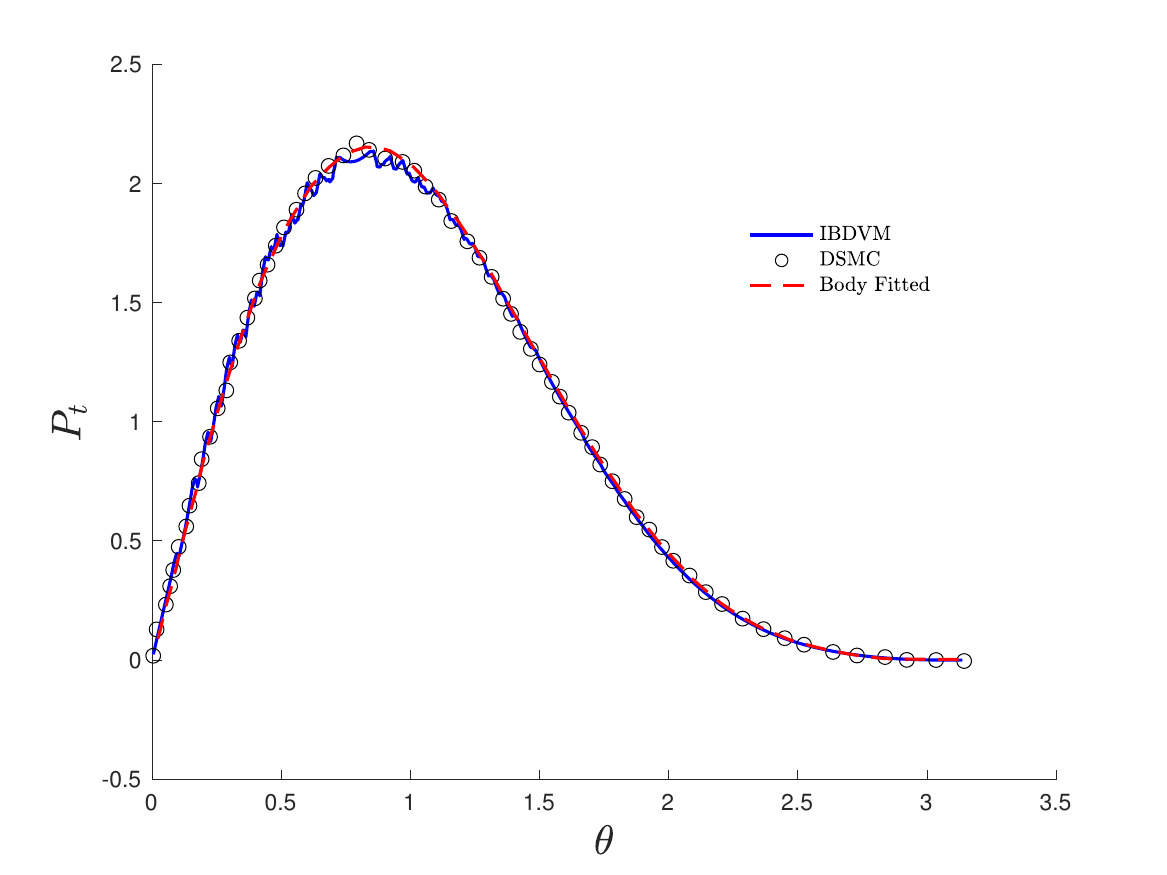}
        \caption{Shear stress along the surface of the cylinder}
        \label{fig:cylinder Pt}
    \end{subfigure}
    \hfil
    \begin{subfigure}[t]{0.49\textwidth}
        \includegraphics[width=\textwidth]{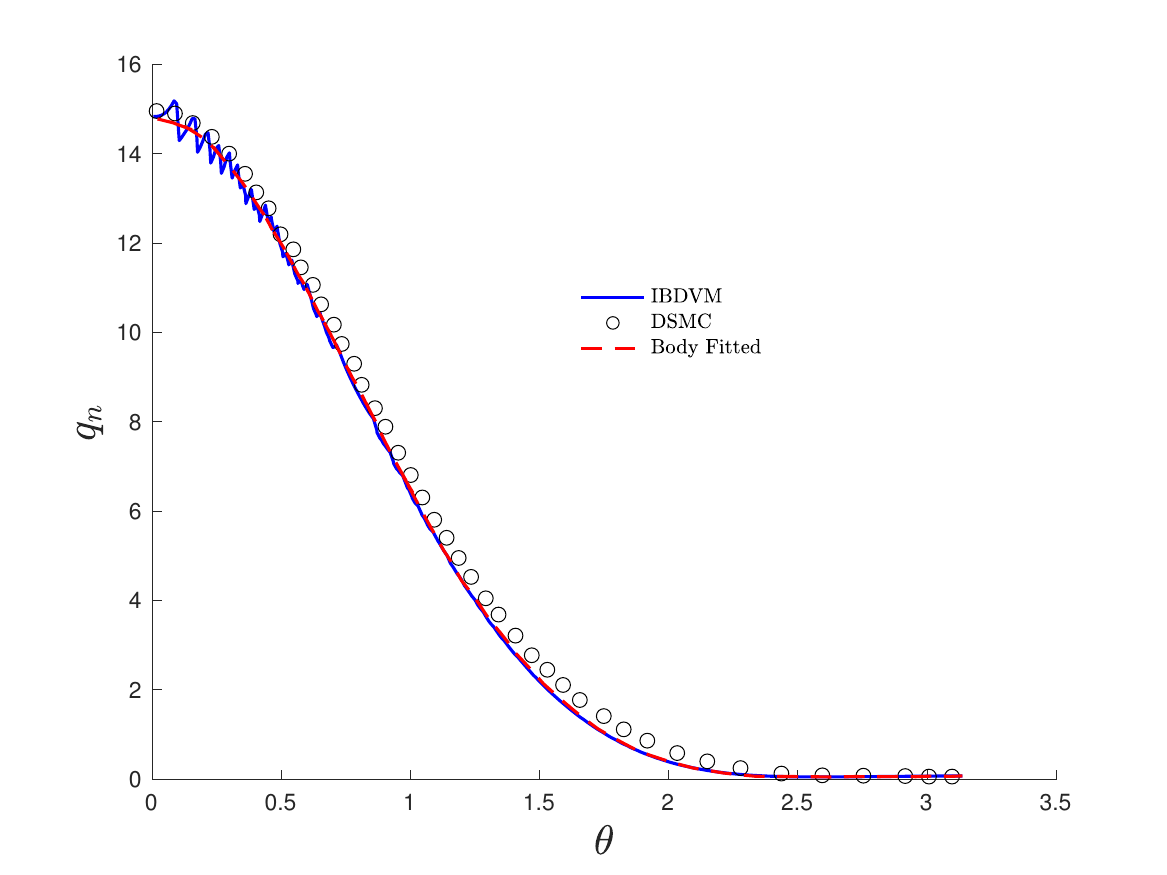}
        \caption{Normal heat flux along the surface of the cylinder}
        \label{fig:cylinder qn}
    \end{subfigure}
    \caption{The distribution of macroscopic flow variables along the stagnation line and surface of the cylinder in the hypersonic flow around the cylinder. The reference results of DSMC and the body-conformal DVM solver are extracted from \cite{yang_efficient_2024}.}
    \label{fig:cylinder quantities}
\end{figure}

The convergent distributions of density, velocity, temperature, and heat flux are presented in \Cref{fig:cylinder field}.
No obvious non-physical oscillations are observed, even in the region immediately adjacent to the cylinder.
The comparison of wall quantities with DSMC and a body-conformal DVM \cite{yang_efficient_2024} is provided in \Cref{fig:cylinder quantities}. 
The simulation results exhibit excellent consistency with the reference solution.

\begin{figure}[htbp!]
    \centering
    \includegraphics[width=0.6\textwidth]{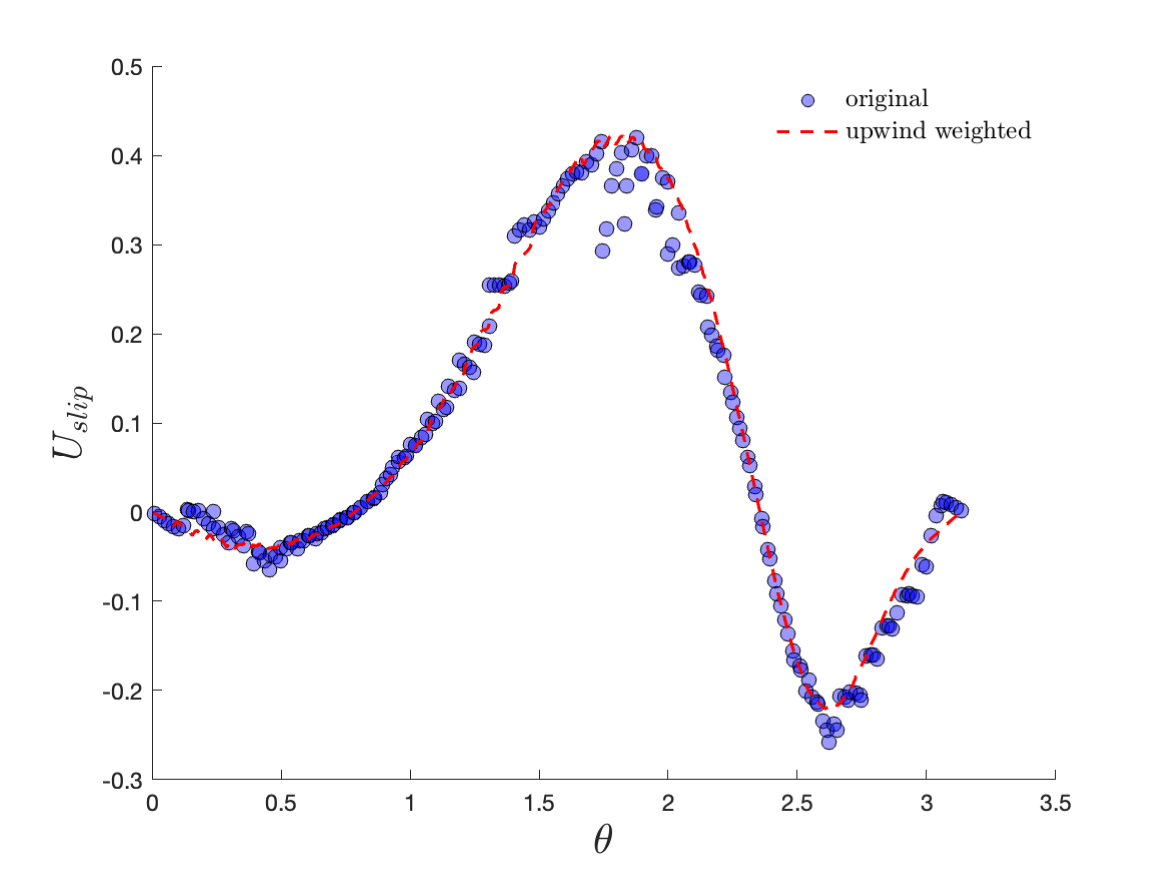}
    \caption{The comparison of slip velocity along the cylinder surface with and without the upwind-weighted interpolation method using the same physical mesh in the hypersonic flow around the cylinder.
    }
    \label{fig:cylinder slip}
\end{figure}

The slipping velocities along the wall surface obtained by the current method with and without the upwind weighting strategy in \Cref{eq:upwind} are shown in \Cref{fig:cylinder slip}. 
It can be seen that the results without upwind weighted average exhibit severe numerical oscillations, especially at the top of the cylinder where a longer horizontal mesh trail is used to represent the boundary. Such trail introduces prominent spurious geometric effects. 
To address this issue, the upwind weighted average utilizes the information from both $x$ and $y$ directions.
This can be understood as the reasonable incorporation of artificial viscosity, which substantially improve the numerical stability.

\begin{figure}[htbp!]
    \centering
    \begin{subfigure}[t]{0.44\textwidth}
        \includegraphics[width=\textwidth]{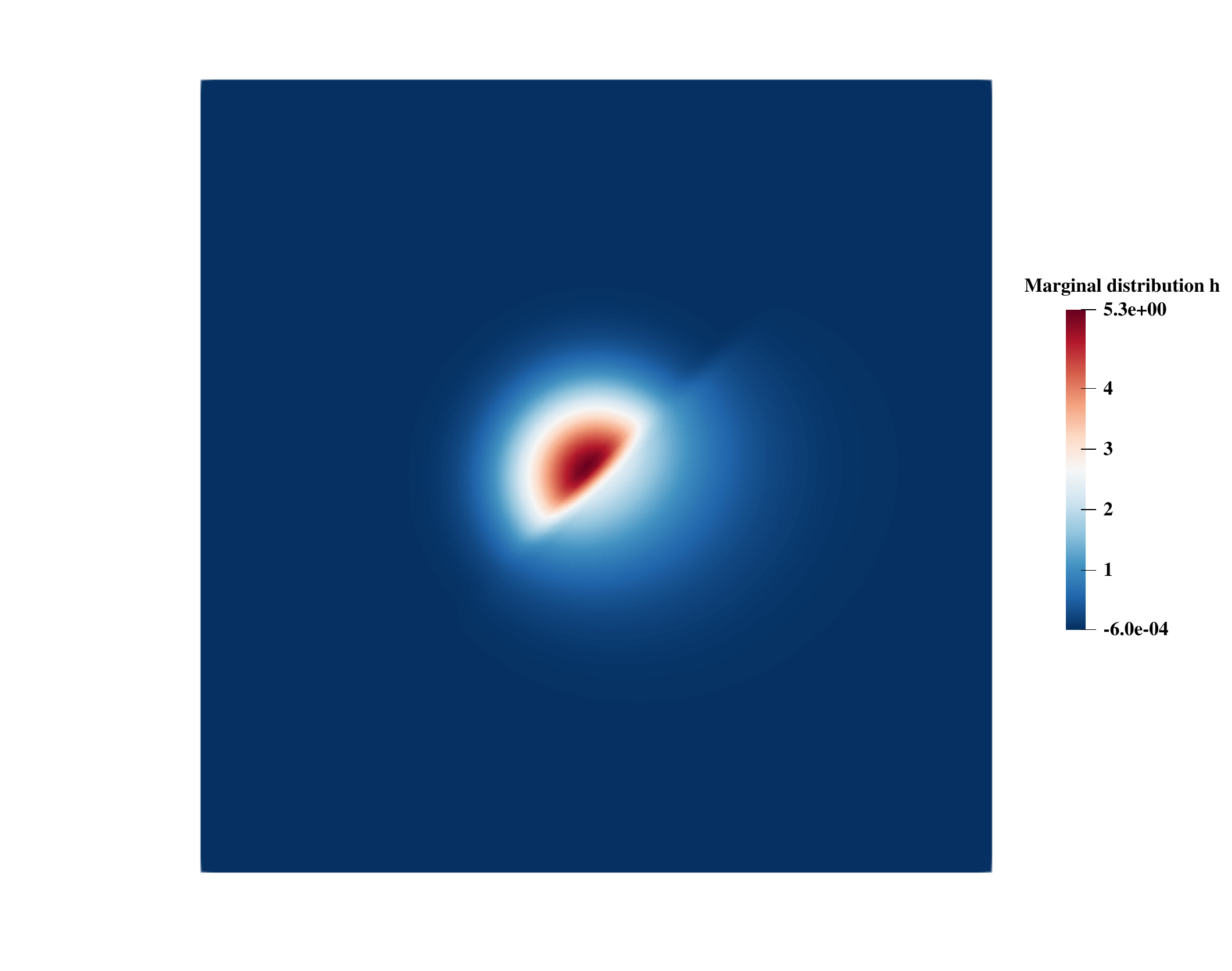}
        \caption{Two-dimensional view}
        \label{fig:cylinder velocity space 2D}
    \end{subfigure}
    \hfil
    \begin{subfigure}[t]{0.45\textwidth}
        \includegraphics[width=\textwidth]{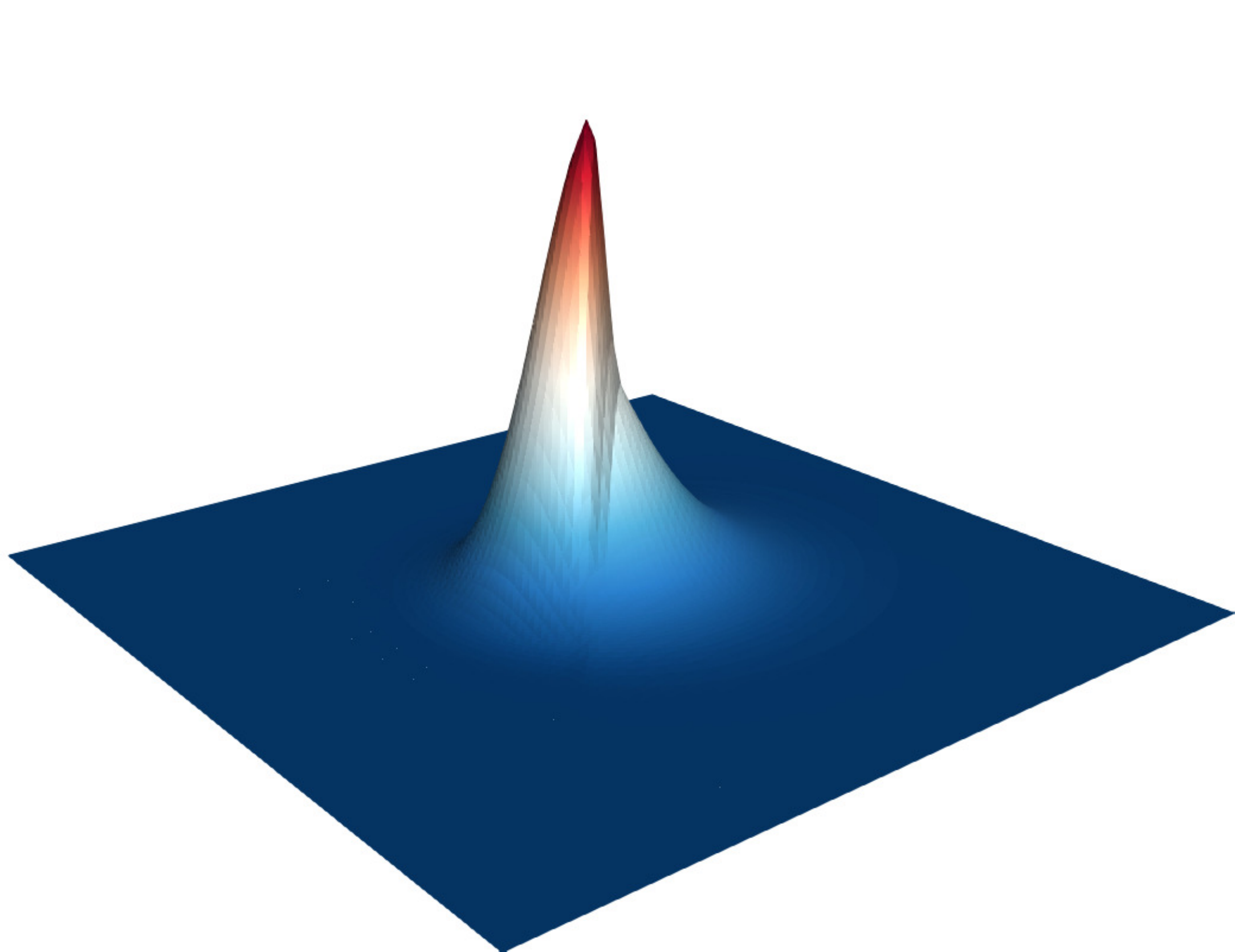}
        \caption{Three-dimensional view}
        \label{fig:cylinder velocity space 3D}
    \end{subfigure}
    \caption{The particle distribution function in velocity space at $\theta=\frac{\pi}{4}$ on the cylindrical surface.}
    \label{fig:cylinder velocity space}
\end{figure}

To elucidate the nature of the Maxwell gas-surface interaction model, the distribution function in the velocity space at the wall boundary is shown in \Cref{fig:cylinder velocity space}.
The three-dimensional plot in \Cref{fig:cylinder velocity space 3D} clearly explains the discontinuity in the velocity space, which results in a reduction in order of accuracy and necessitates the cut-cell correction.

\subsection{Supersonic flow around NACA0012 airfoil}

The next numerical experiment is the flow around a NACA0012 airfoil \cite{fan_computation_2001}. The chord length serves as the characteristic length, i.e., the dimensionless chord length $L_c=1.0$. The Mach number and the Knudsen number are $2.0$ and $0.026$, respectively. The density, velocity and temperature are respectively nondimensionalized by the incoming flow. The dimensionless temperature of the airfoil surface $T_{\rm w}$ is fixed at $1.0$. 
The angle of attack is $0$.
The physical domain and the mesh size used for resolving the boundary flows follows the settings in \cite{shoja-sani_investigation_2014}.
The domain is $[-3,7]\times[-7,7]$, where the leading edge of the airfoil is positioned at $(0,0)$.
The number of surface meshes in the refined region is around $600$.
A uniform Cartesian mesh with the resolution of $60\times 60$ in the range of $[-4,8]\times [-6,6]$ is employed in the velocity space.

\begin{figure}[htbp!]
    \centering
    \begin{subfigure}[t]{0.49\textwidth}
        \includegraphics[width=\textwidth]{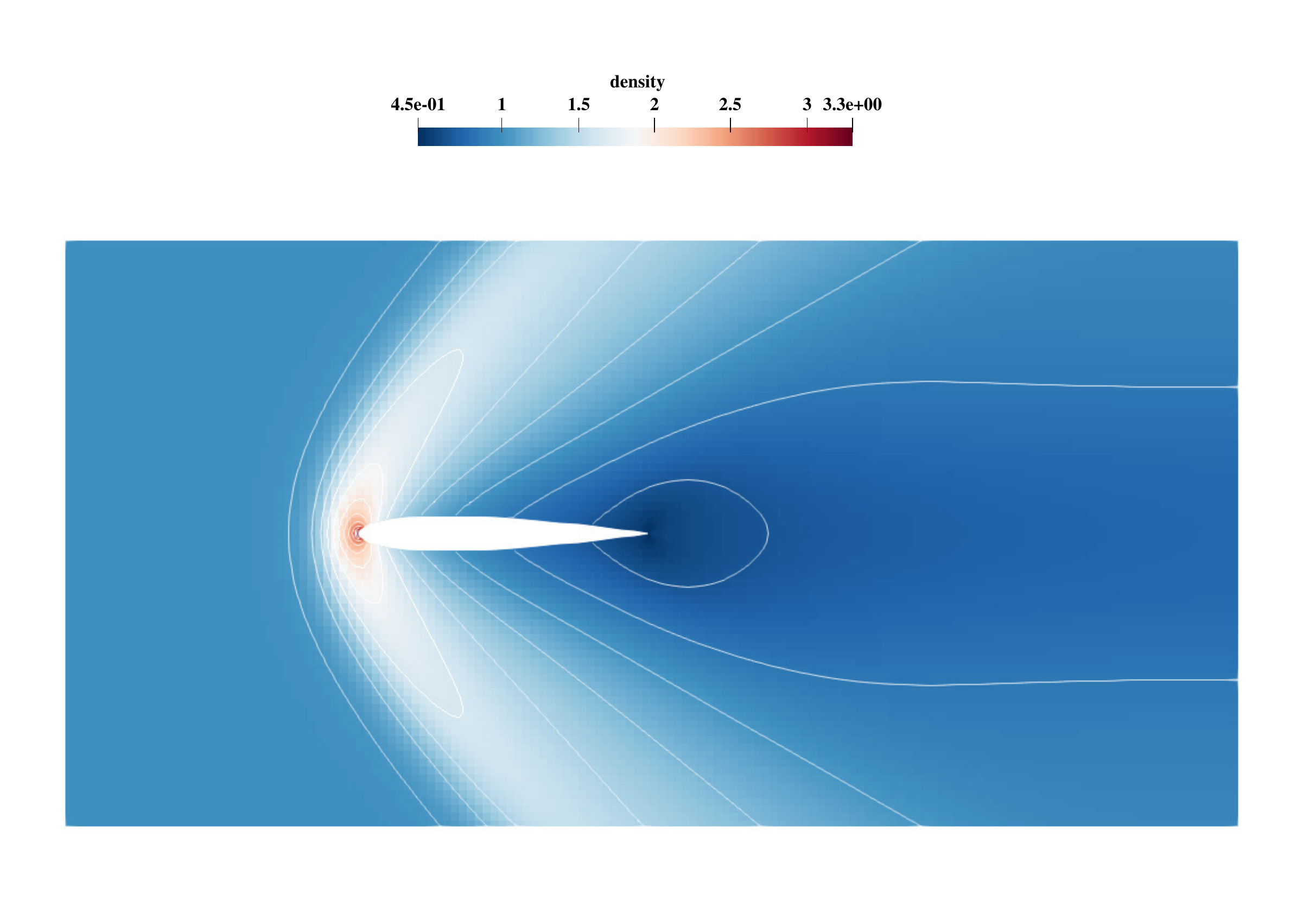}
        \caption{Density}
        \label{fig:airfoil density}
    \end{subfigure}
    \hfil
    \begin{subfigure}[t]{0.49\textwidth}
        \includegraphics[width=\textwidth]{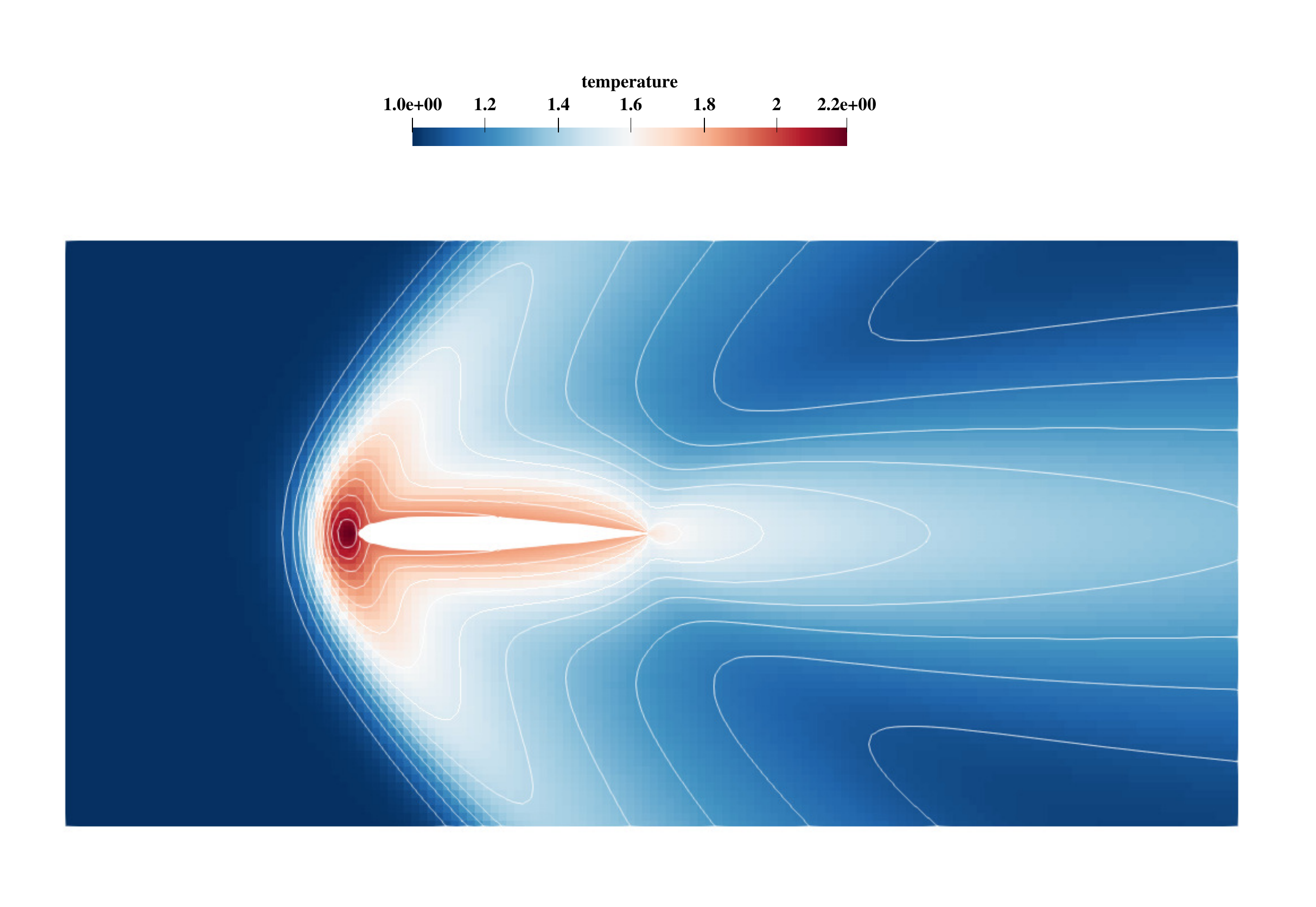}
        \caption{Temperature}
        \label{fig:airfoil temperature}
    \end{subfigure}
    \\
    \begin{subfigure}[t]{0.49\textwidth}
        \includegraphics[width=\textwidth]{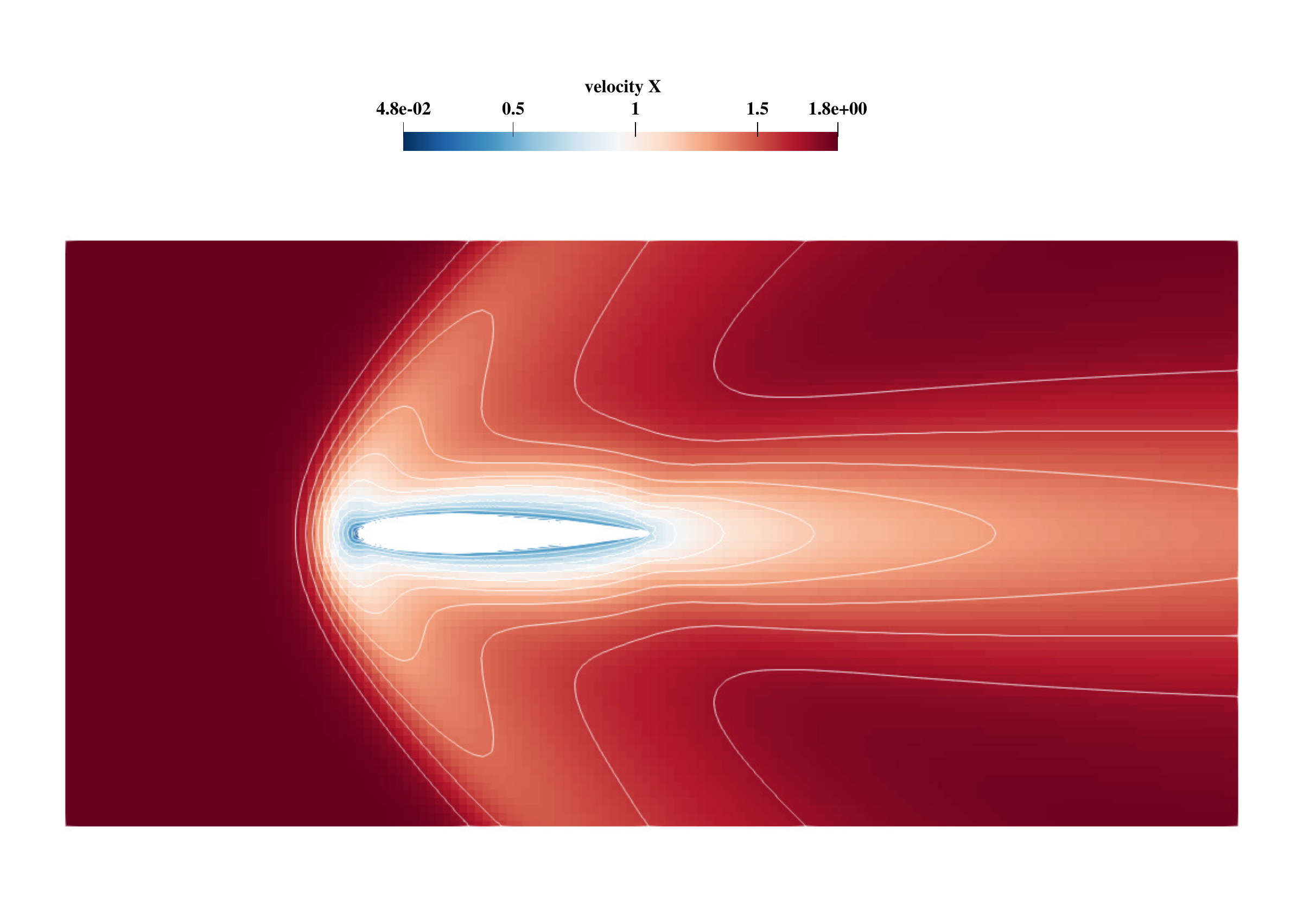}
        \caption{X-component of velocity}
        \label{fig:airfoil U}
    \end{subfigure}
    \hfil
    \begin{subfigure}[t]{0.49\textwidth}
        \includegraphics[width=\textwidth]{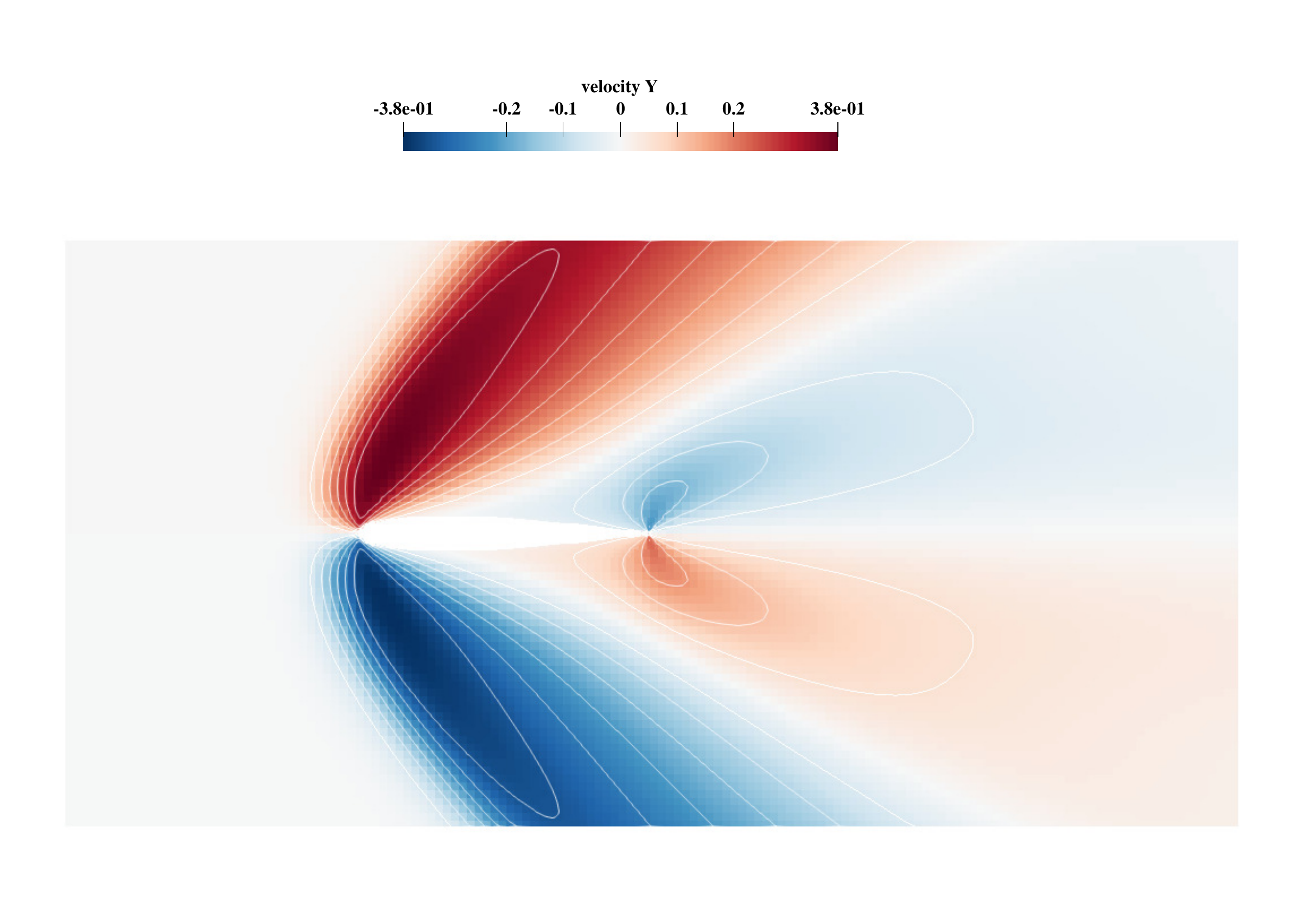}
        \caption{Y-component of velocity}
        \label{fig:airfoil V}
    \end{subfigure}
    \\
    \begin{subfigure}[t]{0.49\textwidth}
        \includegraphics[width=\textwidth]{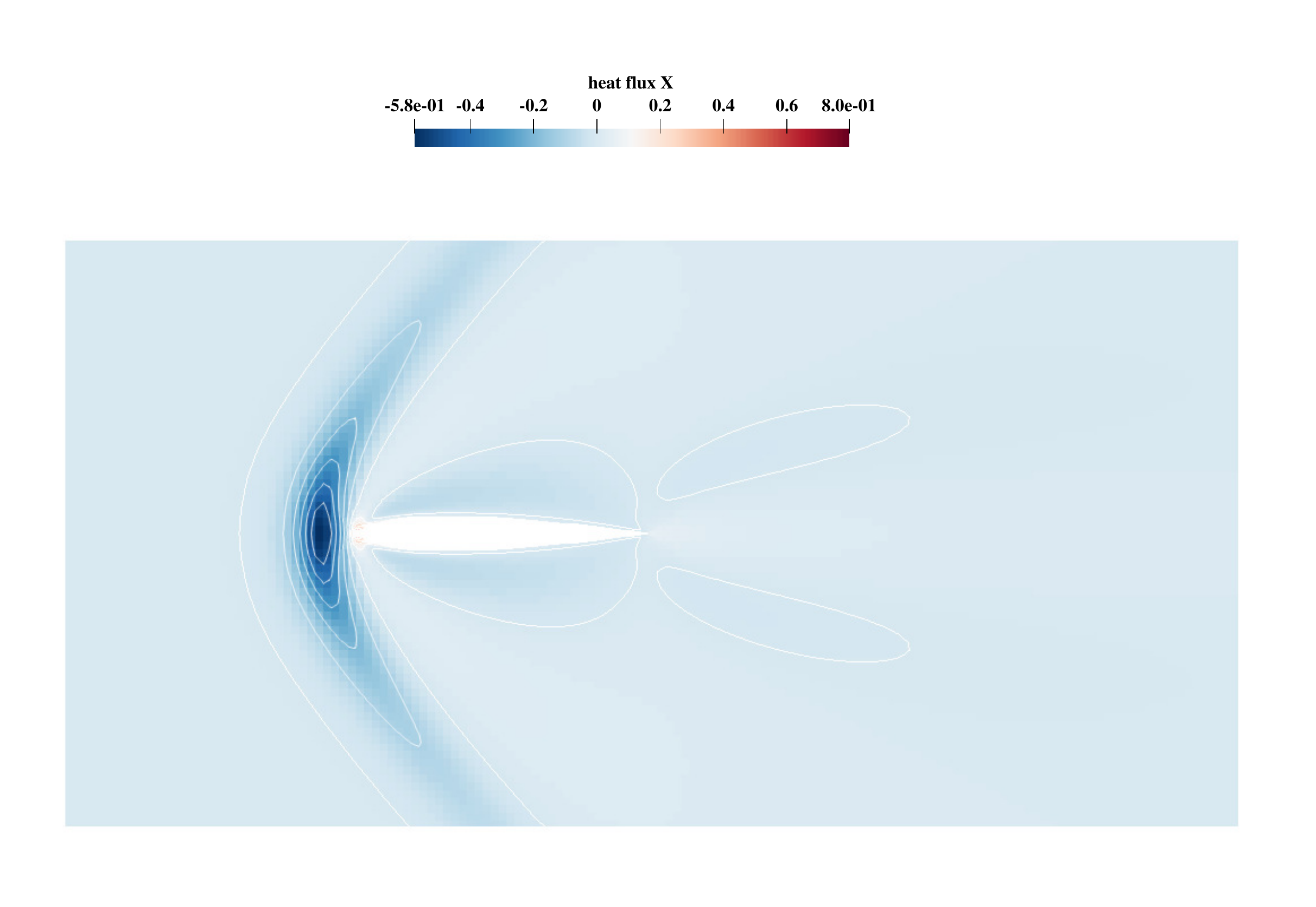}
        \caption{X-component of heat flux}
        \label{fig:airfoil qfx}
    \end{subfigure}
    \hfil
    \begin{subfigure}[t]{0.49\textwidth}
        \includegraphics[width=\textwidth]{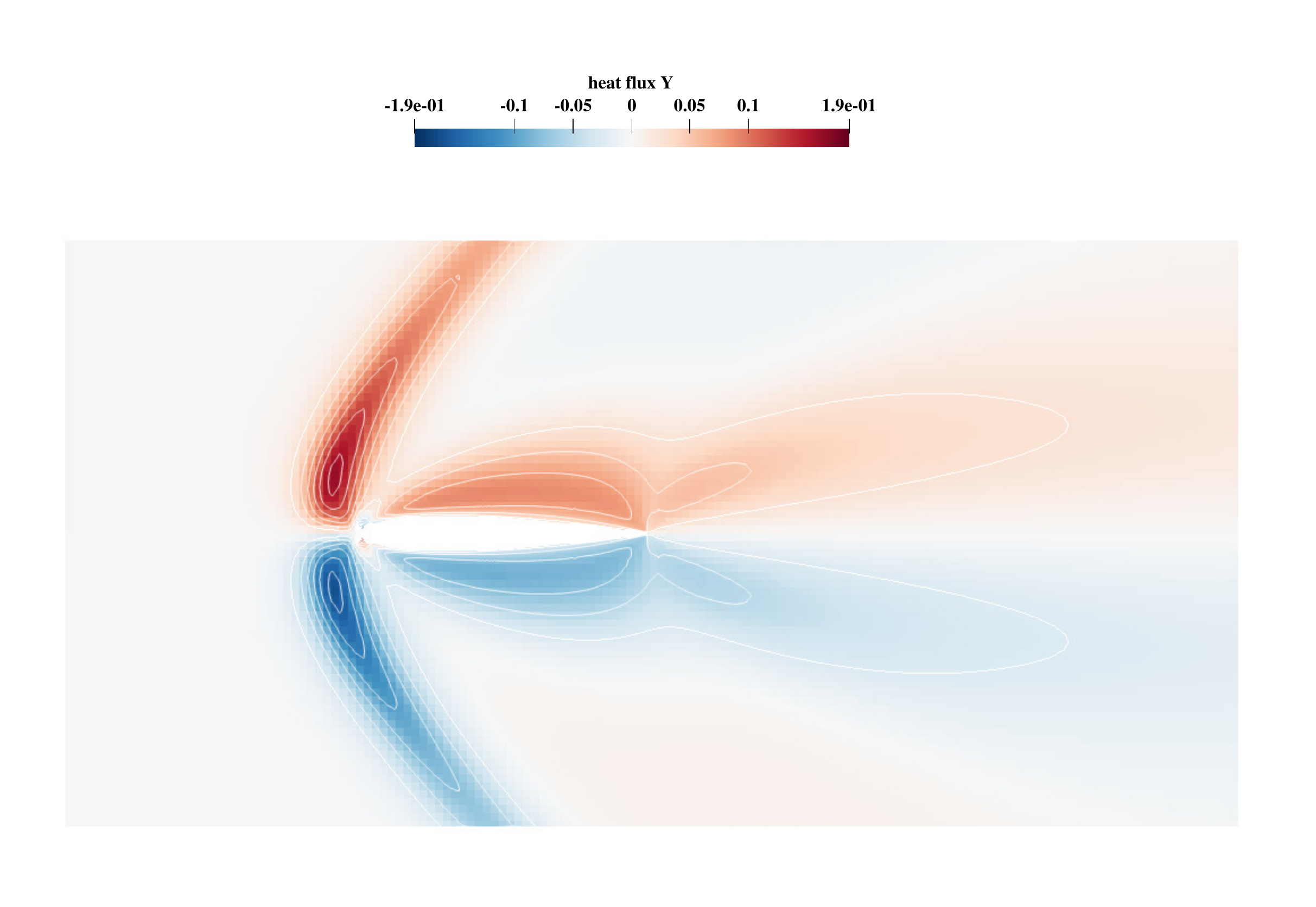}
        \caption{Y-component of heat flux}
        \label{fig:airfoil qfy}
    \end{subfigure}
    \caption{The converged contours of macroscopic flow variables in the supersonic flow around the NACA0012 airfoil.}
    \label{fig:airfoil field}
\end{figure}
\begin{figure}[htbp!]
    \centering
    \begin{subfigure}[t]{0.49\textwidth}
        \includegraphics[width=\textwidth]{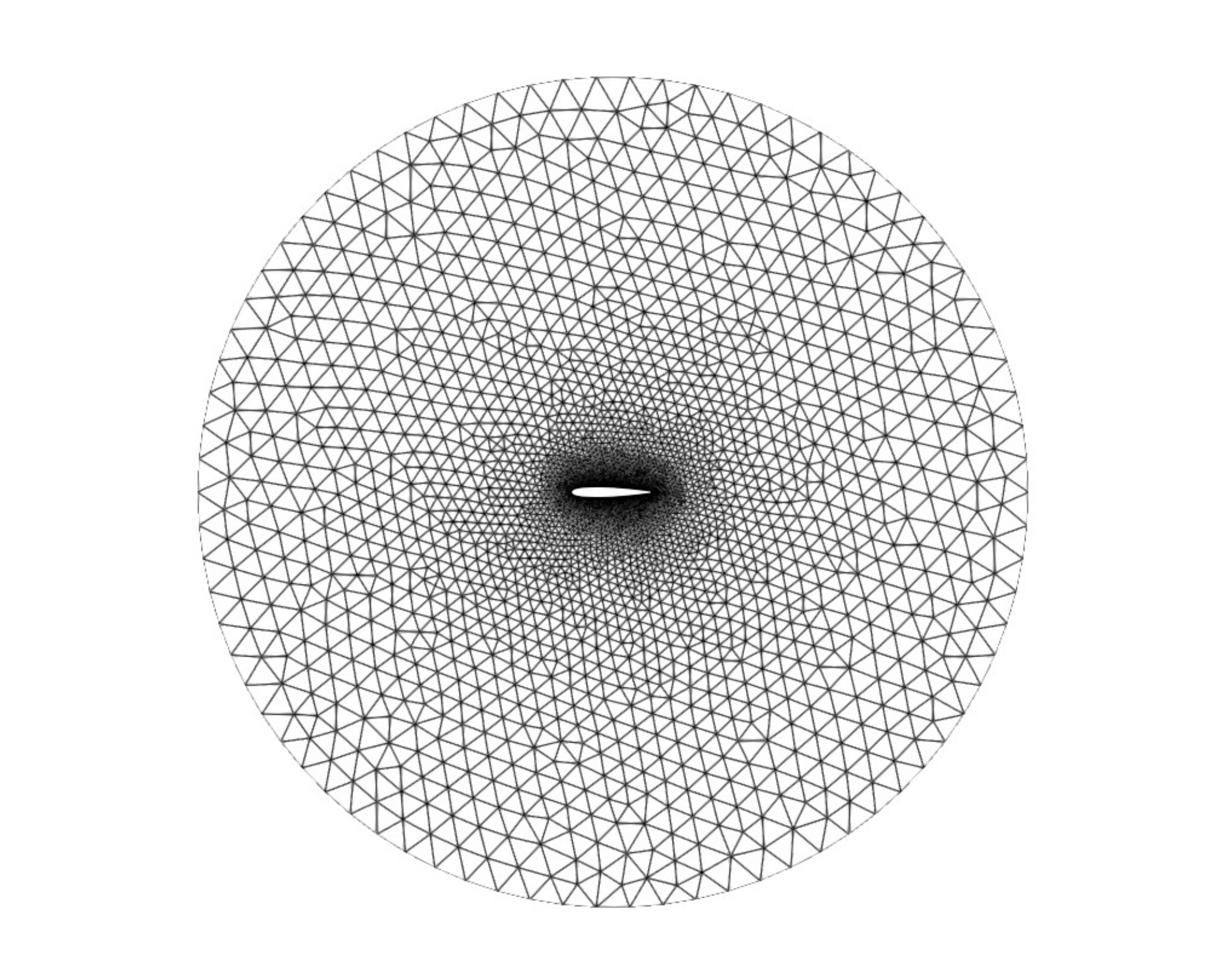}
        \caption{Unstructured mesh}
        \label{fig:airfoil unstructured grid}
    \end{subfigure}
    \hfil
    \begin{subfigure}[t]{0.49\textwidth}
        \includegraphics[width=\textwidth]{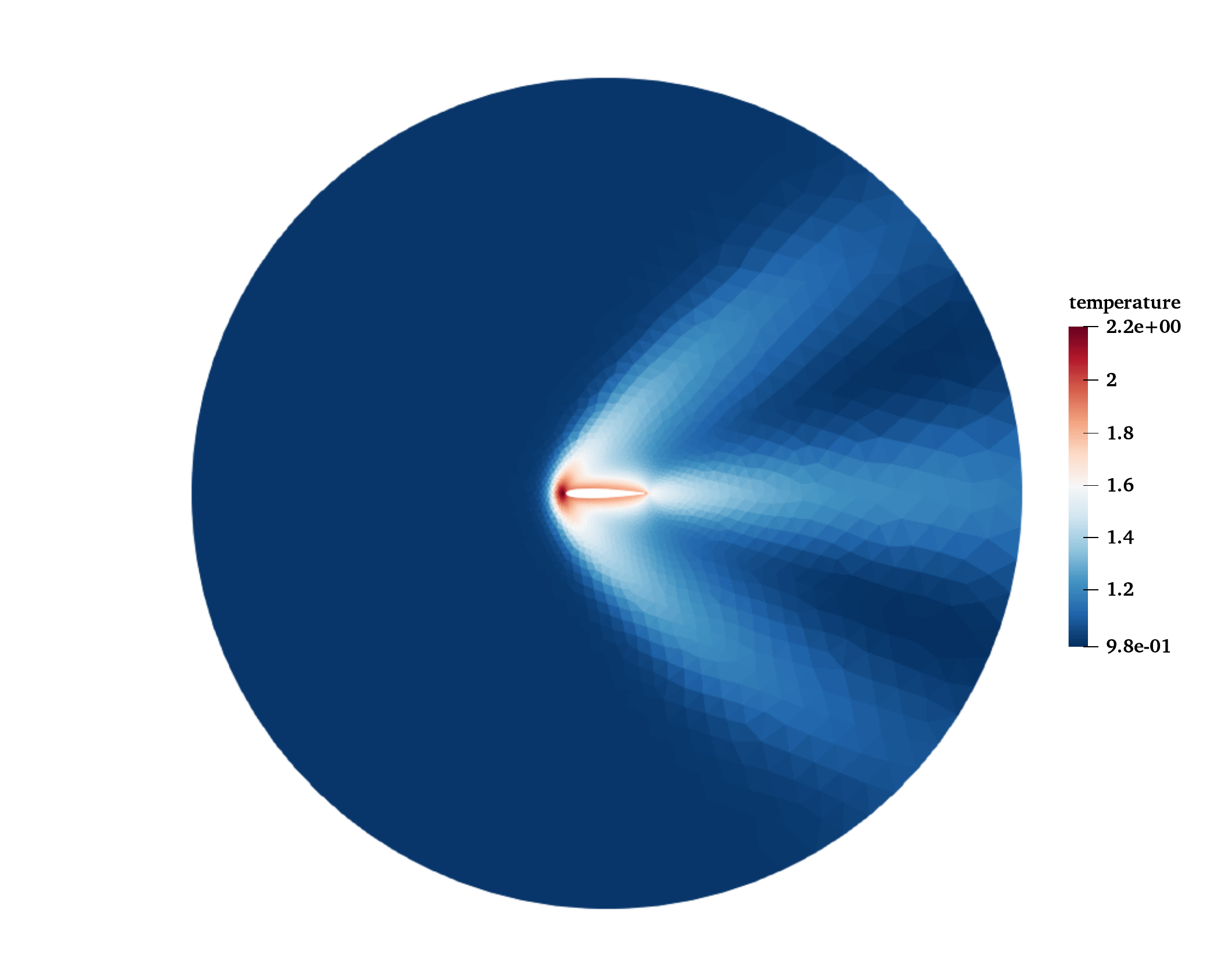}
        \caption{Temperature}
        \label{fig:airfoil unstructured temperature}
    \end{subfigure}
    \caption{The unstructured mesh and benchmark temperature contour used for validation in the supersonic flow around the NACA0012 airfoil.}
    \label{fig:airfoil unstructured}
\end{figure}
\begin{figure}[htbp!]
    \centering
    \begin{subfigure}[t]{0.49\textwidth}
        \includegraphics[width=\textwidth]{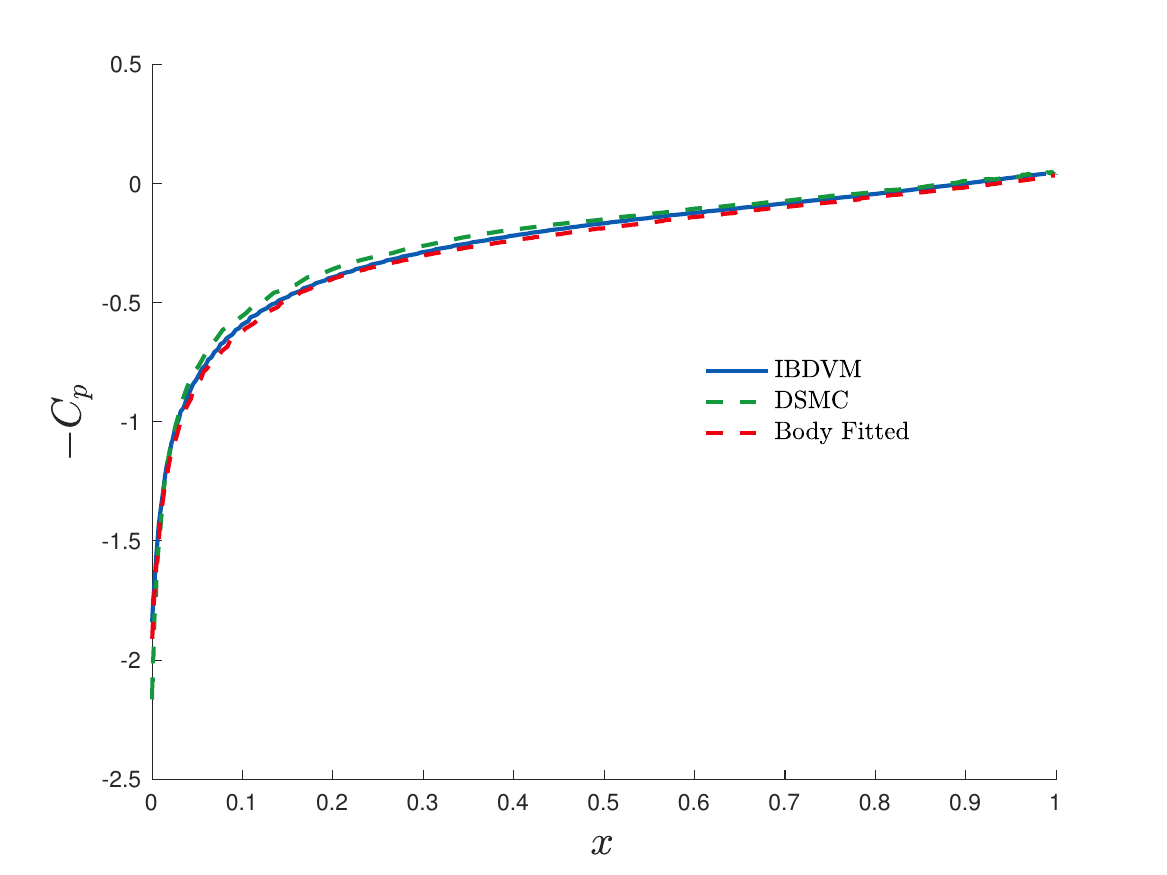}
        \caption{Pressure coefficient}
        \label{fig:airfoil Cp}
    \end{subfigure}
    \hfil
    \begin{subfigure}[t]{0.49\textwidth}
        \includegraphics[width=\textwidth]{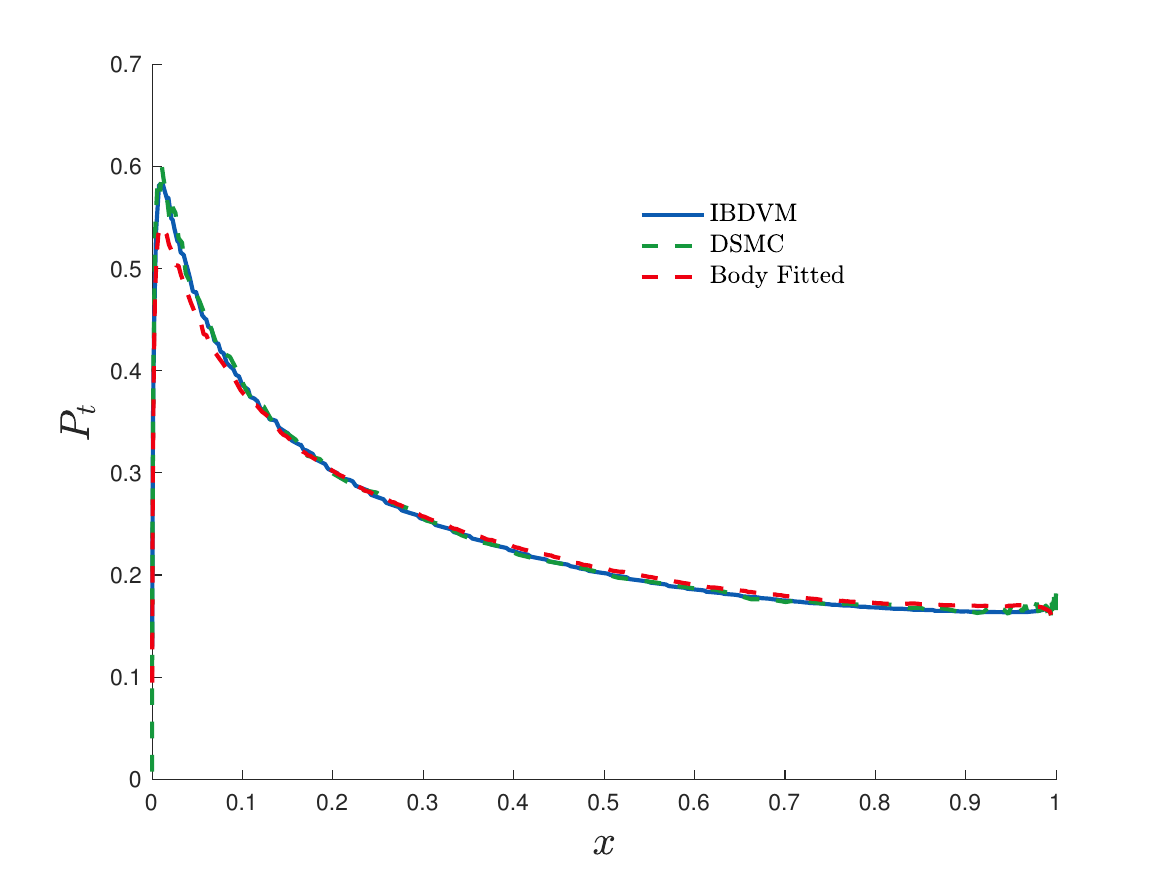}
        \caption{Shear stress}
        \label{fig:airfoil Pt}
    \end{subfigure}
    \\
    \begin{subfigure}[t]{0.49\textwidth}
        \includegraphics[width=\textwidth]{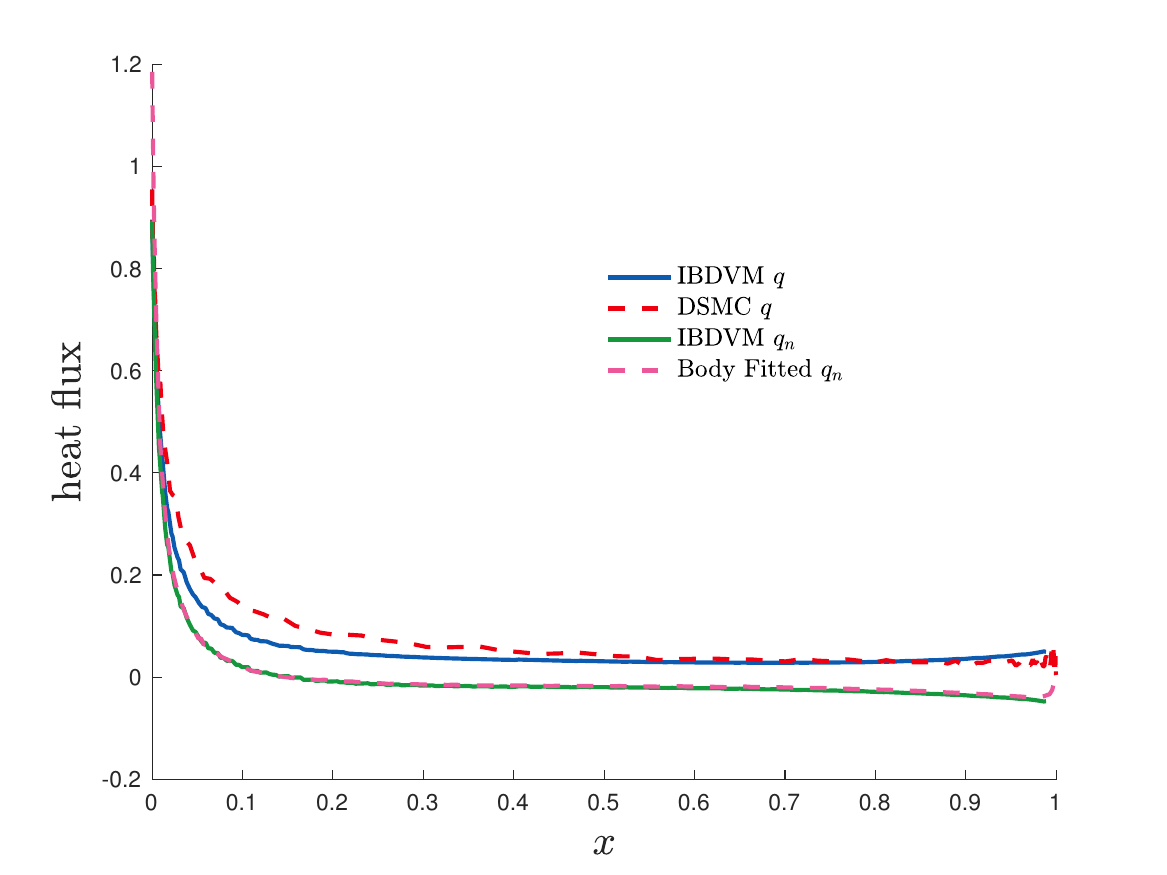}
        \caption{Heat flux}
        \label{fig:airfoil q}
    \end{subfigure}
    \caption{The comparison of macroscopic flow variables on the airfoil surface with the results simulated by DSMC and body-conformal DVM solvers in the supersonic flow around the NACA0012 airfoil.}
    \label{fig:airfoil boundary}
\end{figure}

The distribution of density, velocity, temperature, and heat flux are shown in \Cref{fig:airfoil field}. The numerical results show good agreement with those reported in \cite{fan_computation_2001}. 
To quantify the computational accuracy of wall variables, we compare the surface pressure coefficient defined as
\begin{equation}
    C_p=\frac{2(p_{\rm w}-p_0)}{\rho_0U_0^2}.
\end{equation}
The reference solutions are computed using a body-conformal DVM solver \cite{xiao2020velocity,xiao2021kinetic} based on an unstructured mesh and the SPARTA DSMC simulator \cite{plimpton_direct_2019}.
The minimum mesh size in the unstructured mesh is identical to that for the Cartesian mesh, and the converged simulation results are shown in \Cref{fig:airfoil unstructured}, where the minimum mesh size is $0.003$.
The comparison of wall physical quantities is shown in \Cref{fig:airfoil boundary}, which confirms the consistency of results obtained through different methods.

\subsection{Supersonic flow around a sphere}

\begin{figure}[htbp!]
    \centering
    \begin{subfigure}[t]{0.49\textwidth}
        \includegraphics[width=\textwidth]{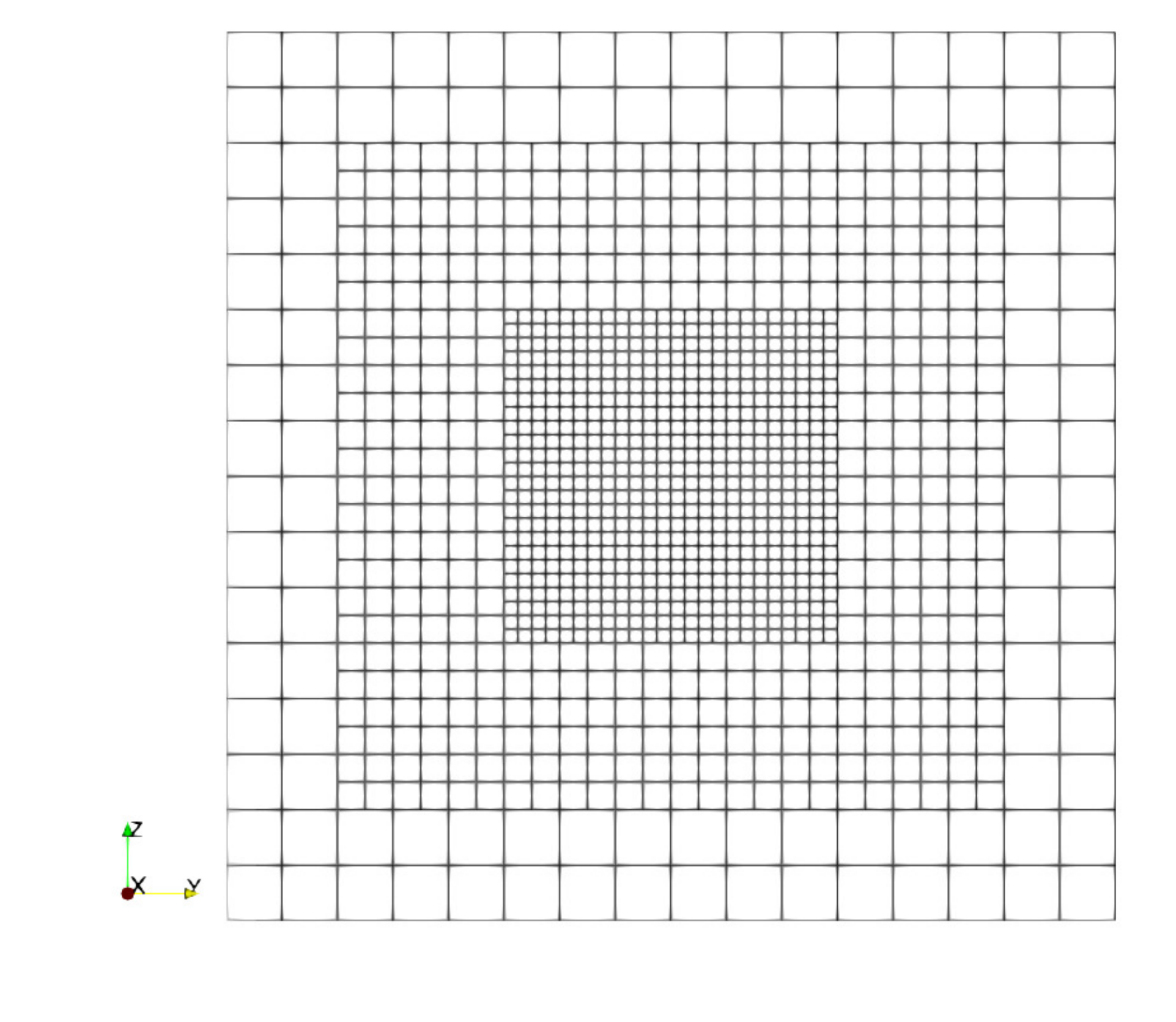}
        \caption{Cross-section of the velocity space mesh at $x=0$}
        \label{fig:sphere vmesh x}
    \end{subfigure}
    \hfil
    \begin{subfigure}[t]{0.49\textwidth}
        \includegraphics[width=\textwidth]{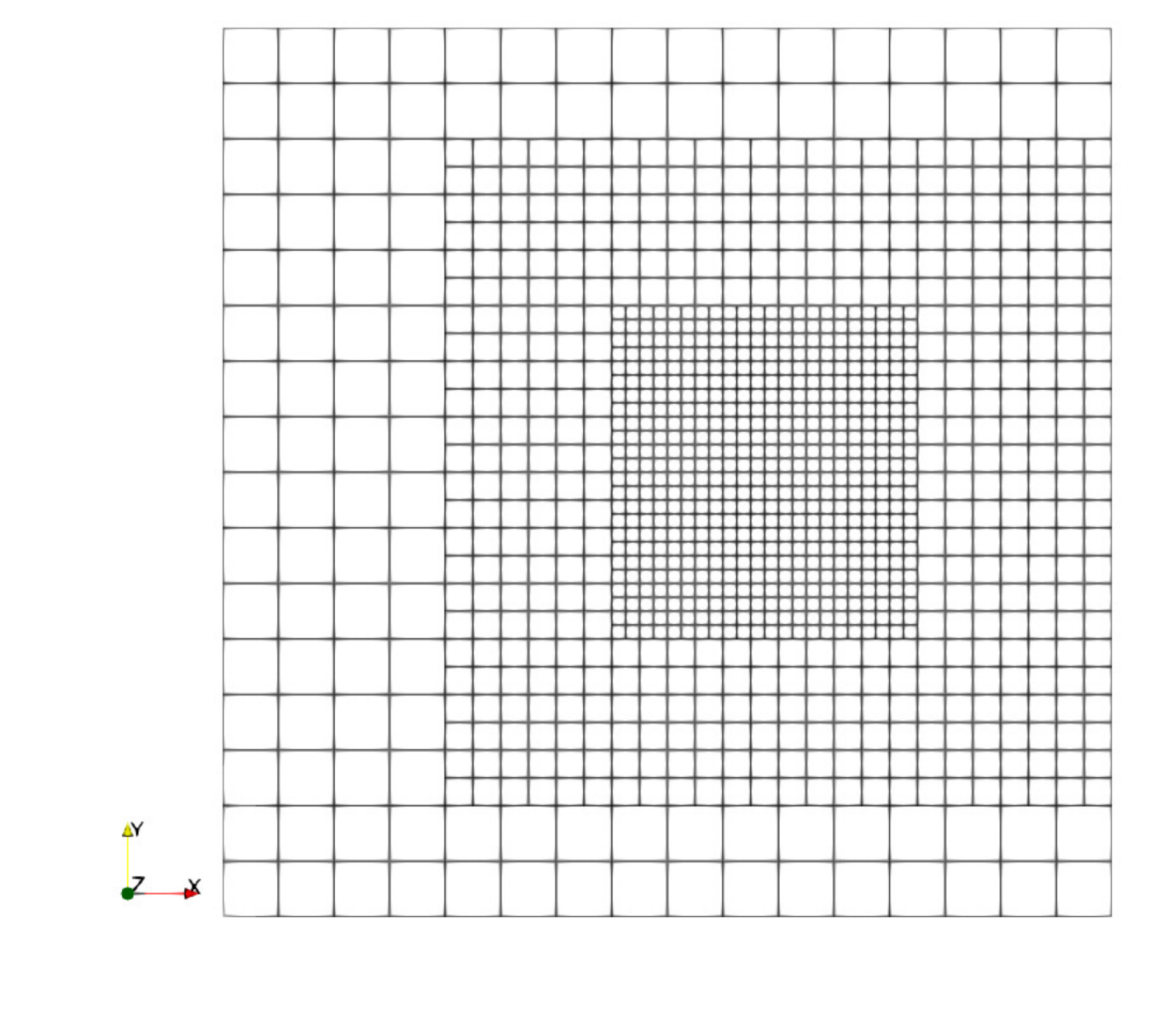}
        \caption{Cross-section of the velocity space mesh at $z=0$}
        \label{fig:sphere vmesh z}
    \end{subfigure}
    \caption{The velocity space mesh used in the supersonic flow around the sphere.}
    \label{fig:sphere vmesh}
\end{figure}

The last numerical experiment aims to validate the performance of the solution algorithm in three-dimensional problems.
In this case, supersonic flow around a sphere with the Mach number $\rm Ma=3.834$ and Knudsen number $\rm Kn=0.03$ is simulated. The diameter of the sphere serves as the characteristic length. The reference density and temperature are from the incoming flow. 
The temperature at the sphere is fixed at $T_\mathrm{w}=[1+(\gamma-1)Ma^2/2]T_0$, where $T_0$ is the reference temperature. 
The physical domain is set as $[-4,4]^3$, and the minimum interval of the refined mesh is $\Delta x=0.02$. To further reduce the computational cost, a locally refined Cartesian mesh is employed in the truncated velocity space $[-3\sqrt{T_\mathrm{w}},3\sqrt{T_\mathrm{w}}]^3$, which is shown in \Cref{fig:sphere vmesh}, and the minimum mesh size is $\Delta \xi=0.2275$.

\begin{figure}[htbp!]
    \centering
    \begin{subfigure}[t]{0.49\textwidth}
        \includegraphics[width=\textwidth]{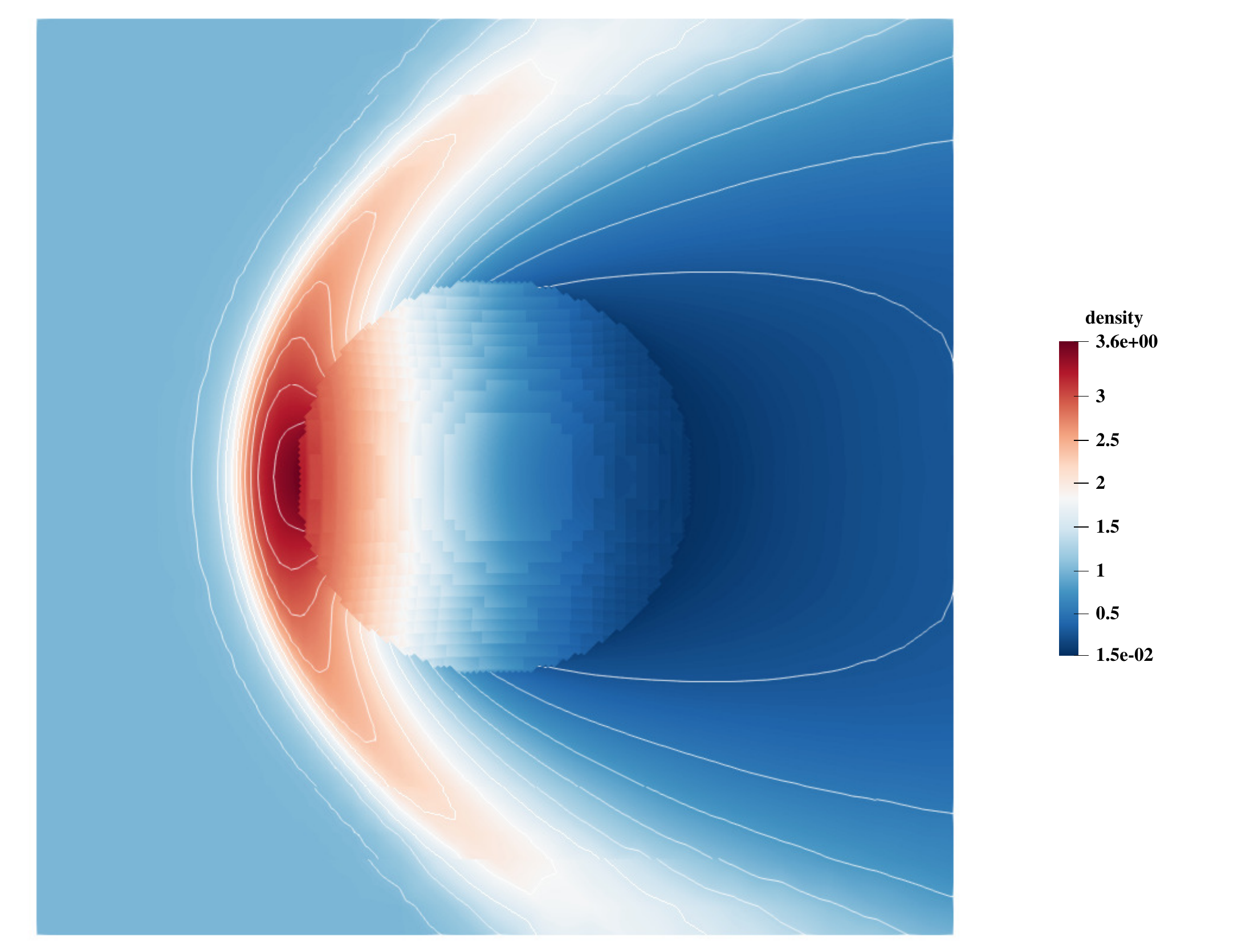}
        \caption{Density}
        \label{fig:sphere density}
    \end{subfigure}
    \hfil
    \begin{subfigure}[t]{0.49\textwidth}
        \includegraphics[width=\textwidth]{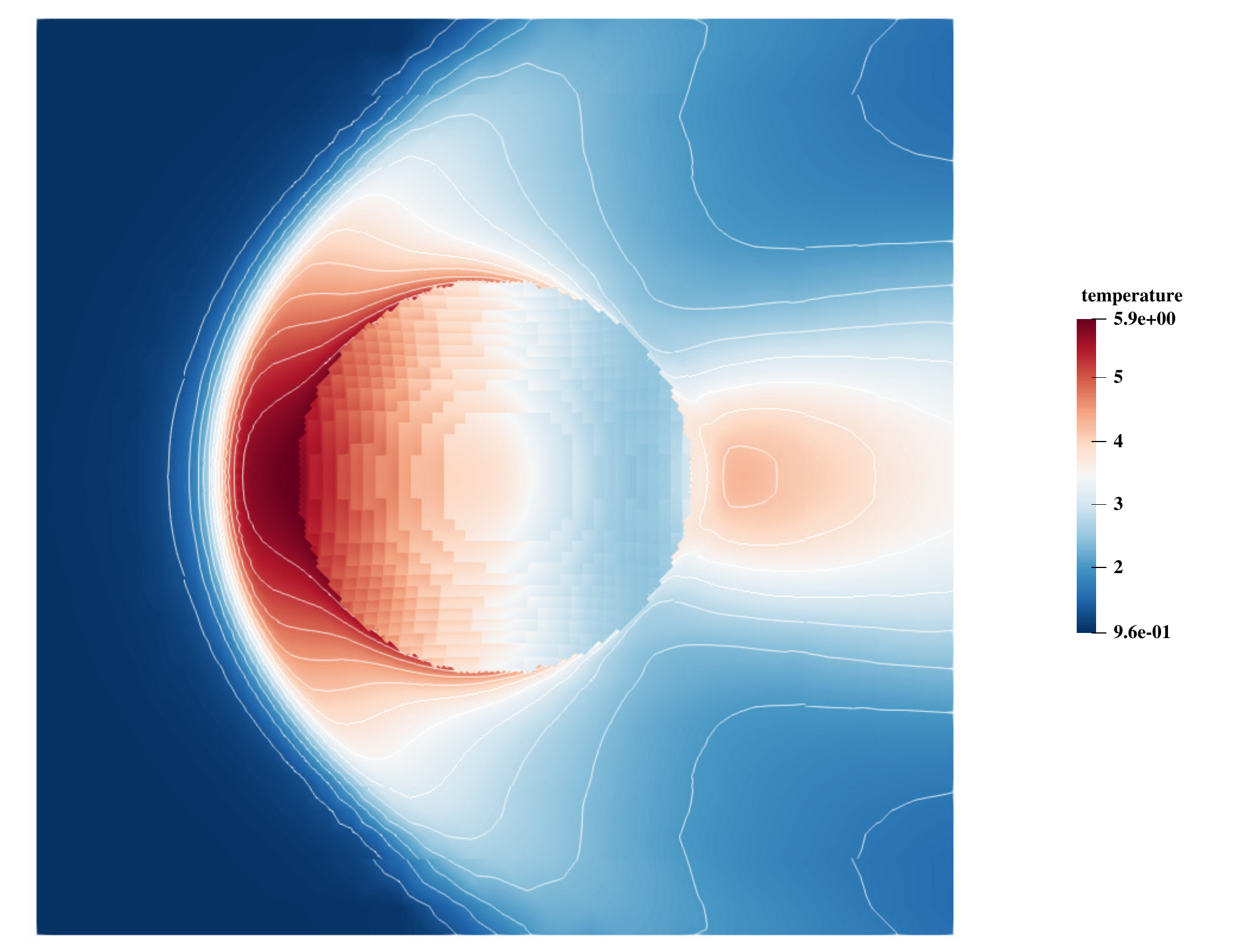}
        \caption{Temperature}
        \label{fig:sphere temperature}
    \end{subfigure}
    \\
    \begin{subfigure}[t]{0.49\textwidth}
        \includegraphics[width=\textwidth]{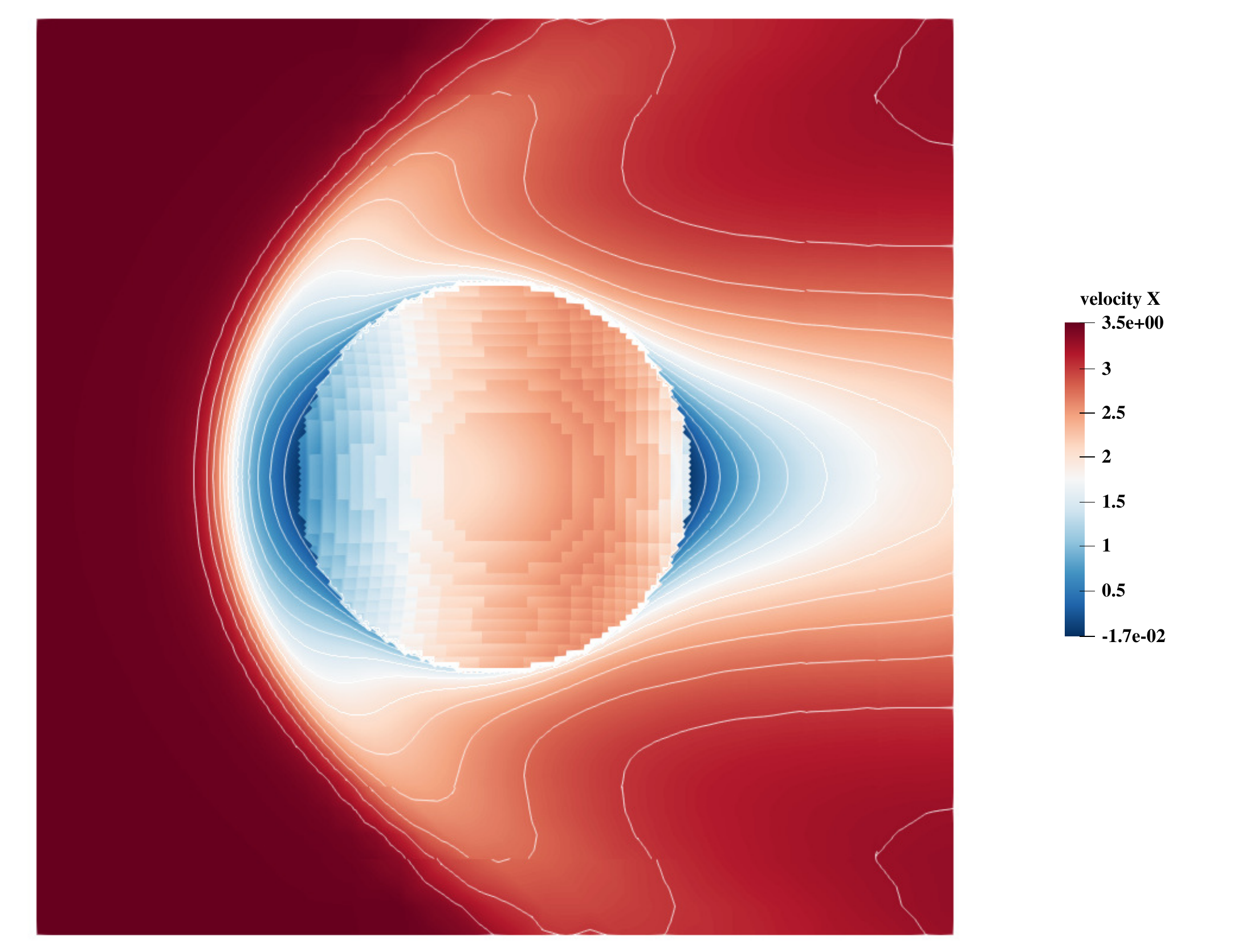}
        \caption{X-component of velocity}
        \label{fig:sphere U}
    \end{subfigure}
    \hfil
    \begin{subfigure}[t]{0.49\textwidth}
        \includegraphics[width=\textwidth]{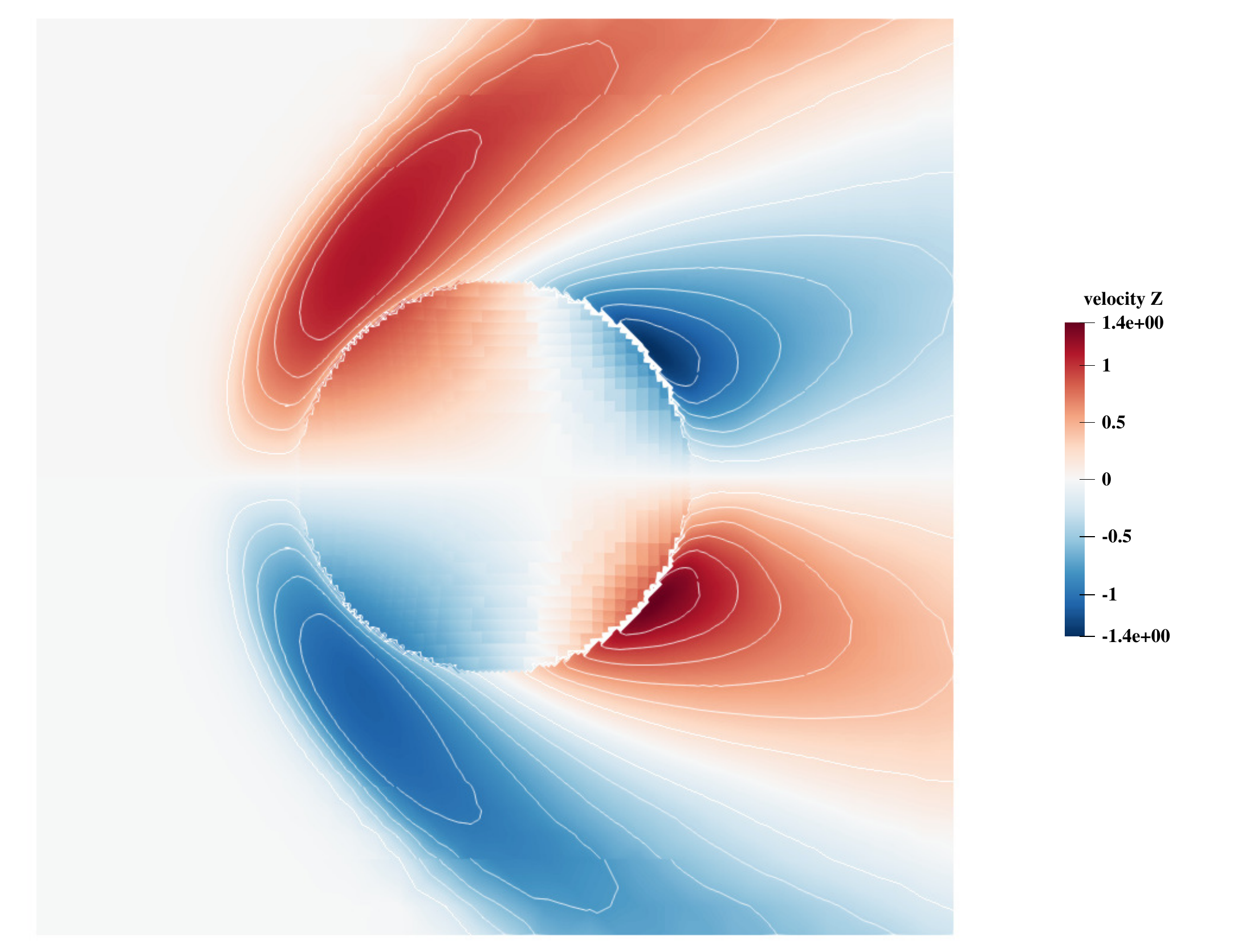}
        \caption{Z-component of velocity}
        \label{fig:sphere W}
    \end{subfigure}
    \caption{The converged contours of macroscopic flow variables in the supersonic flow around the sphere.}
    \label{fig:sphere}
\end{figure}

\begin{figure}
    \centering
    \begin{subfigure}[t]{0.49\textwidth}
        \includegraphics[width=\textwidth]{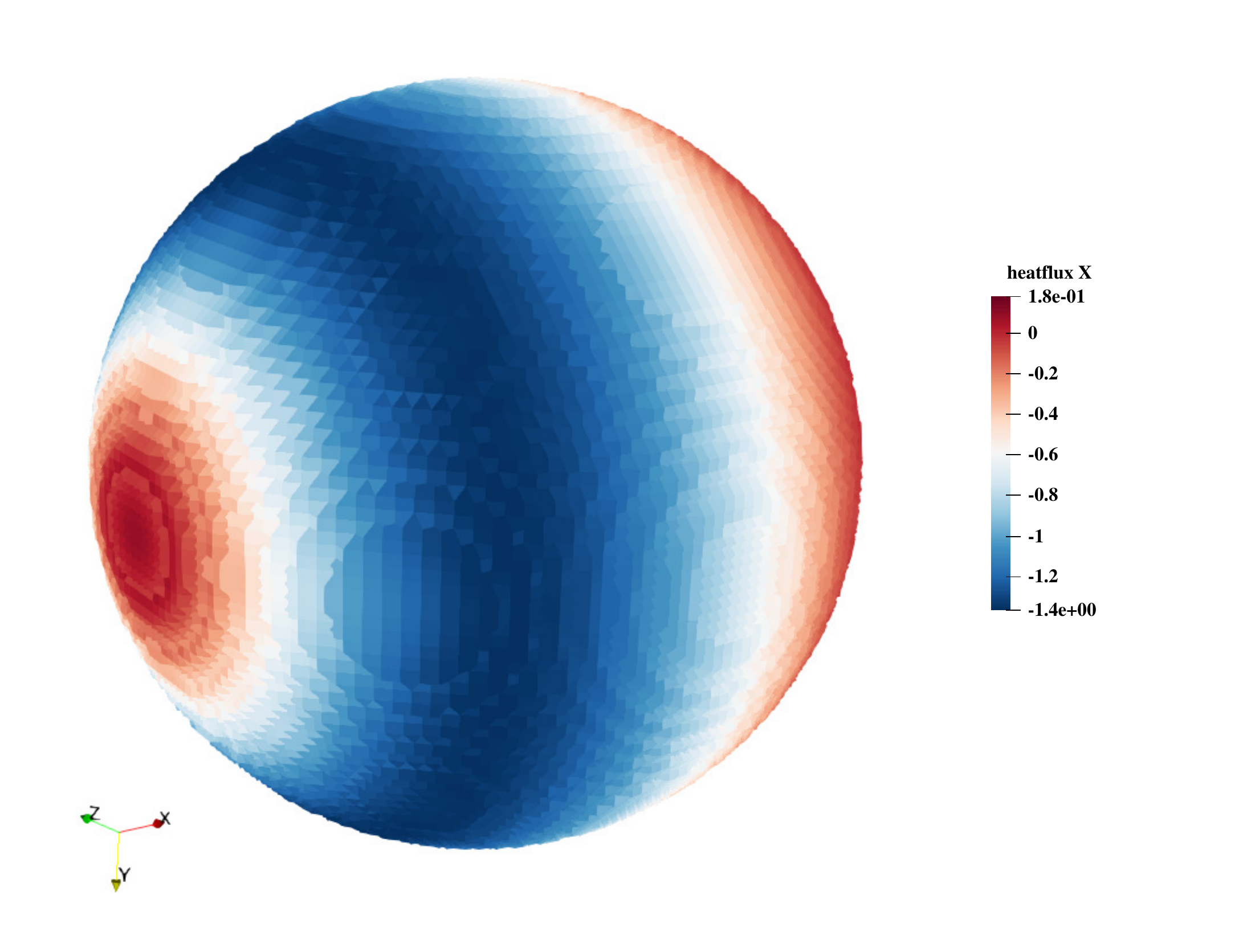}
        \caption{X-component of heat flux}
        \label{fig:surface qx}
    \end{subfigure}
    \hfil
    \begin{subfigure}[t]{0.49\textwidth}
        \includegraphics[width=\textwidth]{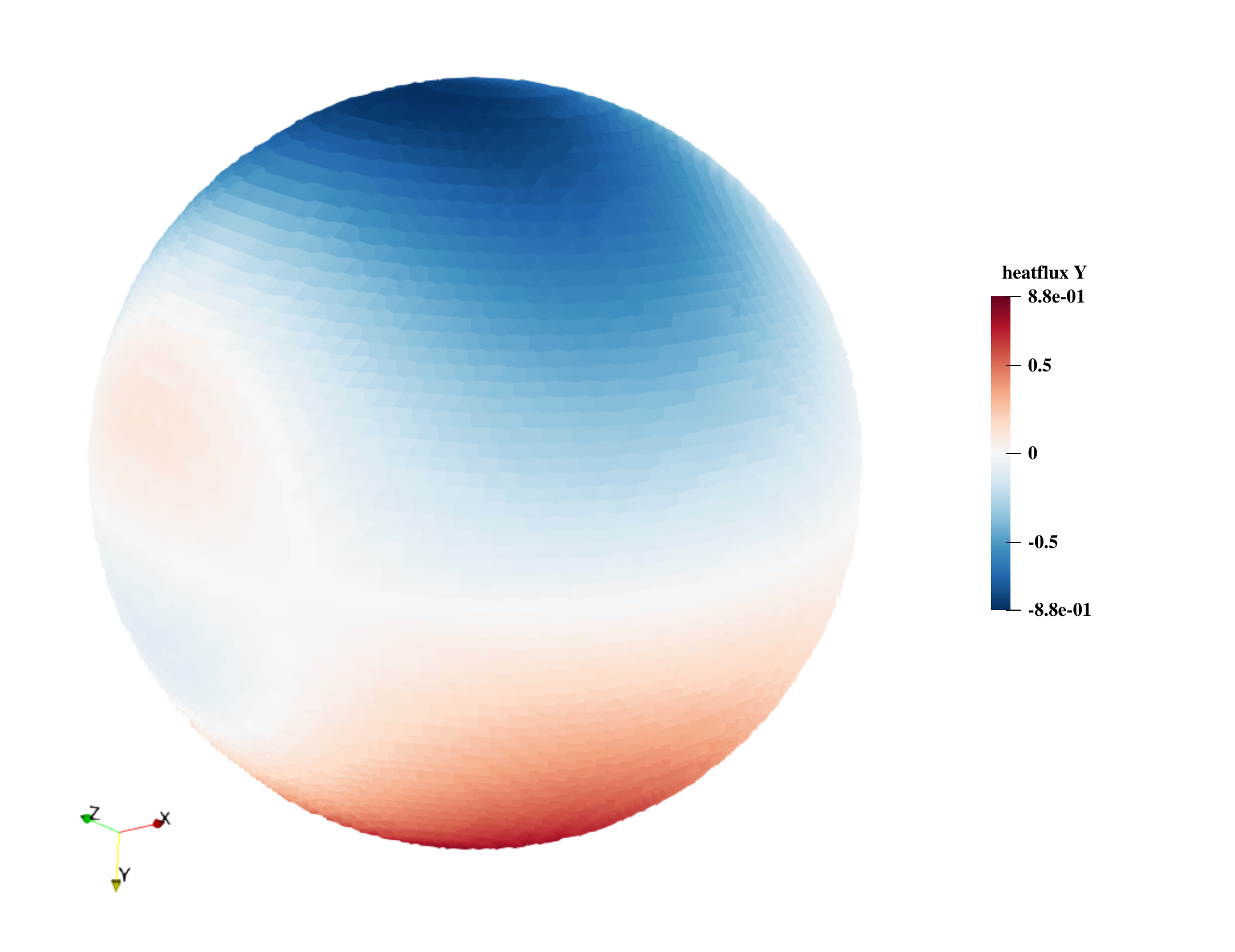}
        \caption{Y-component of heat flux}
        \label{fig:surface qy}
    \end{subfigure}
    \caption{The heat flux near the surface in the supersonic flow around the sphere.}
    \label{fig:surface}
\end{figure}

\begin{figure}[htbp!]
    \centering
    \begin{subfigure}[t]{0.49\textwidth}
        \includegraphics[width=\textwidth]{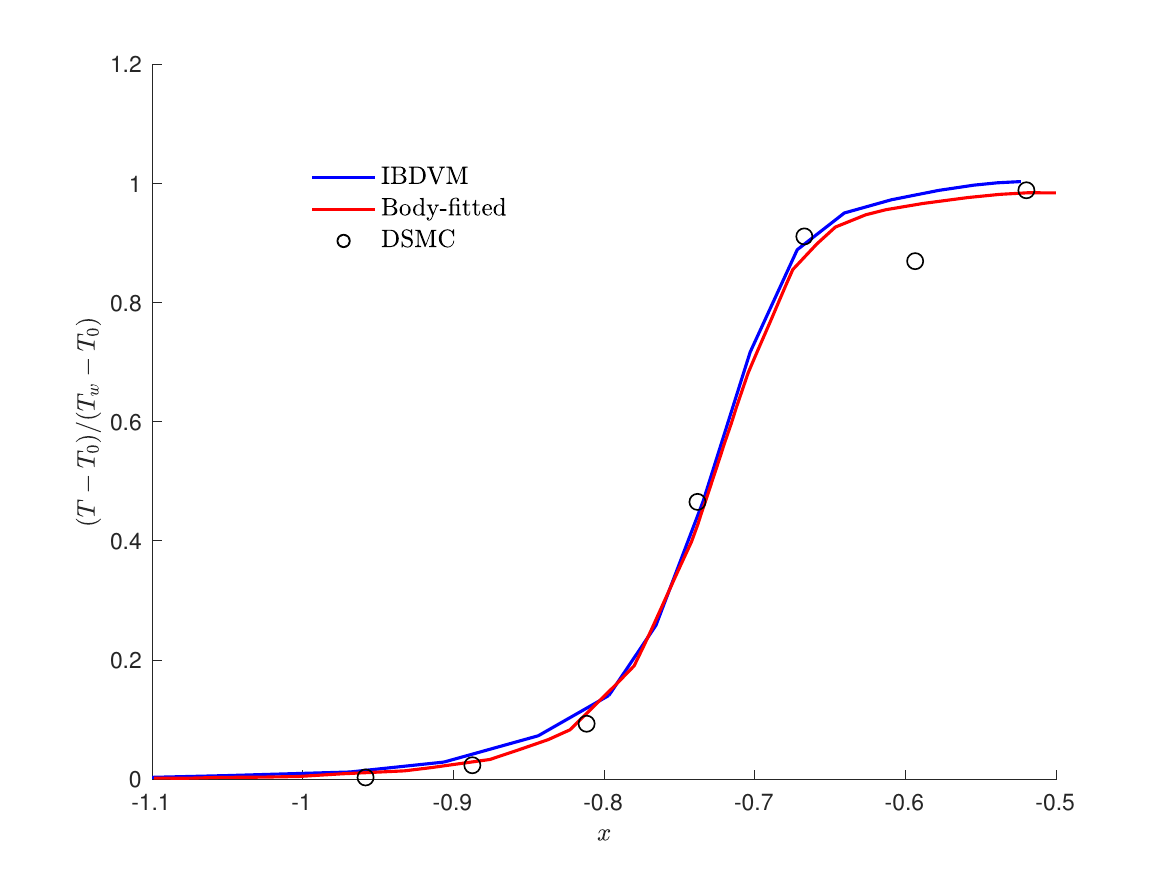}
        \caption{Temperature}
        \label{fig:sphere slT}
    \end{subfigure}
    \hfil
    \begin{subfigure}[t]{0.49\textwidth}
        \includegraphics[width=\textwidth]{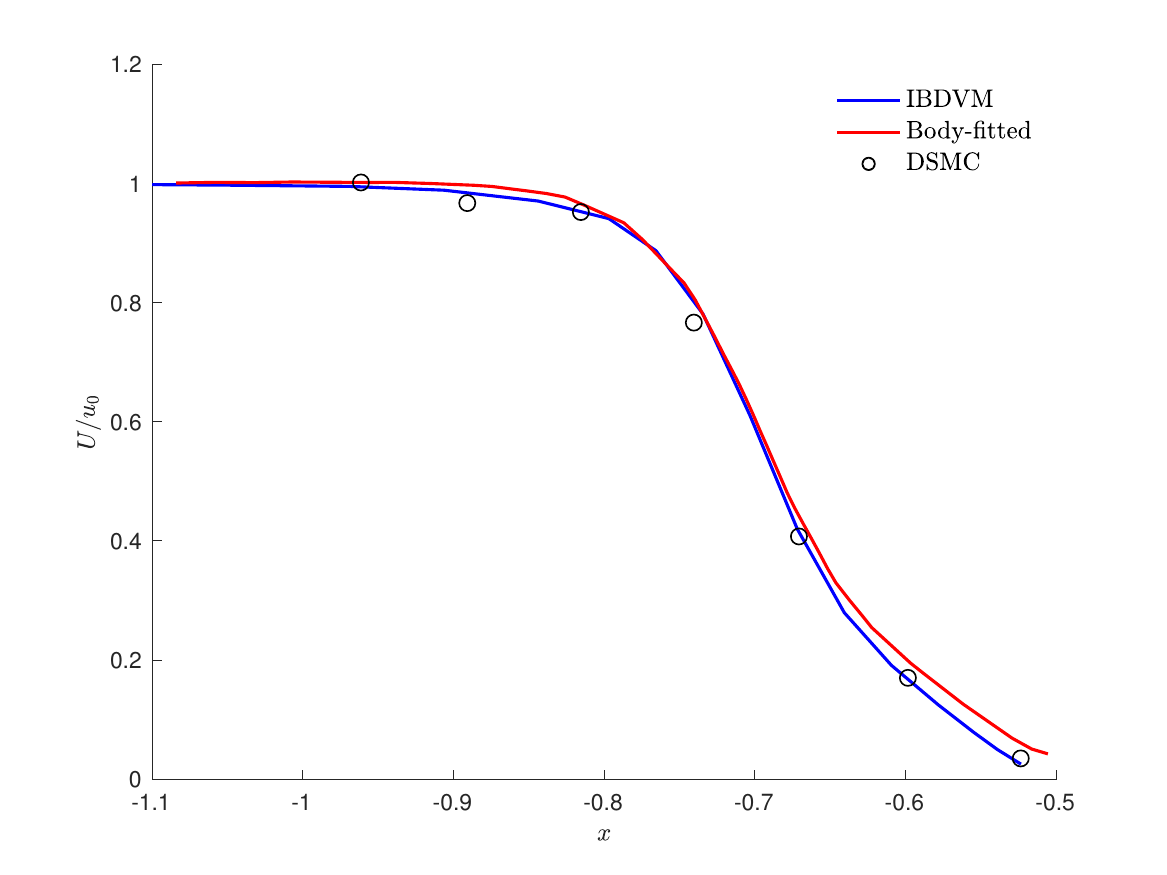}
        \caption{X-component of velocity}
        \label{fig:sphere_slU}
    \end{subfigure}
    \caption{The distribution of macroscopic flow variables along the stagnation line in the supersonic flow around the sphere. The reference results of DSMC and the body-conformal DVM solver are extracted from \cite{yang_efficient_2024}.}
    \label{fig:sphere sl}
\end{figure}

The distributions of density, velocity, temperature, and heat flux near the wall surface can be found in \Cref{fig:sphere} and \Cref{fig:surface}.
\Cref{fig:sphere sl} provides a quantitative comparison with the results in \cite{yang_efficient_2024}.
The comparative study demonstrates satisfactory consistency and validates the current method for three-dimensional fluid-structure interactions.

\section{Conclusion}

Computational modeling and simulation of dynamic interactions between non-equilibrium flows and structural components serve as indispensable tools for advancing aerospace research.
In this paper, an immersed boundary method is developed for the discrete velocity model of the Boltzmann equation within the Cartesian grid framework.
The proposed approach employs a particle-velocity-based upwind compact interpolation strategy to improve numerical stability for arbitrary immersed boundaries, and incorporates a velocity-space cut-cell correction to ensure quadrature accuracy in the presence of discontinuities of particle distribution function. 
The method accurately characterizes velocity slip and temperature jump effects and achieves reliable prediction of aerodynamic parameters.
Owing to its dimension-independent formulation, the method can be equally applied to two- and three-dimensional problems.
Numerical results for a variety of benchmark problems demonstrate that the proposed solution algorithm achieves accuracy comparable to body-conformal solvers while retaining the simplicity, flexibility and scalability of the Cartesian grid method. 
This investigation confines its scope to fixed rigid boundaries.
Extensive study of moving and deformable structures will be considered in future work.

\section*{Acknowledgements}

The current research is funded by Strategic Priority Research Program of Chinese Academy
of Sciences (XDB0620403), Chinese Academy of Sciences Project for Young Scientists in Basic Research (YSBR107), National Science Foundation of China (12302381, 12572340), and Beijing Natural Science Foundation (L252039).
The computing resources provided by Hefei Advanced Computing Center are acknowledged.

\bibliographystyle{unsrt}
\bibliography{main}

\end{document}